\documentclass[seceq]{ptptex}
\usepackage{epsfig}

\def\Journal#1#2#3#4{{#1} {#2} (#4) #3 }

\def\NPA{{\em Nucl. Phys.} A}

\def\NPB{{\em Nucl. Phys.} B}

\def\PLB{{\em Phys. Lett.} B}

\def\PRL{\em Phys. Rev. Lett.}
\def\PREV{\em Phys. Rev.}
\def\PREP{\em Phys. Rep.}

\def\PRD{{\em Phys. Rev.} D}
\def\PRC{{\em Phys. Rev.} C}

\def \ketv #1>{\mbox{$|{#1}\rangle$}} 

\makeatletter
\def\simleq{\mathrel{\mathpalette\gl@align<}}
\def\simgeq{\mathrel{\mathpalette\gl@align>}}
\def\gl@align#1#2{\lower.6ex\vbox{\baselineskip\z@skip\lineskip\z@
     \ialign{$\m@th#1\hfill##\hfil$\crcr#2\crcr\sim\crcr}}}
\makeatother

\newcommand{\bq}{\mbox{\boldmath $q$}}
\newcommand{\bp}{\mbox{\boldmath $p$}}
\newcommand{\bk}{\mbox{\boldmath $k$}}
\newcommand{\br}{\mbox{\boldmath $r$}}
\newcommand{\bx}{\mbox{\boldmath $x$}}
\newcommand{\by}{\mbox{\boldmath $y$}}
\newcommand{\bz}{\mbox{\boldmath $z$}}

\newcommand{\bL}{\mbox{\boldmath $L$}}

\newcommand{\bS}{\mbox{\boldmath $S$}}
\newcommand{\bR}{\mbox{\boldmath $R$}}
\newcommand{\bsigma}{\mbox{\boldmath $\sigma$}}
\newcommand{\btau}{\mbox{\boldmath $\tau$}}
\newcommand{\brho}{\mbox{\boldmath $\rho$}}

\newcommand{\brs}{\mbox{\scriptsize \boldmath $r$}}
\newcommand{\bxs}{\mbox{\scriptsize \boldmath $x$}}

\newcommand{\bks}{\mbox{\scriptsize \boldmath $k$}}

\newcommand{\be}{\begin{equation}}
\newcommand{\ee}{\end{equation}}
\newcommand{\bea}{\begin{eqnarray}}
\newcommand{\eea}{\end{eqnarray}}
\newcommand{\nn}{\nonumber}
\newcommand{\Pu}{p_{\uparrow}}

\newcommand{\Nu}{n_{\uparrow}}
\newcommand{\Nd}{n_{\downarrow}}

\title{Lattice QCD approach to Nuclear Physics}
\author{
Sinya \textsc{Aoki}$^{1,2}$,  Takumi \textsc{Doi}$^3$, Tetsuo \textsc{Hatsuda}$^{3,4}$,  Yoichi \textsc{Ikeda}$^{5}$, Takashi \textsc{Inoue}$^{6}$, Noriyoshi \textsc{Ishii}$^2$,  Keiko \textsc{Murano}$^{3}$, Hidekatsu \textsc{Nemura}$^{2}$, Kenji \textsc{Sasaki}$^{2}$ (HAL QCD Collaboration)%
}
\inst{
$^1$Graduate School of Pure and Applied Sciences, University of Tsukuba, \\ Tsukuba 305-8571, Japan\\
$^2$Center for Computational Sciences, University of Tsukuba, \\ Tsukuba, 305-8577, Japan\\
$^3$Theoretical Research Division, Nishina Center, RIKEN, Wako 351-0198, Japan \\
$^4$IPMU, The University of Tokyo, Kashiwa 277-8583, Japan \\
$^5$Department of Physics, Tokyo Institute of Technology, Tokyo 152-8551, Japan \\
$^6$Nihon University,  College of Bioresource Sciences, Fujisawa 252-0880, Japan \\
}
\abst{
We review recent progress of the HAL QCD method which was recently 
proposed to investigate hadron interactions in lattice QCD. 
 The strategy to extract the energy-independent non-local potential
 in lattice QCD is explained in detail.
The method is applied to study  nucleon-nucleon, nucleon-hyperon, hyperon-hyperon and
 meson-baryon interactions.  Several extensions of the method   are also discussed.
}

\begin{document}

\maketitle

\section{Introduction}
 One of the ultimate goals in nuclear physics is to 
 describe hadronic many-body problems on the basis of the 
 hadronic S-matrices calculated from first principle QCD. 
 In particular, the nuclear forces are the 
  most  fundamental quantities: Once they are obtained from QCD,
 one can solve finite nuclei, hypernuclei, nuclear matter and 
 hyperon matter by employing various many-body techniques developed in nuclear physics.  
  
\begin{figure}[tb]
\begin{center}
\includegraphics[width=0.4\textwidth]{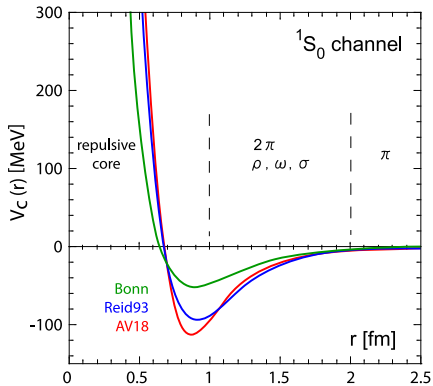}
\caption{Three examples of the modern $NN$ potential in $^1S_0$ (spin-singlet and $S$-wave) channel: Bonn\protect\cite{Machleidt2001a}, Reid93\protect\cite{Stoks:1994wp} and Argonne $v_{18}$\protect\cite{Wiringa:1994wb}. Taken from Ref.~\protect\citen{Ishii:2006ec}.}
\label{fig:NNpotential}
\end{center}
\end{figure}

  Phenomenological nucleon-nucleon ($NN$) potentials, which are designed to 
 reproduce a large number of proton-proton and neutron-proton scattering data
  as well as deuteron properties have been constructed in 90's and are called
  high-precision $NN$ potentials. 
      Some of the examples are shown 
  in Fig.~\ref{fig:NNpotential}, which
  reflect characteristic features of the $NN$ interaction for 
  different values of the relative distance $r$ as reviewed in 
 \citen{Taketani1967,Hoshizaki1968,Brown1976,Machleidt1989,Machleidt2001}:\\
  The long range part of the $NN$ force  ($r >  2$  fm) 
 is dominated by one-pion exchange originally introduced by Yukawa\cite{Yukawa1935}.
  Since the pion is the Nambu-Goldstone boson associated with the spontaneous
   breaking of chiral symmetry, it
  couples to the nucleon's spin-isospin density  and 
  leads to not only the central force but also the tensor force.
  The medium range part ($1\ {\rm fm} < r < 2$ fm) of the $NN$ force receives 
  significant contributions from   
  two-pion ($\pi\pi$) exchange \cite{TMO52}
    and/or heavy meson ($\rho$, $\omega$, and $\sigma$) exchanges.
  In particular, the spin-isospin independent attraction
 of about 50 -- 100 MeV in this region plays an essential role
  to bind the atomic nuclei and nuclear matter.
 The short  range part ($r < 1$ fm) of the $NN$ force is best described by
  a phenomenological repulsive core introduced by Jastrow \cite{Jastrow1951}.

 The nuclear saturation, the nuclear shell structure,
  the nuclear superfluidity and the structure of  neutron stars are all
  related to the properties of the nuclear force \cite{Tamagaki1993,Heiselberg2000,Lattimer2000}.
 Furthermore, the hyperon-nucleon  ($YN$) and hyperon-hyperon ($YY$) forces, whose
 information is still quite limited experimentally, are crucial to understand the
  structure of hypernuclei and the core of the neutron stars.
 The three-nucleon forces (and the three-baryon forces in general) 
  are also important to understand the binding energies of finite nuclei
   and the equation of state of dense hadronic matter. 

  It has been a long-standing 
 challenge    in theoretical particle and nuclear physics 
 to extract the hadron-hadron interactions from first principle.
 A  framework suitable for such a purpose in lattice QCD 
 was first proposed by L\"{u}scher\cite{Luscher:1990ux}: For two hadrons in a finite box with the size $L \times L \times L$ under periodic boundary conditions,  an exact relation between  the energy spectra in the box
and the elastic scattering phase shift at these energies has been derived. If the range of the hadron interaction $R$  is sufficiently smaller than the size of the box $R<L/2$, the behavior of the 
two-particle Nambu-Bethe-Salpeter (NBS) wave function $\varphi ({\bf r})$ in the interval $R < \vert {\bf r} \vert < L/2 $ is sufficient
 to relate the phase shift and the two-particle spectrum.  This 
 L\"{u}scher's finite volume 
 method bypasses the difficulty to treat the real-time scattering process on the Euclidean lattice.  Furthermore, it utilizes the finiteness of the lattice box effectively to extract the information of the on-shell scattering matrix and the phase shift.  

A closely related but a new approach to the hadron
 interactions from lattice QCD has been proposed recently
  by three of the present authors \cite{Ishii:2006ec,Aoki:2008hh,Aoki:2009ji}  and has been developed extensively by the HAL QCD Collaboration. (Therefore the approach is now called the HAL QCD method.) 
Its starting point is the same NBS wave function  $\varphi (\br)$ as discussed in Ref.~\citen{Luscher:1990ux}.  Instead of looking at the wave function outside the range of the interaction, the authors consider the internal region $ |\br | < R$ and define an
 integral kernel (or the non-local ``potential" in short) 
 $U(\br, \br')$ from $\varphi (\br)$ so that  it  obeys the Schr\"{o}dinger type equation in a finite box. This potential can be shown to be energy-independent by construction.
Since $U(\br, \br')$ for strong interactions is localized in its spatial coordinates due to confinement of quarks and gluons, it receives only weak finite volume effect in a large box. Therefore, once $U$ is determined and is appropriately extrapolated to  $L \rightarrow \infty$, one may simply use the Schr\"{o}dinger type equation in infinite space to calculate the scattering phase shifts and bound state spectra to compare the results  with  experimental data. 
 Since $U$ is a smooth function of the quark masses,
 it is relatively easy to handle on the lattice. This is in sharp contrast to the scattering length, which shows a singular behavior in the quark mass corresponding to the formation of the hadronic bound state. A further advantage of the HAL QCD method is that it can be 
  generalized directly to the many-body forces and also to the case of inelastic
  scattering.

 Studying structure of $S({\rm strangeness}) =-1$ and $S=-2$ hypernuclei is one of the key 
 challenges
  in modern 
  nuclear physics. Also,  the central core of the neutron stars will have hyperonic matter
  if the neutron beta-decays to hyperons become possible at high density.
  The hyperon-nucleon ($YN$) and hyperon-hyperon ($YY$) interactions are crucial to
  determine the level structures of hypernuclei as well as onset-density of hyperonic matter
  in neutron stars \cite{Hashimoto:special}.  
   By generalizing the $NN$ scattering in the flavor SU(2) space
  to the baryon-baryon ($BB$) scatterings 
   in the flavor SU(3) space, the HAL QCD method can give   the $YN$ and $YY$ potentials 
  as natural extension of the $NN$ potentials.  
  Such extension is also useful for identifying the origin of the 
  short-range repulsive core of the $NN$ potential and  for studying possible $S=-2$ six-quark
   state such as the $H$-dibaryon.
  
  In this article, we review the basic ideas and  recent progress of the HAL QCD method
  to hadron interactions. (As for the L\"{u}scher's finite volume method, see a recent review
   Ref.~\citen{Beane:2010em}.)
In Sec.~\ref{sec:strategy}, the basic strategy to define the $NN$ potential in QCD is explained.
In Sec.~\ref{sec:lattice}, we introduce lattice formulations of the time-independent
 HAL QCD method originally proposed in Refs.\citen{Ishii:2006ec,Aoki:2008hh,Aoki:2009ji}
 as well as its time-dependent  generalization.
In Sec.~\ref{sec:NNpotential}, some recent results of lattice QCD calculations for the 
$NN$ potential are given in both quenched and full QCD. Magnitude of the non-locality in
 $U$ is also discussed in the section. 
In Sec.~\ref{sec:hyperon}, the method 
 is applied to the hyperon-nucleon interactions such as $N\Xi$ and $N\Lambda$ systems.
In Sec.~\ref{sec:su3limit},
 interactions between octet baryons are investigated in the flavor SU(3) limit, where up, down and  strange quark masses are all equal. 
 In Sec.~\ref{sec:inelastic},  a generalization of the HAL QCD method to the case of 
 inelastic scattering is given.
 In Sec.~\ref{sec:3NF}, we show results of the three-nucleon potential, especially its 
  short distant structure. 
In Sec.~\ref{sec:MBint},  an application to the kaon-nucleon scattering is considered.
Sec.~\ref{sec:conclusion} is devoted to summary and concluding remarks.

\section{Defining the potential in QCD}
\label{sec:strategy}
\subsection{Nambu-Bethe-Salpeter (NBS) wave function}
A key quantity to define the baryon-baryon($BB$)
``potential" in QCD is the equal-time Nambu-Bethe-Salpeter wave function, 
\bea
\varphi^W(\bx) e^{-Wt}&=& \langle 0 \vert T\{B(\br + \bx,t ) B(\br, t)\} \vert 2B,W, s_1 s_2\rangle,
\eea
where $\vert 2B, W, s_1s_2\rangle$ is a QCD eigenstate  for two baryons  with equal mass
 $m_B$, helicity $s_1$ and $s_2$,  total energy $W=2\sqrt{\bk^2+m_B^2}$, the relative momentum $\bk$, and  
 the total momentum $\bp$ (we take $\bp=0$ in this paper).
 Generalization to the unequal mass can be formulated 
 in a similar manner.
In the case of two nucleons, 
the local interpolating operator $B(x)$ is taken as 
\bea
B_{\alpha}(x) &\equiv& \left(\begin{array}{c}
p_\alpha (x) \\
n_\alpha (x) \\
\end{array}\right) =
\varepsilon_{abc} \left( u_a^T (x) C\gamma_5 d_b(x) \right) q_{c,\alpha}(x), \quad
q(x)=\left(\begin{array}{c}
u (x) \\
d (x) \\
\end{array}\right),
\eea
where $x=(\bx,t)$, $a,b,c$ are the color indices,  and $\alpha$ is the spinor index.
The charge conjugation matrix is given by $C=\gamma_2\gamma_4$, and
$p,n$ are proton and neutron operators while $u,d$ denote up and down quark operators.
Here $\varphi^W$ implicitly has two pairs of spinor-flavor indices from $B_\alpha (\br + \bx,t) B_\beta(\br, t)$ as well as two helicity indices $s_1$ and $s_2$.

The most important property of the above NBS wave function is as follows.
If the total energy $W$ lies below the threshold of meson production ({\it i.e.} $W < 2 m_B + m_M$ with the meson mass $m_M$),  it satisfies the Helmholtz equation
 with $k=\vert\bk\vert$  at $r=|\br| \rightarrow \infty $,
\be
\left[ {k^2} +\nabla^2 \right] \varphi^W(\br )\simeq 0.
\ee
Furthermore, the asymptotic behavior of the radial part of the NBS wave function for
 given orbital angular momentum $L$ and total spin $S$ 
 reads \cite{Ishizuka2009a,Aoki:2009ji}
\be
 \varphi^W(r;LS) \propto  
\frac{\sin( k r - L\pi/2 + \delta_{LS}(k))}{k r}  e^{i\delta_{LS}(k)} 
\label{eq:asympt} .
\ee  
Here $\delta_{LS}(k)$ is nothing but the phase shift obtained 
from the baryon-baryon S-matrix in QCD below the inelastic threshold. 
It should be remarked here that only the upper  components of the spinor indices for the NBS wave function ($\alpha=1,2$ and $\beta =1,2$) are enough to reproduce all $BB$ scattering phase shifts $\delta_{LS}(k)$ with $L=0,1,2,3,\cdots$ and $S=0,1$ (See Appendix A of Ref.\citen{Aoki:2009ji} for the precise expression of Eq.(\ref{eq:asympt}) 
and its relation to the S-matrix in QCD.)

\subsection{Non-local potential from the NBS wave function}

From the NBS wave function, 
we define a non-local potential through the relation 
\cite{Ishii:2006ec,Aoki:2008hh,Aoki:2009ji}
\bea
\left( E_k- H_0\right) \varphi^W_{\alpha\beta}(\bx ) &=&
\int U_{\alpha\beta;\gamma\delta}(\bx, \by)  \varphi^W_{\gamma\delta}(\by )d^3y, \ \  
 \left( E_k=\frac{k^2}{2\mu},\  H_0=\frac{-\nabla^2}{2\mu} \right) ,
\label{eq:schroedinger}
\eea
where $U(\bx,\by)$ is expected to be short-ranged because of absence of massless particle exchanges between two baryons.
As mentioned in the previous subsection, 
 it is enough to consider the upper  spinor indices of $\alpha,\beta,\gamma,\delta$: 
Then 16 components of $U_{\alpha\beta;\gamma\delta}$   can be determined from $4$ components of $\varphi_{\alpha\beta}^W$ for 4 different combinations of $(s_1, s_2)$. 
Since the NBS wave function $\varphi^W$ is multiplicatively renormalized, the potential $U(\bx,\by)$ is finite and does not depend on the particular renormalization scheme.   
Note that, while Lorentz covariance is lost by using the equal-time NBS wave function and 
Eq. (\ref{eq:schroedinger}) is written as a Schr\"{o}dinger type equation, 
no non-relativistic approximation is employed here to define $U(\bx,\by)$.

  The non-local potential $U(\bx,\by)$ has been shown to be
 energy-independent\cite{Aoki:2008hh,Aoki:2009ji}. To see this, 
 let ${\cal V}_{\rm th}$ be the space spanned by the wave function at 
 $W\le W_{\rm th} \equiv 2m_B+m_M$: ${\cal V}_{\rm th}=\{\varphi^W_c\vert W\le W_{\rm th}\}$ where $c$ represents quantum numbers of the NBS wave function other than energy $W$.
 Then the projection operator to ${\cal V}_{\rm th}$ is given by
\bea
P^{W_{\rm th}}(\bx,\by) &=&\sum_{W_1,W_2\le W_{\rm th}}\, \sum_{c_1,c_2}
  \varphi^{W_1}_{c_1}(\bx) N^{-1}_{c_1,c_2}(W_1,W_2)\varphi^{W_2}_{c_2}(\by)^\dagger    \nonumber \\
&\equiv& \sum_{W \le W_{\rm th} }\, \sum_c \, P_c^{W_{\rm th}}(W;\bx,\by) 
\eea
where $N^{-1}_{c_1,c_2}(W_1,W_2)$ is defined as the inverse of the Hermitian operator
\be
N_{c_1,c_2}(W_1,W_2) = \int  \varphi^{W_1}_{c_1}(\br)^\dagger \varphi^{W_2}_{c_2}(\br)\, d^3 r,
\quad W_{1,2}  \le W_{\rm th},
\ee 
which satisfies
\be
 \sum_{W \le W_{\rm th} }\, \sum_c N_{c_1,c}(W_1,W) N^{-1}_{c,c_2}(W,W_2) =\delta_{c_1,c_2}\delta_{W_1, W_2}
\ee
in the restricted indices that $W_{1,2} \le W_{\rm th}$. (We here assume that $N(W_1,W_2)$ does not have zero eigenvalues in this restricted space.)

Using these, the non-local potential is defined by
\bea
U^{W_{\rm th}}(\bx,\by) &=&  \sum_{W_{1,2}\le W_{\rm th} }\sum_{c_1,c_2}\rho(W_1)\,  \left[E_k - H_0\right]\varphi^{W_1}_{c_1}(\bx) N^{-1}_{c_1,c_2}(W_1,W_2)\varphi^{W_2}_{c_2}(\by)^\dagger \nonumber \\
&=&  \sum_{W \le W_{\rm th} }\sum_c \,   \left[E_k - H_0\right] P_c^{W_{\rm th}}(W;\bx,\by).
\label{eq:non-local}
\eea
Then, it is easy to observe
 that the above non-local potential satisfies Eq.(\ref{eq:schroedinger})  at 
$W\le W_{\rm th}$:
\bea
\int U^{W_{\rm th}}(\bx,\by)\varphi^W_c(\by)\, d^3y &=&
\sum_{W_1\le W_{\rm th} }\, \sum_{c_1} \,  \left[E_k - H_0\right]\varphi^{W_1}_{c_1}(\bx) 
\delta_{c.c_1}\delta_{W_1,W}\nonumber \\
&=& \theta(W_{\rm th}-W)\, \left[E_k - H_0\right]\varphi^{W }_c(\bx) .
\eea
This  non-local potential  $U(\bx,\by)$  is  energy independent by construction. 
It is also easy to see that we can make the potential local but energy-dependent.
Similar trade-off between non-locality and energy-dependence has been also discussed
long time ago in Ref.\citen{KR56} in a different context.
Note however that the non-local potential $U(\bx,\by)$ which satisfied 
Eq.~(\ref{eq:schroedinger})  at  $W\le W_{\rm th}$  is not unique. For example, one may add a term such as
$
 f(\bx) [1- P^{W_{\rm th}}(\bx,\by) ]
$
with arbitrary functions $f(\bx)$ to the non-local potential $U(\bx,\by)$ without affecting Eq.(\ref{eq:schroedinger})  at  $W\le W_{\rm th}$.

We remark that
One may define a non-local potential different from Eq.(\ref{eq:non-local}) as
\be
U^{\infty}(\bx,\by) =  \sum_{ W\le \infty}\, \sum_c\,    \left[E_k - H_0\right] P_c^{\infty}(W;\bx,\by),
\ee 
which satisfies Eq. (\ref{eq:schroedinger}) for all $W$. This potential, however, becomes long-ranged, due to the presence of inelastic contributions above $W_{\rm th}$.
An extension of the HAL QCD  method, which keeps the short-range nature of the potential
while inelastic channels open,  will be discussed in Sec.~\ref{sec:inelastic}. 

The most general form of the Schr\"odinger 
type equation for the NBS wave function has energy-dependent and non-local potential
 as shown in Ref.\citen{Luscher:1990ux}.  
 However, one can always  remove its energy-dependence as demonstrated in the above derivation.

\subsection{Velocity expansion of the non-local potential}
If one knows NBS wave functions $\varphi^W$ for all $W\le W_{\rm th} $, 
the non-local potential $U$ can be constructed according to Eq. (\ref{eq:non-local}). 
In lattice QCD simulations in a finite box, however, 
 only a limited number of wave functions at low energies 
 (ground state and possibly a few low-lying excited  states) can be obtained. In such a situation, 
 it is useful to expand the non-local potential  in terms of the velocity (derivative) with 
 local coefficient functions\cite{TW67};
\be
U(\bx,\by) = V(\bx,\nabla) \delta^3(\bx-\by) .
\ee 
In the lowest few orders we have
\bea
V(\br,\nabla) &=&\underbrace{V_0(r) + V_\sigma(r) \bsigma_1\cdot\bsigma_2 + V_T(r) S_{12}}_{\rm LO} + \underbrace{V_{\rm LS} (r){\bf L}\cdot{\bf S}}_{\rm NLO} + O(\nabla^2),
\label{eq:velocity_exp}
\eea
where $r=\vert\br\vert$, $\bsigma_i$ is the Pauli-matrix acting on the spin index of the $i$-th baryon, ${\bf S}=(\bsigma_1+\bsigma_2)/2$ is the total spin, ${\bf L} = \br\times \bp$ is the angular momentum, and
\be
S_{12} = 3 \frac{(\br\cdot \bsigma_1) (\br\cdot \bsigma_2) }{r^2} -\bsigma_1\cdot\bsigma_2
\ee
is the tensor operator. Each coefficient function is further decomposed into its flavor components.
In the case of nucleons (i.e. $N_f=2$ ), we have
\be
V_X(r) = V_X^0(r) + V_X^\tau(r) \btau_1\cdot \btau_2, \quad
X=0, \sigma, {\rm T}, {\rm LS},\cdots,
\ee
where $\btau_i$ is the Pauli-matrix acting on the flavor index of the $i$-th nucleon. 
The form of the velocity expansion (\ref{eq:velocity_exp}) agrees with the form determined by symmetries\cite{okubo1958}.

At the leading order of the velocity expansion, the local potential is given by
\be
V^{\rm LO}(\br) = V_0(r) +  V_\sigma(r) \bsigma_1\cdot\bsigma_2 + V_T(r) S_{12},
\ee
which is obtained from the NBS wave function at one value of $W$. Since $S_{12}=0$ for the spin-singlet state, for example, one has
\be
V_C(r, S=0)\equiv V_0(r) - 3V_\sigma(r) = \frac{\left(E_k -H_0\right)\varphi^W(\br)}{\varphi^W(\br)} .
\label{eq:singlet_LO}
\ee

\subsection{Remarks on the ``scheme"-dependence of the potential}
\label{sec:remark}

We emphasize that
the potential itself is not a physical observable, and is therefore not a unique quantity
in quantum mechanics and in field theory. 
 In fact, the baryon-baryon
  potential in QCD depends on the choice of the interpolating 
  baryon operator to define the NBS wave function.
 Among others, the local baryon operator used in HAL QCD method
  is a most convenient choice, since 
 the reduction formula for composite particles can be
  derived in a simplest way for this choice \cite{Nishijima,Zimmermann,Haag}. 
 
  Nevertheless, one may adopt  other interpolating operators (such as  higher dimensional
   operators and non-local operators): 
  Particular choice of the baryon operator and associated potential 
  may be considered as a "scheme" 
  to describe  physical observables such as the scattering phase shift and the binding energies.
  The potential, although being ``scheme"-dependent, is still useful
  to understand physical phenomena as we know well in quantum mechanics.
 The repulsive core of the nucleon-nucleon potential in the 
 coordinate space, which is known to be the best way to summarize
 the $NN$ scattering phase shift at high energies, 
 is one of such examples.\footnote{Although in a different sense of the ``scheme", 
 analogous situation 
   in quantum field theory is the running coupling constant.
    It is scheme-dependent quantity but  is   quite
  useful to understand the high energy processes such as the deep inelastic scattering data.}

Among different schemes,  
 good convergence of the velocity expansion is an important
 check of the choice of our present scheme. 
Such a check can be carried out by examining the $W$ dependence of the lower order  potentials. For example, if we have $\varphi^{W_n}$ for $n=1,2,\cdots N$, we can determine the $N-1$ unknown local functions of the velocity expansion in $N$ different ways.
The variation among $N$ different determinations gives an estimate of the size of the higher order terms. Furthermore one of these higher order terms can be determined from   $\varphi^{W_n}$ for $n=1,2,\cdots N$. The convergence of the velocity expansion will be  investigated explicitly in Sec.~\ref{sec:NNpotential}.

The analysis in this section shows that 
the use of Schr\"odinger type equation
 with non-local potential is justified to describe the $BB$ scattering
  in QCD. The key quantity is 
   the NBS wave function, whose asymptotic behavior encodes phases of the S-matrix for the $BB$ scattering. 
If the velocity expansion of the non-local 
 potential is reasonably good at low energies, 
 one can use the LO and NLO  potentials  to investigate various nuclear many-body problems.

\section{Lattice formulation}
\label{sec:lattice}
We now discuss procedures to extract the NBS wave function from lattice QCD simulations.
For this purpose, we consider the correlation function on the lattice defined by
\be
F(\br,t-t_0)=\langle 0\vert T \{B(\bx+\br,t) B(\bx,t)\} \overline{\cal J}(t_0)\vert 0 \rangle
\label{eq:4-pt}
\ee
where $\overline{\cal J}(t_0)$ is  a source operator which creates two-baryon  states. 
Inserting a  complete set and considering  baryon number conservation,
we have
\bea
F(\br,t-t_0) &=&\langle 0\vert T \{B(\bx+\br,t) B(\bx,t)\} \sum_{n,s_1,s_2} \vert 2B, W_n, s_1,s_2\rangle \langle 2B, W_n,s_1,s_2 \vert  \overline{\cal J}(t_0)\vert 0 \rangle \nonumber \\
&+&\cdots
= \sum_{n, s_1,s_2} A_{n, s_1,s_2} \varphi^{W_n}(\br) e^{-W_n (t-t_0)}+\cdots,
\eea
where $A_{n,s_1,s_2} =\langle 2B, W_n,s_1,s_2 \vert  \overline{\cal J}(t_0)\vert 0 \rangle $
and ellipses represent contributions from inelastic states such as $NN\pi$, $NNN\bar N$, etc.
At  large time separation $(t-t_0)\rightarrow \infty$, we obtain
\be
\lim_{(t-t_0)\rightarrow\infty} F(\br,t-t_0) = A_0 \varphi^{W_0}(\br) e^{-W_0 (t-t_0)}
+ O(e^{- W_{n\not=0} (t-t_0)})
\label{eq:ground}
\ee 
where $W_0$ is the lowest energy of $BB$ states.
Since the source dependent term $A_0$ is just a multiplicative constant to the NBS wave function $\varphi^{W_0}(\br)$, the potential defined from $\varphi^{W_0}(\br)$ is manifestly source-independent. 
For this extraction of  the wave function to work, the ground state saturation for $F$ in Eq. (\ref{eq:ground}) must be satisfied by taking large $t-t_0$.
In practice, however,  $F$ becomes very noisy at large $t-t_0$.
In Sec.~\ref{sec:t-dep}, we will discuss more on this point.

\subsection{Choice of source operators}
We choose the source operator $\bar{\cal J}$ to fix quantum numbers of  $\vert 2B, W, s_1,s_2\rangle $.  Since lattice QCD simulations are
 usually performed on a hyper-cubic lattice,  the cubic transformation group $SO(3,{\bf Z})$ instead of  $SO(3,{\bf R})$ is considered as the symmetry of 
 3-dimensional space. Therefore the quantum number is classified in terms of the irreducible representation of $SO(3,{\bf Z})$,  which is denoted by
  $A_1$, $A_2$, $E$, $T_1$ and $T_2$ whose dimensions are $1,1,2,3$ and 3, respectively. 
Relation of irreducible representations between $SO(3,{\bf Z})$ and $SO(3,{\bf R})$ is given in Table~\ref{tab:cubic} for $L\le 6$, where $L$ denotes the angular momentum  for the  irreducible representation of $SO(3,{\bf R})$. For example,  the source operator $\bar{\cal J}(t_0)$ in the $A_1$ representation with positive parity generates states with $L=0,4,6,\cdots$ at $t=t_0$, while 
the operator in the $T_1$ representation with negative parity produces states with $L=1,3,5,\cdots$.  For two octet-baryons, the total spin $S$ becomes 
$1/2\otimes 1/2 = 1 \oplus 0$, which corresponds to $T_1$($S=1$) and $A_1$($S=0$) of 
 $SO(3,{\bf Z})$.
The total representation $J$ for a two baryon system is thus determined by the product $R_1\otimes R_2$, where $R_1=A_1,A_2,E,T_1,T_2$ for the orbital "angular momentum" while $R_2=A_1, T_1$ for the total spin. In Table~\ref{tab:product}, the product $R_1\otimes R_2$ is decomposed into the direct sum of irreducible representations. 

\begin{table}
\begin{center}
\caption{The number of each representation of  $SO(3,{\bf Z})$ which appears in the angular momentum $L$ representation of $SO(3,{\bf R})$. $P=(-1)^L$ denotes the eigenvalue under parity transformation.}
\label{tab:cubic}
\vspace{0.3cm}
\begin{tabular}{|cc|ccccc|}
\hline
$L$ & $P$ & $A_1$ & $A_2$ & $E$ & $T_1$ & $T_2$ \\
\hline
0 (S)&  $+$ & 1 & 0 &0 &0 & 0 \\
1 (P)&  $-$  & 0 & 0 &0 & 1 & 0\\
2 (D)&  $+$ & 0 & 0 &1 & 0 & 1 \\
3 (F)&  $-$  & 0 & 1 &0 & 1 & 1\\
4 (G)&  $+$ & 1 & 0 &1 & 1 & 1 \\
5 (H)&  $-$  & 0 & 0 &1 & 2 & 1\\
6 (I) &  $+$ & 1 & 1 &1 & 1 & 2 \\
\hline
\end{tabular}
\end{center}
\end{table}     

\begin{table}
\begin{center}
\caption{The decomposition of a product of two irreducible representations, $R_1\otimes R_2$, into irreducible representations in $SO(3,{\bf Z})$.  Note that $R_1\otimes R_2= R_2\otimes R_1$ by definition. }
\label{tab:product}
\vspace{0.3cm}
\begin{tabular}{|c||ccccc|}
\hline
  & $A_1$ & $A_2$ & $E$ & $T_1$ & $T_2$ \\
\hline
\hline
$A_1$ & $A_1$ & $A_2$ & $E$ & $T_1$ & $T_2$ \\
$A_2$   & $A_2$  & $A_1$ & $E$ & $T_2$ &  $T_1$\\
$E$       &  $E$ & $E$  &$A_1\oplus A_2\oplus E$ & $T_1\oplus T_2$ & $T_1\oplus T_2$ \\
$T_1$    & $T_1$ & $T_2$& $T_1\oplus T_2$ & $A_1\oplus E \oplus T_1\oplus T_2$ & $A_2\oplus E \oplus T_1\oplus T_2$\\
$T_2$  & $T_2$ & $T_1$  &$T_1\oplus T_2$& $A_2\oplus E \oplus T_1\oplus T_2$ &$A_1\oplus E \oplus T_1\oplus T_2$  \\
\hline
\end{tabular}
\end{center}
\end{table}

We often use the wall source at $t=t_0$ defined by
\be
{\cal J}^{\rm wall}(t_0)_{\alpha\beta, fg} = B^{\rm wall}_{\alpha,f} (t_0) B^{\rm wall}_{\beta,g} (t_0)
\ee
where $\alpha,\beta=1,2$ are upper component of the 
spinor indices while $f,g$ are flavor indices.
Here $B^{\rm wall}(t_0)$ is obtained by replacing the local quark field $q(x)$ of $B(x)$ by the wall source,
\be
q^{\rm wall}(t_0) \equiv \sum_{\bxs} q(\bx,t_0) 
\ee
with the Coulomb gauge fixing at $t=t_0$. Note that this gauge-dependence of the source operator disappears for the potential.
All states created by the wall source have zero total momentum. Among them 
the state with zero relative momentum has the largest magnitude.
A reason for employing the wall source here is that the ground state saturation for the potential at long distance is better achieved for the wall source  than for 
other sources. 

Let us consider the case of the two nucleons. The source  operator $\bar{\cal  J}^{\rm wall}(t_0)$
has zero orbital angular momentum at $t=t_0$, which corresponds to the
$A_1$  representation with  positive parity.
Therefore, the total  angular momentum can be fixed  by using the spin
recoupling   matrix  $M^{(S,S_z)}$,   e.g.,   $M^{(S=0,S_z=0)}  \equiv
\sigma_2/\sqrt{2}$     and    $M^{(S=1,S_z=m)}     \equiv    (\sigma_2
\sigma_m)/\sqrt{2}$ for $m=0,\pm 1$ as
\be
{\cal J}(t_0;J^{P=+},J_z=m,I) = M^{(S)}_{\beta\alpha} {\cal J}^{\rm wall}(t_0)_{\alpha\beta,fg}  .
\ee
Here $P=\pm $ is the parity and $I=1,0$ is the total isospin of the system. Since the  nucleon is a fermion, exchange of the nucleon operators in the source should give a minus sign. This fact fixes the total isospin given the total spin: $(S,I)=(0,1)$ or $(1,0)$.
(Note that $S,I=0$ are antisymmetric while $S,I=1$ are symmetric under the exchange.)
Since $A_1^+\otimes A_1(S=0) = A_1^+$  and $A_1^+\otimes T_1(S=1) = T_1^+$, the state with either $(J^P,I) =(A_1^+, 1)$ for the spin-singlet or $(J^P,I) =(T_1^+, 0)$ for the spin-triplet is created at $t=t_0$ by the corresponding source operator.  The NBS wave function extracted at $t > t_0$ has the same  quantum numbers $(J^P,I)$  as they are conserved under QCD interactions. 
In addition the total spin $S$ is conserved at $t > t_0$ for the two nucleon system with equal up and down quark masses: Under the exchange of the two particles, the constraint $(-1)^{S+1+I+1}P=-1$ must be satisfied due to the fermionic nature of the nucleon. Also, 
the parity $P$ and the isospin $I$ are conserved in this system. Therefore $S$ is conserved.
 However, $L$ is not conserved in general. While the state with
 $(J^P,I) =(A_1^+, 1)$  always has $L=A_1^+$ even at $t > t_0$, the one with  
 $(J^P,I) =(T_1^+, 0)$  has both $L=A_1^+$ and $L=E^+, T_2^+$ components\footnote{This can be seen from Table~\ref{tab:product} for $R_2=T_1$(spin-triplet), 
 which also tells us the  existence of $L= T_1^+$ component in addition. The extra component is expected to be small since it appears as a consequence of the violation of
  $SO(3,{\bf R})$ on the hyper-cubic lattice.} at $t> t_0$, which corresponds 
to $L=0$ and $L=2$ in $SO(3,{\bf R})$, respectively.
Note that  $J$ and $L$ are used to represent the total and orbital  quantum numbers respectively
 for $SO(3,{\bf Z})$ as well as for $SO(3,{\bf R})$. 

The orbital angular momentum $L$ of the NBS wave function for $NN$ can be fixed
to a particular value by the projection operator $P^{(L)}$ as
 \be
 \varphi^W(\br; J^P,I, L,S) = P^{(L)} P^{(S)} \varphi^W(\br; J^P,I) 
 \ee
 where $\varphi^W(\br; J^P,I) $ is extracted from
 \bea
 F(\br,t-t_0;J^P,I) &\simeq& A(J^P,I) \varphi^{W}(\br; J^P,I)e^{-W(t-t_0)} ,\\
 A(J^P,I) &=&\langle 2B,W  \vert \bar{\cal J}(t_0;J^P,I)\vert 0\rangle \nn
 \eea
 for large $t-t_0$.
The total spin projection operator is 
$
  (P^{(S=0)})_{\alpha\beta;\alpha'\beta'}
  \equiv
  \frac1{2}
  (\sigma_2)_{\alpha\beta}
  (\sigma_2)_{\beta'\alpha'}
$ for spin-singlet
and
$P^{(S=1)} \equiv \mathbb{I} - P^{(S=0)}$
for spin-triplet,
but this is  redundant since the total spin $S$,  already fixed by the
source,  is conserved  as  mentioned before.
The  projection  operator $P^{(L)}$  of  the orbital  angular momentum for   an arbitrary
function  $\varphi(\br)$  is
defined in general by
\be
P^{(L)} \varphi^W(\br)
\equiv
\frac{d_L}{24}
\sum_{g\in SO(3,{\bf Z})}
\chi^L(g)^* \varphi^W(g^{-1}\cdot\br)
\ee
for $L=A_1,A_2,E,T_1,T_2$,  where $\chi^L$ denotes the character of the representation $L$
in  $SO(3,{\bf  Z})$, $^*$ is its complex conjugate, $g$  is one  of 24  elements in
$SO(3,{\bf Z})$ and $d_L$ is the dimension of $L$.

\subsection{Leading order $NN$ potential: spin-singlet case}
We present the procedure to determine potentials at the leading order(LO):
\be
V^{\rm LO}(\br ) = V_0(r) +V_\sigma(r)(\bsigma_1\cdot\bsigma_2)+ V_T(r) S_{12}.
\ee
Since $S_{12}=0$ and $\bsigma_1\cdot\bsigma_2= - 3$ for the spin-singlet case, 
the LO  central potential for the spin-singlet case is extracted from the $(J^P,I)=(A_1^+,1)$ state as
\be
V_C(r)^{(S,I)=(0,1)} \equiv V^{I=1}_0 (r)-3 V^{I=1}_\sigma (r)
= \frac{\left[E_k- H_0\right]\varphi^W(\br; A_1^+,I=1,L=A_1,S=0)}{\varphi^W(\br; A_1^+,I=1,L=A_1,S=0)},
\ee
where $V_X^{I=1} = V_X^0+V_X^\tau$ in isospin space.
The potential $V_C(\br)^{(S,I)=(0,1)}$ in the above
is often referred to as the central potential for the $^1S_0$ state, where the notation $^{2S+1}L_J$ represents the orbital angular momentum $L$ (see Table~\ref{tab:cubic}), the total spin $S$ and the total angular momentum $J$ of ${\bf J}={\bf L}+{\bf S}$. It is noted, however, that in the leading order of the velocity expansion, the potential does not depend on the quantum number of the state $J=L=A_1$. Moreover the $A_1$ state may contain $L=4,6,\cdots$ components other than $L=0$, though the $L=0$ component may dominate. Therefore it is more precise to refer to $V_C(\br)^{(S,I)=(0,1)}$ as the spin-singlet (isospin-triplet) central potential determined from the state with $J=L=A_1$. A possible difference of spin-singlet central potentials between this determination and others such as the one determined from $J=L=E$ gives an estimate for contributions from higher order terms in the velocity expansion. 

\subsection{Leading order potential: spin-triplet case}
Both the tensor potential $V_T$ and central potential $V_C$ appear in the LO
for the spin-triplet case.  Let us consider the determination from 
 $(J^P,I)=(T_1^+,0)$ state. The 
Schr\"odinger equation for this state becomes
\be
\left[H_0+V_C(r)^{(S,I)} + V_T(r) S_{12}\right]\varphi^W(\br;J^P=T_1^+,I) = E_k \varphi^W(\br;J^P=T_1^+,I)
\label{eq:schroedinger-tensor}
\ee 
with $(S,I)=(1,0)$,
where the spin-triplet central potential is given by
\be
V_C(r)^{(1,0)} \equiv  V_0^{I=0}(r) + V_\sigma^{I=0}(r), \qquad
V_X^{I=0} = V_X^0-3V_X^\tau .
\ee
We         separate        the         Schr\"odinger        equation
Eq.~(\ref{eq:schroedinger-tensor})  into  the  $A_1$  and  non-$A_1$
components by using projection operators ${\cal P} \equiv P^{(A_1)}$
and ${\cal Q} \equiv \mathbb{I} - {\cal P}$ as
\begin{eqnarray}
  \nonumber
  (V_{C}(r)^{(1,0)} - E_k)
  {\cal P}
  \varphi^W(\br)
  +
  V_{T}(r)
  {\cal P}
  S_{12}
  \varphi^W(\br)
  &=&
  -H_0
  {\cal P}
  \varphi^W(\br)
  \\
  (V_{C}(r)^{(1,0)} - E_k)
  {\cal Q}
  \varphi^W(\br)
  +
  V_{T}(r)
  {\cal Q}
  S_{12}
  \varphi^W(\br)
  &=&
  -H_0
  {\cal Q}
  \varphi^W(\br).
\end{eqnarray}
Note that ${\cal  P}$ and ${\cal Q}$ commute  with $H_0$, $V_C(r)$ and
$V_T(r)$, whereas they do not commute with $S_{12}$.
Non-$A_1$ component receives contributions  from $E$, $T_1$ and $T_2$,
among  which only  $E$  and  $T_2$ contribute  to  the D-wave.  Since  the
contribution from  $T_1$ component turns  out to be negligible  in the
numerical simulation,
the non-$A_1$ component is dominated by D-wave contributions.

Using these projections, $V_C$ and $V_T$ can be extracted as
\bea
V_C(r)^{(1,0)} &=& E_k -\frac{1}{\Delta(\br)}\left( [{\cal Q} S_{12}\varphi^W]_{\alpha\beta}(\br)
H_0[{\cal P}\varphi^W]_{\alpha\beta}(\br) \right. \nn \\
&-&\left.  [{\cal P} S_{12}\varphi^W]_{\alpha\beta}(\br)
H_0[{\cal Q}\varphi^W]_{\alpha\beta}(\br)
\right) \label{eq:central}
\eea
\bea
V_T(r) &=& \frac{1}{\Delta(\br)}\left( [{\cal Q} \varphi^W]_{\alpha\beta}(\br)
H_0[{\cal P}\varphi^W]_{\alpha\beta}(\br) - [{\cal P} \varphi^W]_{\alpha\beta}(\br)
H_0[{\cal Q}\varphi^W]_{\alpha\beta}(\br)
\right) \label{eq:tensor}
\\
\Delta(\br) &\equiv& [{\cal Q} S_{12}\varphi^W]_{\alpha\beta}(\br)
[{\cal P}\varphi^W]_{\alpha\beta}(\br) - [{\cal P} S_{12}\varphi^W]_{\alpha\beta}(\br)
[{\cal Q}\varphi^W]_{\alpha\beta}(\br) .
\eea
In numerical simulations, $(\alpha,\beta)=(2,1)$  in $J_z = 0$ state
is mainly employed.

One may focus only on the $A_1$ component of the wave function and 
define so-called the effective central potential for the spin-triplet (isospin-singlet),
 often used in nuclear physics:
\be
V_C^{\rm eff}(r)^{(1,0)} =  \frac{\left[E_k- H_0\right] {\cal P}\varphi^W_{\alpha\beta}(\br)  }{{\cal P}\varphi^W_{\alpha\beta}(\br) } . 
\label{eq:effective_central}
\ee
The effect of $V_T$, which leads to a transition from the $A_1$ component to the 
 non-$A_1$ component of the wave-function,  is implicitly included in
 this effective central potential:  For small $V_T$,
 the difference between $V_C$ and $V_C^{\rm eff}$ is $O(V_T^2)$ 
 as  the second order perturbation tells us.

\subsection{Time-dependent HAL QCD method}
\label{sec:t-dep}
One of the practical difficulties to extract the NBS wave function and
the potential  from the correlation function  Eq.(\ref{eq:4-pt}) is to
achieve the ground state  saturation in numerical simulations at large
but finite  $t-t_0$ with  reasonably small statistical  errors.
While  the  stability  of  the  potential  against  $t-t_0$  has  been
confirmed within statistical  errors in numerical 
simulations\cite{Ishii:2006ec,Aoki:2009ji},  the determination of $W$ for 
 the ground state suffers
from  systematic  errors due  to  contaminations  of possible  excited
states.
There exist three different methods  to determine $W$. The most well-known
method  is  to  determine  $W$  from the  $t-t_0$  dependence  of  the
correlation function  Eq.(\ref{eq:4-pt}) summed over $\br$  to pick up
the zero momentum state.
On the  other hand, one may determine $\bk^2$  of $W$ by  fitting the
$\br$ dependence of the NBS wave function with its expected asymptotic
behavior at  large $r$  or by  reading off the  constant shift  of the
Laplacian part of the potential  from zero at large $r$.
Although the latter two methods usually give consistent results within
statistical  errors,  the first  method 
sometimes leads  to a  result different from  those determined  by the
latter two  at the value of $t-t_0$  employed  in numerical simulations.
Although, in principle, the increase  of $t-t_0$ is needed in order to
see  an agreement  among three methods, it  is
difficult  in practice  due  to larger  statistical  errors at  large
$t-t_0$.

The problem above is  common 
 in various applications of  lattice QCD.
Fortunately,
the  original HAL QCD method can be improved to
overcome this difficulty as follows.
Let  us   consider  the  normalized  correlation  function defined from
Eq.(\ref{eq:4-pt}) as
\begin{eqnarray}
  R(\br,t)
  &\equiv&
  \frac{F(\br, t)}{e^{-2m_B t}}  
  =
  \sum_{n}
  A_{n}
  \varphi^{W_n}(\br)
  \exp\left(-t \Delta W_n \right)
  +
  O(e^{-\Delta W_{\rm th}t}), 
 \label{eq:RF1}
  \end{eqnarray}
where  $\Delta W_{n}  = W_n -2 m_B  $ and $\Delta W_{\rm th} = W_{\rm th} -2m_B= m_M$.
By neglecting 
the inelastic contributions  above  the meson production threshold, represented by $O(e^{-\Delta W_{\rm th}t})$, 
for large enough $t$\footnote{This limitation for $t$ can be removed if the coupled
  channel potentials are introduced as in Sec.~\ref{sec:inelastic}.},
 non-relativistic approximation
$W_{n} - 2m_B \simeq {k_n^2}/m_B$ leads us to
\bea
  R(\br, t)
  \simeq
  \sum_{n}
  A_{n}
  \varphi^{W_n}(\br)
  \exp\left(-t \frac{k_n^2}{m_B} \right)
  &=&
  e^{-t (H_0 + U)}
  \sum_{n}
  A_{n}
  \varphi^{W_n}(\br) \nn \\
  &=&
  e^{-t (H_0 + U)} R(\br, t=0),
\eea
where  the  Schr\"odinger  equation  Eq.~(\ref{eq:schroedinger}),  the
defining relation of  the non-local potential $U$, is  used to replace
$e^{-t{k_n}^2/m_N}$ by $e^{-t (H_0 + U)}$.
By applying a time derivative on both side, we have the time-dependent
Sch\"odinger equation in imaginary time
\begin{equation}
  \left(
    -\frac{\partial}{\partial t} - H_0
  \right)
  R(\br, t)
  \simeq
  \int d^3 \br'
  U(\br,\br')
  R(\br', t).
  \label{eq:tdep-nonrela}
\end{equation}
Now the velocity expansion of  the non-local potential leads us to the
formula of the leading order potential
\begin{equation}
  V^{\rm LO}(\br)
  =
  -
  \frac{
    (\partial/\partial t) R(\br, t)
  }{
    R(\br, t)
  }
  -
  \frac{
    H_0 R(\br, t)
  }{
    R(\br, t)
  }.
  \label{eq:tdep-lo}
\end{equation}
Once  the  ground  state   saturation  is  achieved  in  $R(\br,  t)$,
Eq.(\ref{eq:tdep-lo}) reduces to  Eq.(\ref{eq:singlet_LO}), for example for the spin-singlet case.  
Indeed, in this case, $-\partial/\partial
t$ is  safely replaced by  the non-relativistic energy $E_{k}$ of
the ground state under the non-relativistic approximation.

The non-relativistic  formula for $V^{\rm LO}(r)$ above  can be 
generalized to the case that masses of two particles are different by
the replacement, $R(\br, t) = F (\br, t)/e^{- (m_A+m_B) t}$.  Note
also  that  the  potential  extracted  in  this  method  automatically
satisfies   $V^{\rm LO}(r\rightarrow)  \rightarrow  0$ 
without  constant  shift.   This  property can  be  used to  check
whether this extraction works correctly or not.

The non-relativistic approximation used to derive
 Eq.(\ref{eq:RF1}) can  be removed by using the second
order derivative in $t$;
\begin{equation}
  \left(
    \frac1{4m_B}
    \frac{\partial^2}{\partial t^2}
    -
    \frac{\partial}{\partial t}
    -
    H_0
  \right)
  R(\br, t)
  =
  \int d^3 \br'
  U(\br,\br')
  R(\br', t),
\end{equation}
which leads to
\begin{equation}
  V^{\rm LO}(\br)
  =
  \frac1{4m_B}
  \frac{(\partial/\partial t)^2 R(\br, t)}{R(\br,t)}
  -
  \frac{
    (\partial/\partial t) R(\br, t)
  }{
    R(\br, t)
  }
  -
  \frac{
    H_0 R(\br, t)
  }{
    R(\br, t)
  }.
\end{equation}
Here  we have assumed   that  the  inelastic
contributions are  negligibly small and that the  two particles 
have the same  mass.
Since $t$ is discrete on the lattice, the  $t$ derivatives has to be carefully performed.
One has to employ the numerical derivative scheme  which 
reduces statistical as well  as  systematic  errors  of  $V^{\rm LO}(\br)$.

One can generalize Eq.(\ref{eq:RF1}) to the 
 correlation function with two relative coordinates $\bx$ and $\by$;
\begin{equation}
  R(\bx , \by, t)
  = \frac{1}{e^{-2m_B t}}
  \int d^3\bx_1 d^3\by_1
  \langle 0 \vert
  T\{B(\bx_1+\bx,t) B(\bx_1,t) \overline{B}(\by_1+\by,0) \overline{B}(\by_1,0)\}
  \vert 0\rangle ,
\end{equation}
which leads to
\begin{equation}
  \left(
    \frac1{4m_B}
    \frac{\partial^2}{\partial t^2}
    -
    \frac{\partial}{\partial t}
    - H_0
  \right)
  R(\bx,\by, t)
  =
  \int d^3 \bz\, U(\bx,\bz) R(\bz, \by,t).
\end{equation}
Then, we obtain the non-local potential as
\begin{equation}
  U(\bx,\by)
  =
  \int d^3\bz
  \left(
    \frac1{4m_B}
    \frac{\partial^2}{\partial t^2}
    -
    \frac{\partial}{\partial t}
    -
    H_0
  \right)
  R(\bx,\bz, t) \cdot
  \tilde R^{-1}(\bz,\by, t),
\end{equation}
where $\tilde  R^{-1}(\bx,\by, t)$ is  an ``truncated" inverse  of the
Hermitian operator $R (\bx,\by, t)$, 
\be
\tilde R^{-1}(\bx,\by, t) =\sum_{\lambda_n\not= 0 } \frac{1}{\lambda_n(t)} v_n(\bx, t) v_n^\dagger(\by,t)
\ee
with  $\lambda_n(t)$   and  $v_n(\bx,t)$   being  the  eigenvalues
   and 
corresponding eigenvectors of $R  (\bx,\by, t)$, respectively.
Note that  zero
eigenvalues  are  removed  in   the above summation.   
Suppose we introduce a modified potential as
\be
\hat U(\bx,\by) =  U(\bx,\by) + \sum_{\lambda_n=0} c_n v_n(\bx, t) v_n^\dagger(\by,t) .
\ee 
Then it satisfies the same  Schr\"odinger equation for all possible
 values of $c_n $, the non-local  potential  is  not unique as discussed before.

\section{$NN$ potential from lattice QCD}
\label{sec:NNpotential}.

\subsection{Central potential in quenched QCD}
\label{ssec:CPQ}
Let us first 
show results in the quenched QCD, where creations and annihilations of virtual quark-antiquark pairs  are neglected: The standard plaquette gauge action is employed on a 32$^4$ lattice at the bare gauge coupling constant $\beta=6/g^2=5.7$. This corresponds to the lattice spacing $a\simeq 0.137$ fm ($1/a=1.44(2)$ GeV), determined from the $\rho$ meson mass in the chiral limit, and the physical size of the lattice $L\simeq 4.4$ fm\cite{Ishii:2006ec}. As for the quark action, the standard Wilson fermion action is used at three different values of the quark mass corresponding to the pion mass $m_\pi \simeq  731, 529, 380$ MeV and the nucleon mass $m_N \simeq 1560,1330,1200$ MeV, respectively. 

\begin{figure}[tb]
\begin{center}
\includegraphics[width=0.33\textwidth,angle=270]{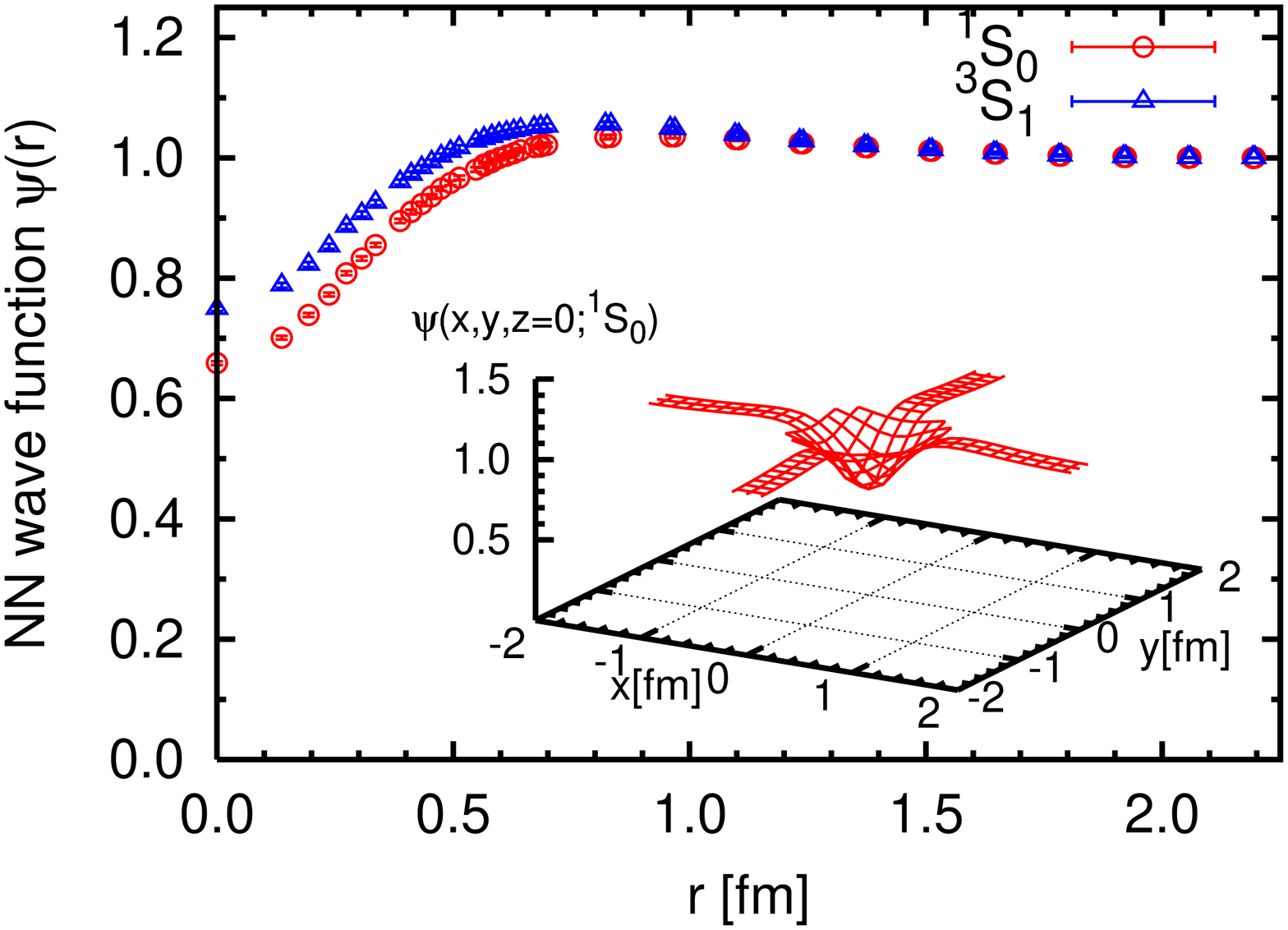}
\includegraphics[width=0.33\textwidth,angle=270]{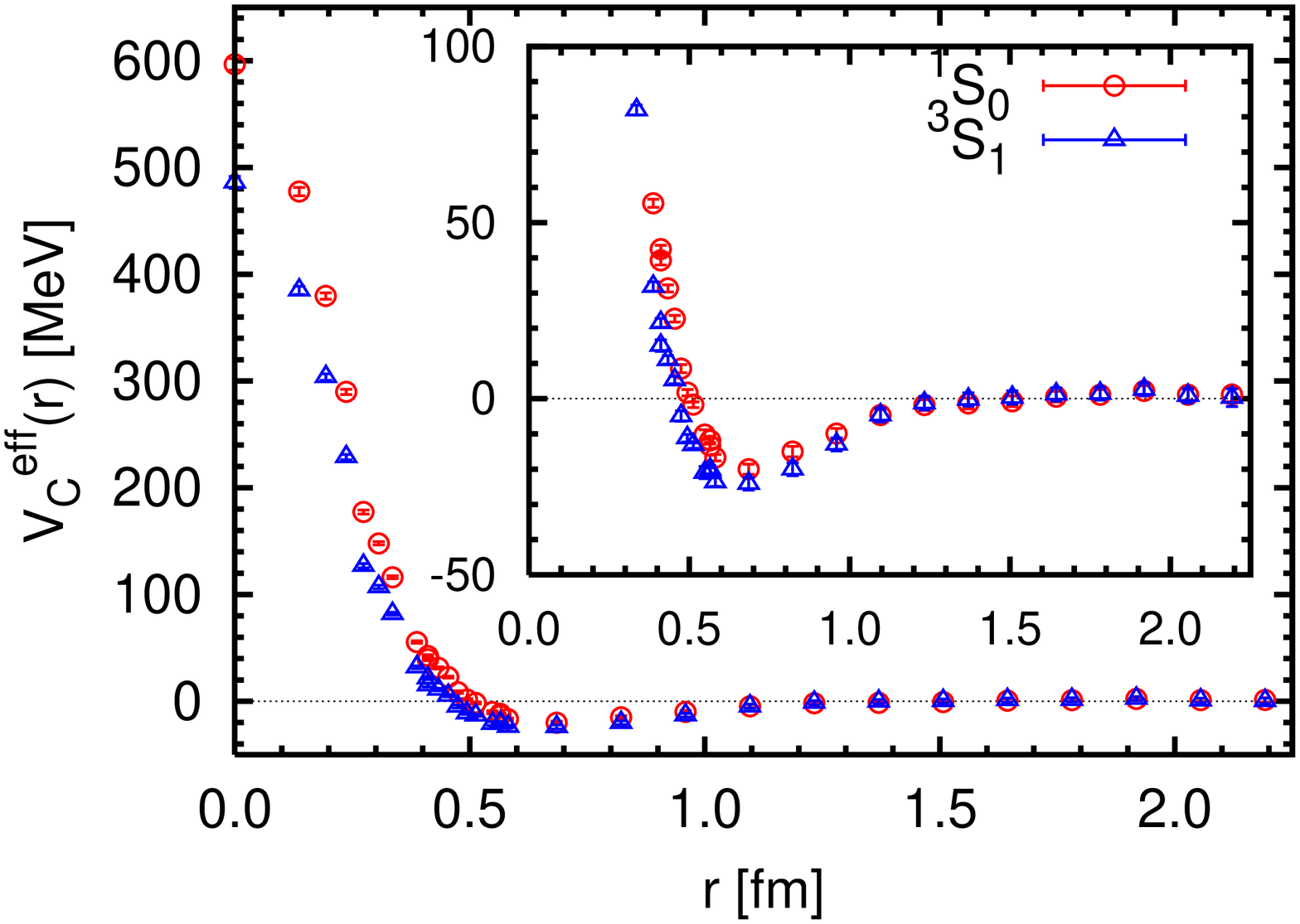}
\caption{(Left)The $NN$ wave function for the spin-singlet and spin-triplet channels in the orbital $A_1^+$ representation at $m_\pi\simeq 529$ MeV and $a\simeq 0.137$ fm in quenched QCD. The insert is a three-dimensional plot of the spin-singlet wave function $\varphi^W(x,y,z=0)$. 
(Right)  The $NN$ (effective) central potential for the spin-singlet (spin-triplet) channel determined from the orbital $A_1^+$ wave function. Both figures are taken from Ref.~\protect\citen{Aoki:2009ji}. }
\label{fig:wave_potential}
\end{center}
\end{figure}

Fig.~\ref{fig:wave_potential}(Left) shows the NBS wave functions for the spin-singlet and the spin-triplet channels in the orbital $A_1$ representation at $m_\pi \simeq 529$ MeV. These wave functions are normalized to be 1 at the largest spatial point $r\simeq 2.2$ fm.
\begin{figure}[tb]
\begin{center}
\includegraphics[width=0.4\textwidth,angle=270]{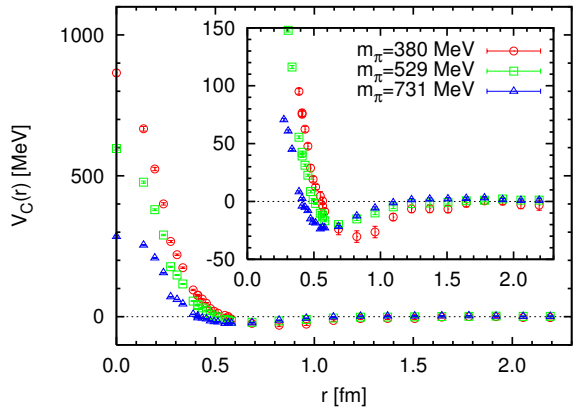}
\caption{The central potentials for the spin-singlet channel from the orbital $A_1^+$ representation at three different pion masses in quenched QCD. Taken from Ref.~\protect\citen{Aoki:2009ji}.}
\label{fig:mass-dep}
\end{center}
\end{figure}
The central potential in the spin-singlet channel and the effective central potential in the spin-triplet channel extracted from the wave functions at $m_\pi\simeq 529$ MeV are shown in Fig.~\ref{fig:wave_potential}(Right). These potentials reproduce the qualitative features of the phenomenological $NN$ potentials, namely  the repulsive core at short distance  surrounded by the attractive well at medium and long distances. From this figure one observes that the interaction range of the potential is smaller than 1.5 fm, showing that the box size $L\simeq 4.4$ fm is large enough for the potential.
  Labels $^1S_0$ and $^3S_1$
of the potentials in the figure represent the fact that potentials are determined from $A_1$ wave functions, which are dominated by the $S$-wave component.
Note here that the lattice artifacts are expected to be large for potentials (as well as wave functions) at short distance such that $ r \simeq O(a)$. 
Therefore our results at short distance should be considered to be qualitative , not quantitative,
and this caution should be applied to all of our results in this paper otherwise stated.  
The continuum extrapolation is necessary to predict short distance behaviors of potentials quantitatively. Indeed, $BB$ potentials in the continuum limit are shown to diverge as $r\rightarrow 0$\cite{Aoki:2010kx,Aoki:2010uz}.
 
In Fig.\ref{fig:mass-dep}, $NN$ central potentials in the spin-singlet channel are shown for three different pion masses. The repulsion at short distance and the attraction at medium distance are simultaneously enhanced as the pion mass decreases.

\subsection{Tensor potential in quenched QCD}
\begin{figure}[tb]
\begin{center}
\includegraphics[width=0.33\textwidth,angle=270]{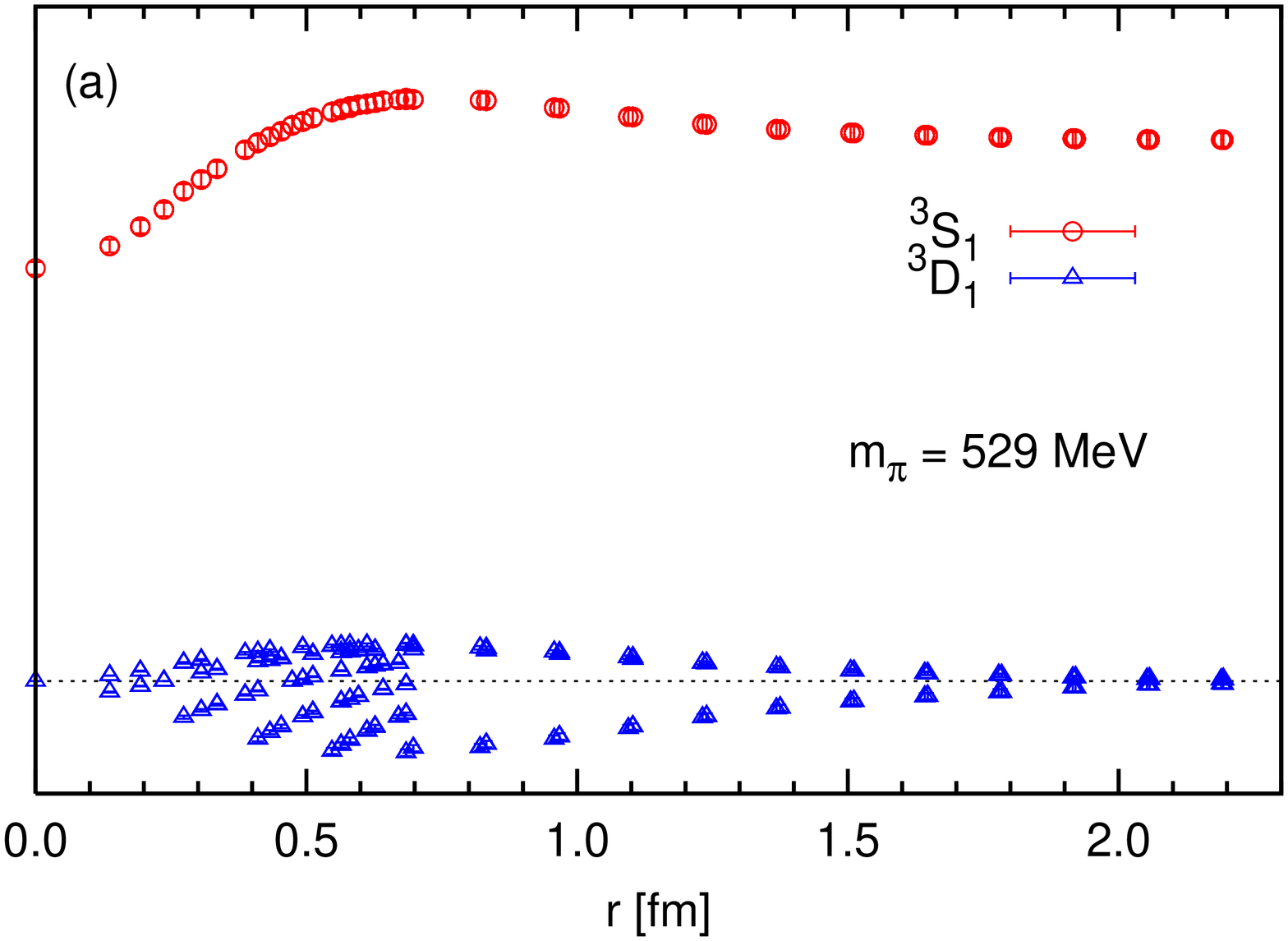}
\includegraphics[width=0.33\textwidth,angle=270]{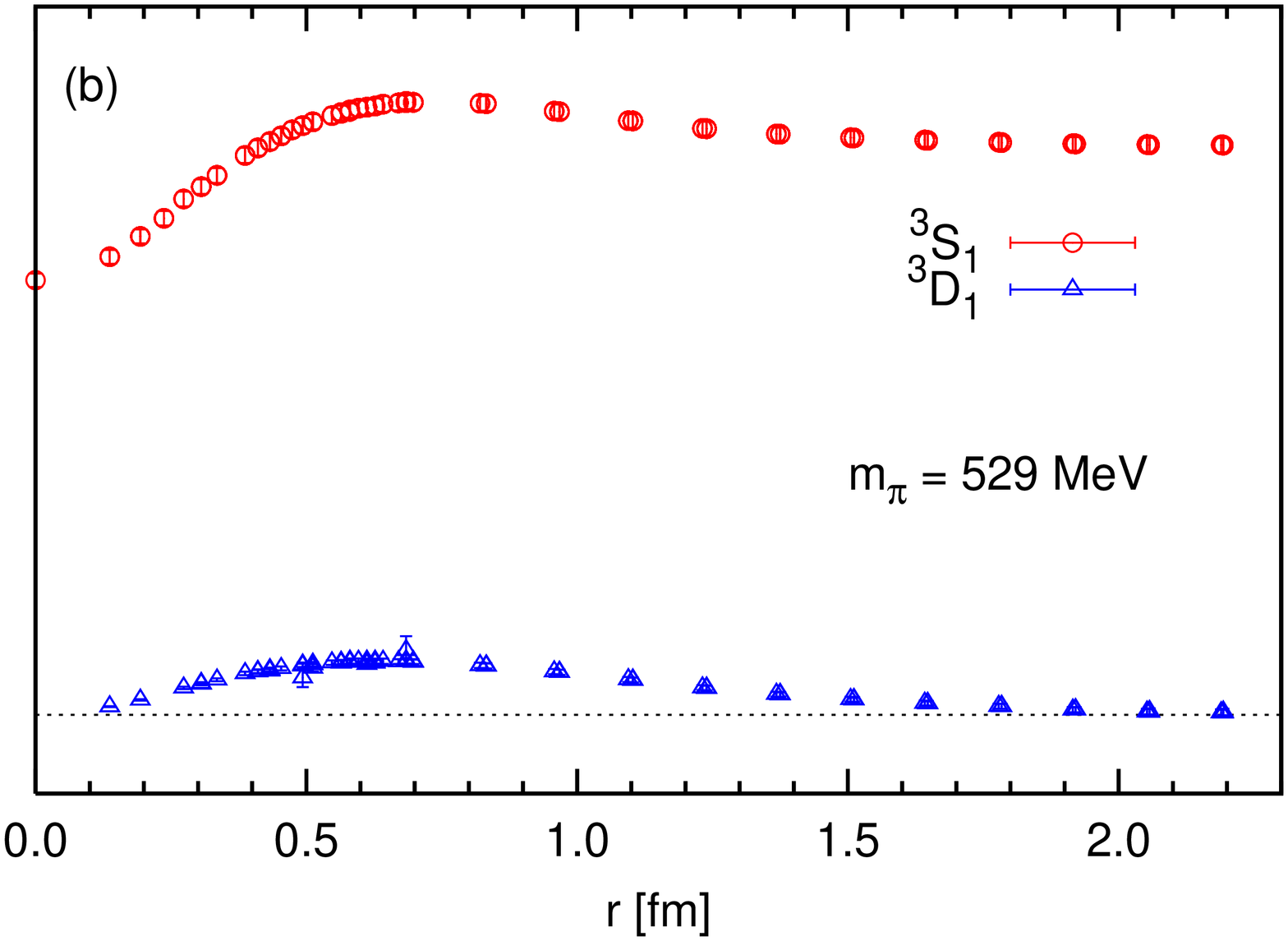}
\caption{(Left) $(\alpha,\beta)=(2,1)$ components of the orbital $A_1^+$ and non-$A_1^+$ wave functions from  $J^P=T_1^+$ (and $J_z=S_z=0$) states at $m_\pi\simeq 529$ MeV. (Right) The same wave functions but the spherical harmonics components are removed from the non-$A_1^+$ part. Taken from Ref.~\protect\citen{Aoki:2009ji}.}
\label{fig:d-wave}
\end{center}
\end{figure}
\begin{figure}[bt]
\begin{center}
\includegraphics[width=0.33\textwidth,angle=270]{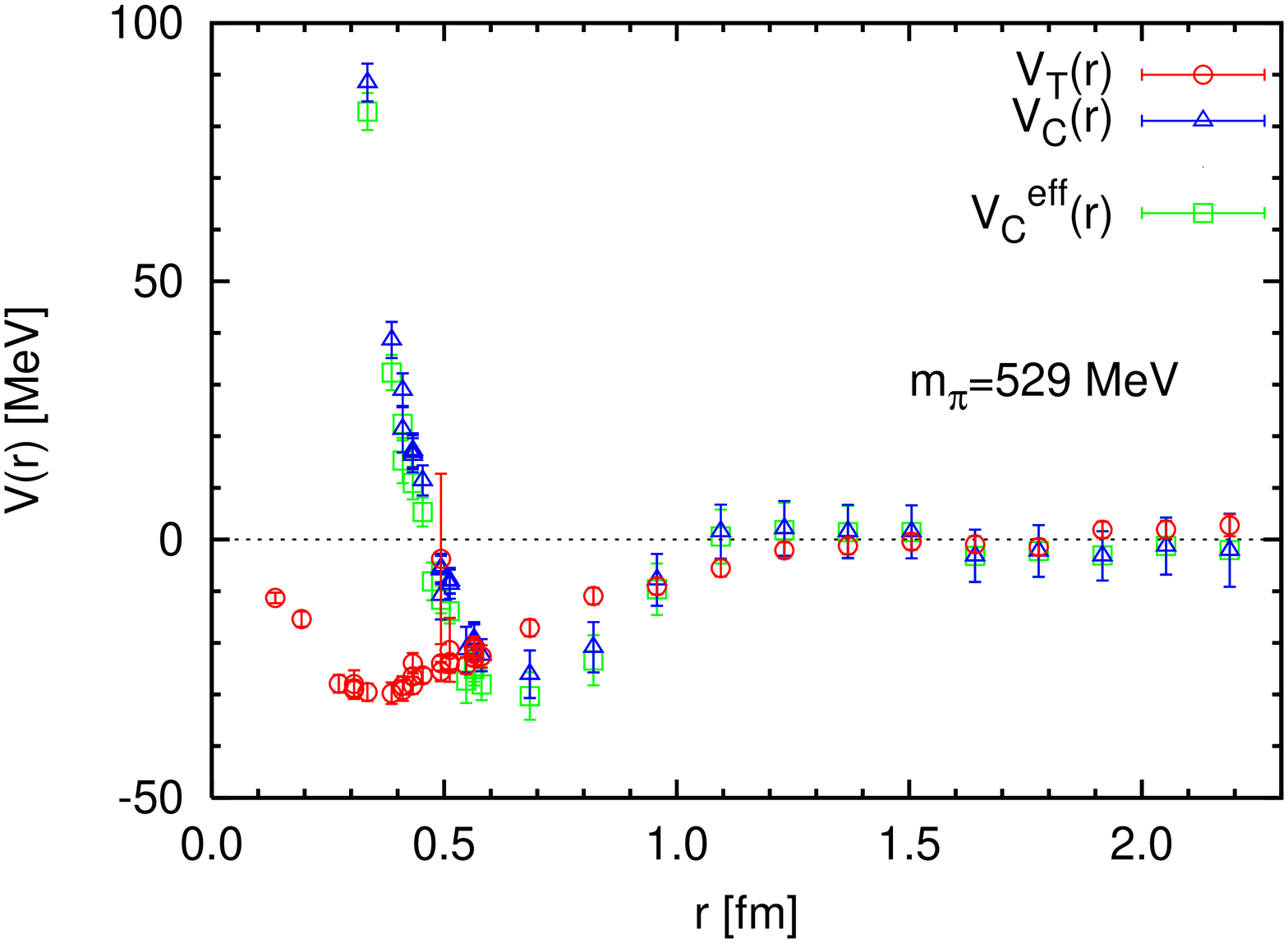}
\includegraphics[width=0.33\textwidth,angle=270]{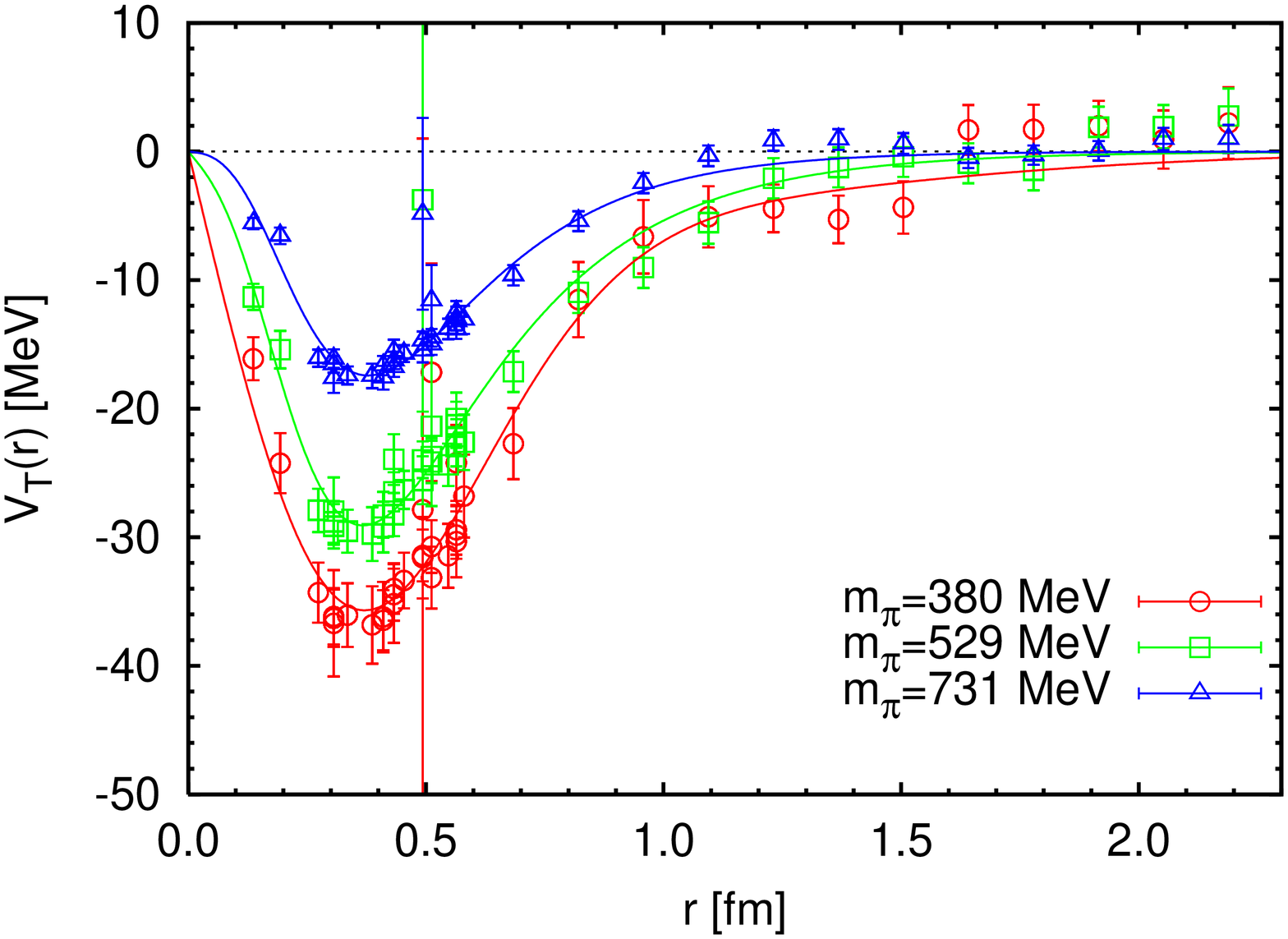}
\caption{(Left) The central potential $V_C(r)^{(1,0)}$ and  the tensor potential $V_T(r)$ obtained from  the $J^P=T_1^+$ NBS wave function, together with the effective central potential $V_C^{\rm eff} (r)^{(1,0)}$, at $m_\pi\simeq 529$ MeV. (Right) Pion mass dependence of the tensor potential. The lines are the four-parameter fit using one-pion-exchange $+$ one-rho-exchange with Gaussian form factor. Taken from Ref.~\protect\citen{Aoki:2009ji}.}
\label{fig:tensor}
\end{center}
\end{figure}

In Fig.~\ref{fig:d-wave}(Left), we show the $A_1$ and non-$A_1$ components of the NBS wave function obtained from the $J^P=T_1^+$ (and $J_z=S_z=0$) states at $m_\pi\simeq 529$ MeV.
The non-$A_1$ wave function is multivalued as a function of $r$ due to its angular dependence. For example, the $(\alpha,\beta) = (2,1)$  component of the
$L=2$ part of the non-$A_1$  wave function is proportional to the spherical harmonics $Y_{20}(\theta,\phi) \propto 3\cos^2\theta -1$.  Fig.~\ref{fig:d-wave}(Right) shows the non-$A_1$ component divided by $Y_{20}(\theta,\phi)$. The non-$A_1$ wave function seems to be dominated by the $D$ ($L=2$) state,
since its multivaluedness is mostly absorbed to $Y_{20}(\theta,\phi)$.  
Fig.~\ref{fig:tensor} (Left) shows the central potential $V_C(r)^{(1,0)}$ and tensor potential $V_T(r)$, together with the effective central potential $V_C^{\rm eff}(r)^{(1,0)}$, at the leading order of the velocity expansion as given in Eqs. (\ref{eq:central}), (\ref{eq:tensor}) and  (\ref{eq:effective_central}), respectively.

Note that $V_C^{\rm eff}(r)$ contains the effect of $V_T(r)$ implicitly as higher order effects through processes such as ${}^3S_1\rightarrow {}^3D_1\rightarrow {}^3S_1$. At the physical pion mass, 
$V_C^{\rm eff}(r)$ is expected to gain sufficient attraction from the tensor potential, which leads to the appearance of a bound deuteron in the spin-triplet (and flavor-singlet) channel while an absence of the bound dineutron in the spin-singlet (and flavor-triplet) channel. The difference between $V_C(r)^{(1,0)}$ and   $V_C^{\rm eff}(r)$ in Fig.~\ref{fig:tensor} (Left) is still small in this quenched simulation due to relatively large pion mass. 

The tensor potential in Fig.~\ref{fig:tensor} (Left) is negative for the whole range of $r$ within statistical errors and has a minimum around 0.4 fm.  If the tensor potential receives a significant contribution from one-pion exchange as  expected from the meson theory, $V_T(r)$ is rather sensitive to the change of the pion mass.  As shown in Fig.~\ref{fig:tensor} (Right), it is indeed the case: Attraction of $V_T(r)$ is substantially enhanced as the pion mass decreases. 

The central and tensor potentials in lattice QCD are given at discrete data points. For practical applications to nuclear physics, however, it is more convenient to parameterize the lattice results by known functions.  Such a fit for $V_T(r)$ is given by the form of one-pion-exchange $+$ one-rho-exchange with Gaussian form factors as
\bea
V_T(r) &=& b_1(1-e^{-b_2r^2})^2\left(1+\frac{3}{m_\rho r}+\frac{3}{(m_\rho r)^2}\right)\frac{e^{-m_\rho r}}{r} \nn \\
&+&
b_3(1-e^{-b_4r^2})^2\left(1+\frac{3}{m_\pi r}+\frac{3}{(m_\pi r)^2}\right)\frac{e^{-m_\pi r}}{r} ,
\eea
where $b_{1,2,3,4}$ are the fitting parameters while $m_\pi$ ($m_\rho$) is taken to be the pion mass (the rho meson mass) calculated at each pion mass.
The fit line for each pion mass is drawn in Fig.~\ref{fig:tensor} (Right).
It may be worth mentioning that the pion-nucleon coupling constant extracted from the parameter $b_3$ in the case of the lightest pion mass ($m_\pi = 380$ MeV) gives $g_{\pi N}^2/(4\pi) = 12.1
(2.7)$, which is encouragingly close to the empirical value.

\subsection{Validity of velocity expansion in quenched QCD}
\label{sec:convergence}
\begin{figure}[bt]
\begin{center}
\includegraphics[width=0.44\textwidth,angle=0]{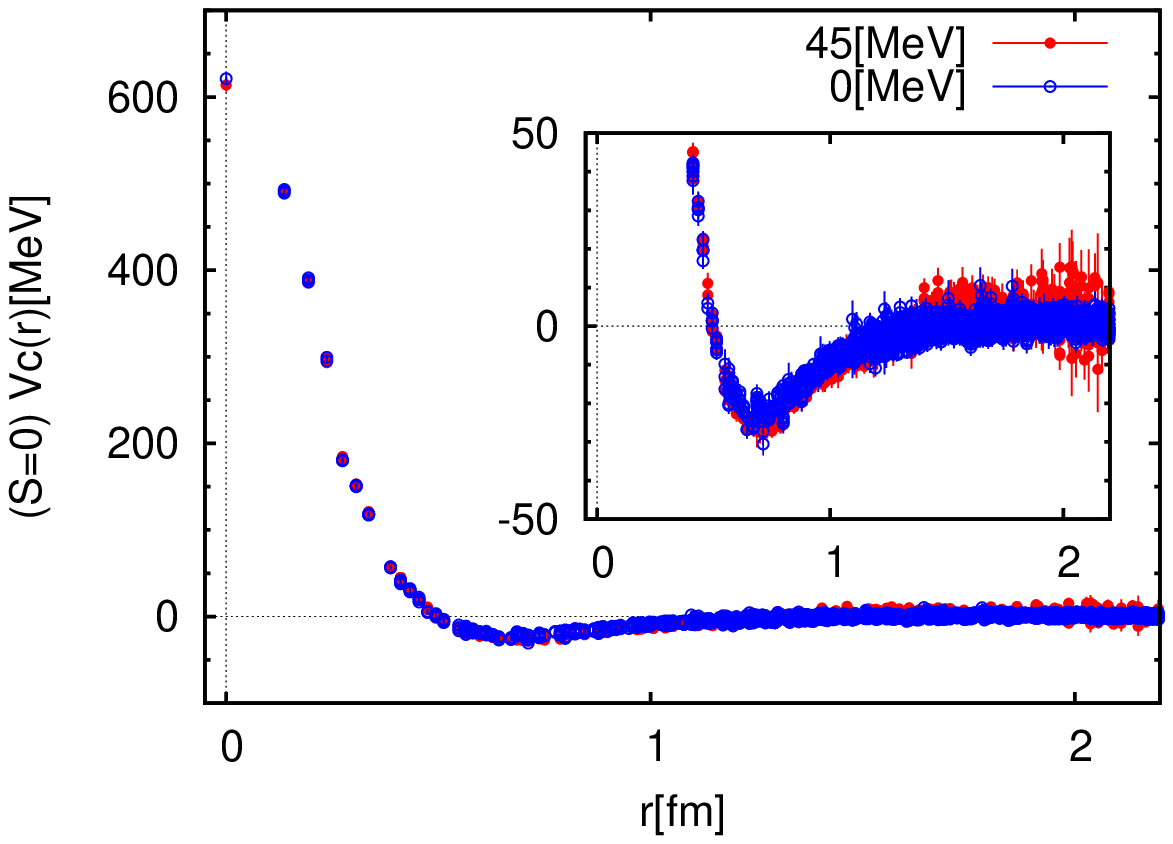}
\includegraphics[width=0.44\textwidth,angle=0]{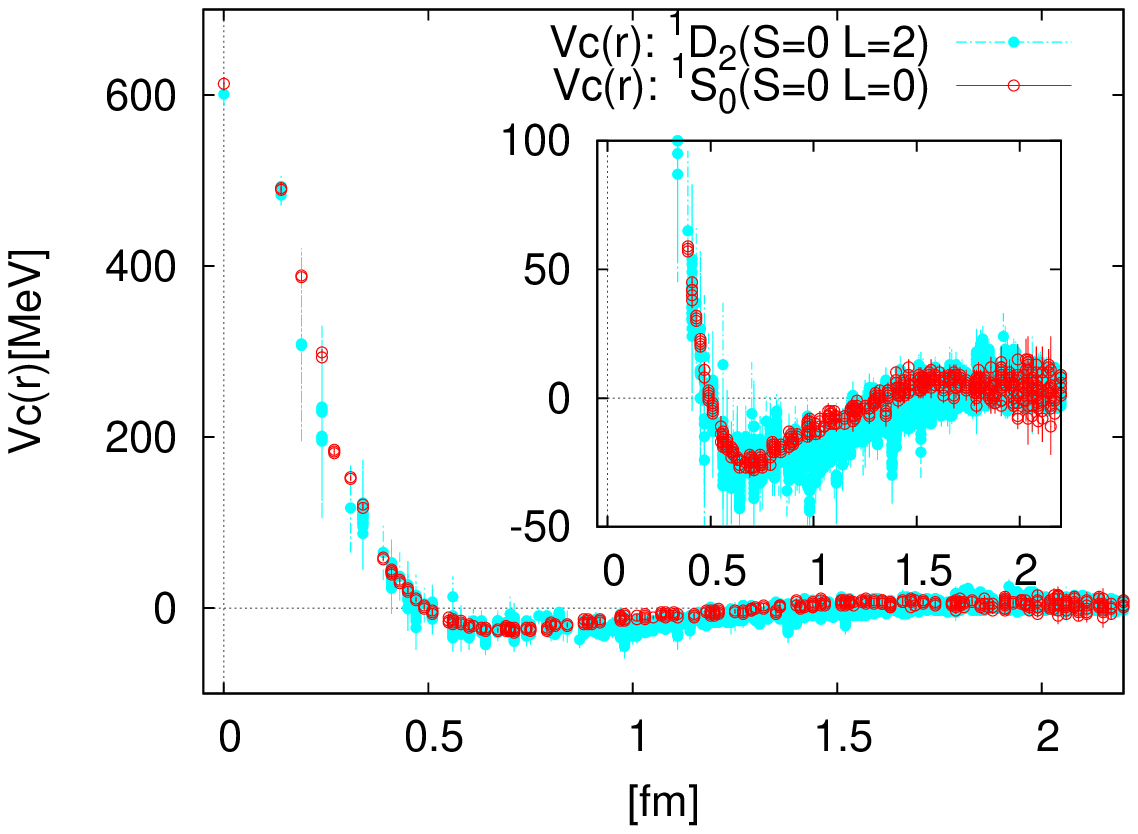}
\caption{(Left)  The spin-singlet  central  potential $V_C(r)^{(0,1)}$
  obtained from the  orbital $A_1^+$ channel at $E\simeq  45$ MeV (red
  solid  circles)  and at  $E\simeq  0$  MeV  (blue open  circles)  in
  quenched QCD at $m_\pi\simeq  529$ MeV.  (Right) The same potentials
  at   $E\simeq   45$  MeV,   obtained   from   the  orbital   $A_1^+$
  representation   (red   open   circles)   and   from   the   $T_2^+$
  representation     (cray     solid     circles).      Taken     from
  Ref.~\protect\citen{Murano:2011nz}.}
\label{fig:E-depA}
\end{center}
\end{figure}
\begin{figure}[bt]
\begin{center}
\includegraphics[width=0.44\textwidth,angle=0]{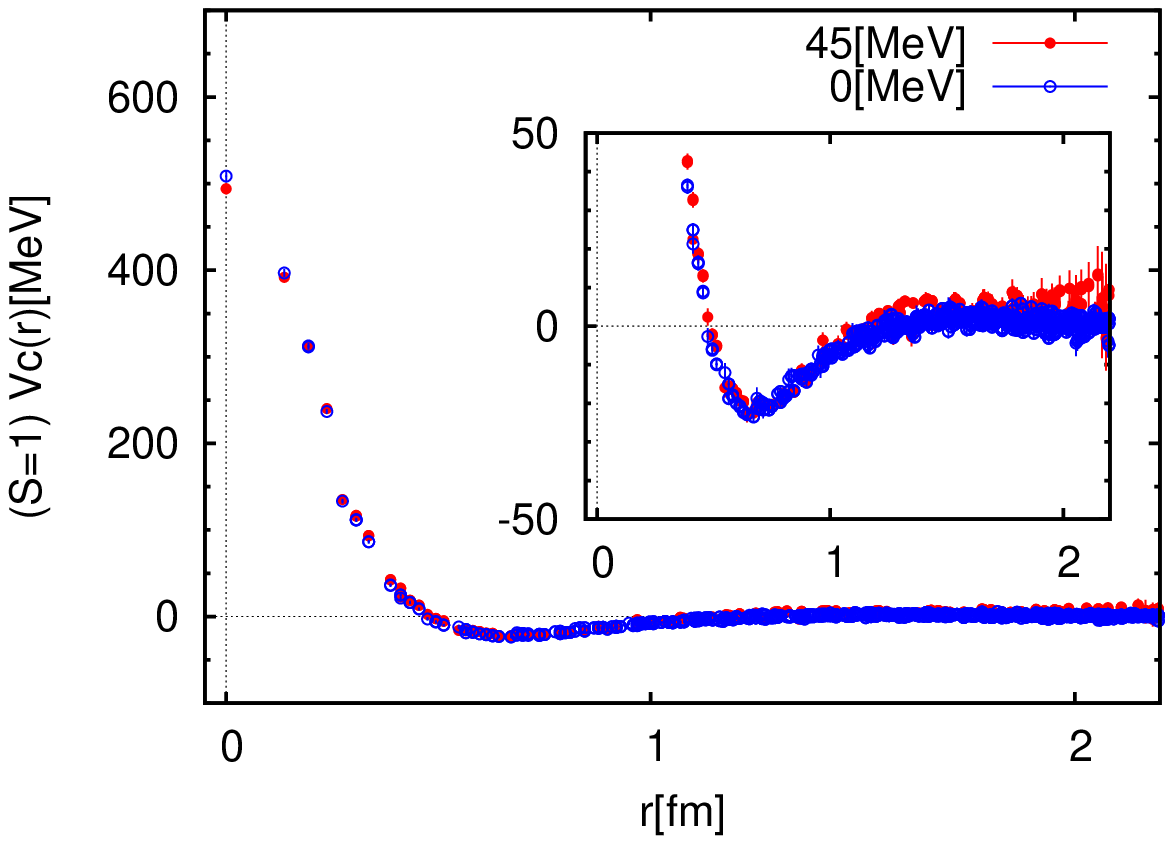}
\includegraphics[width=0.44\textwidth,angle=0]{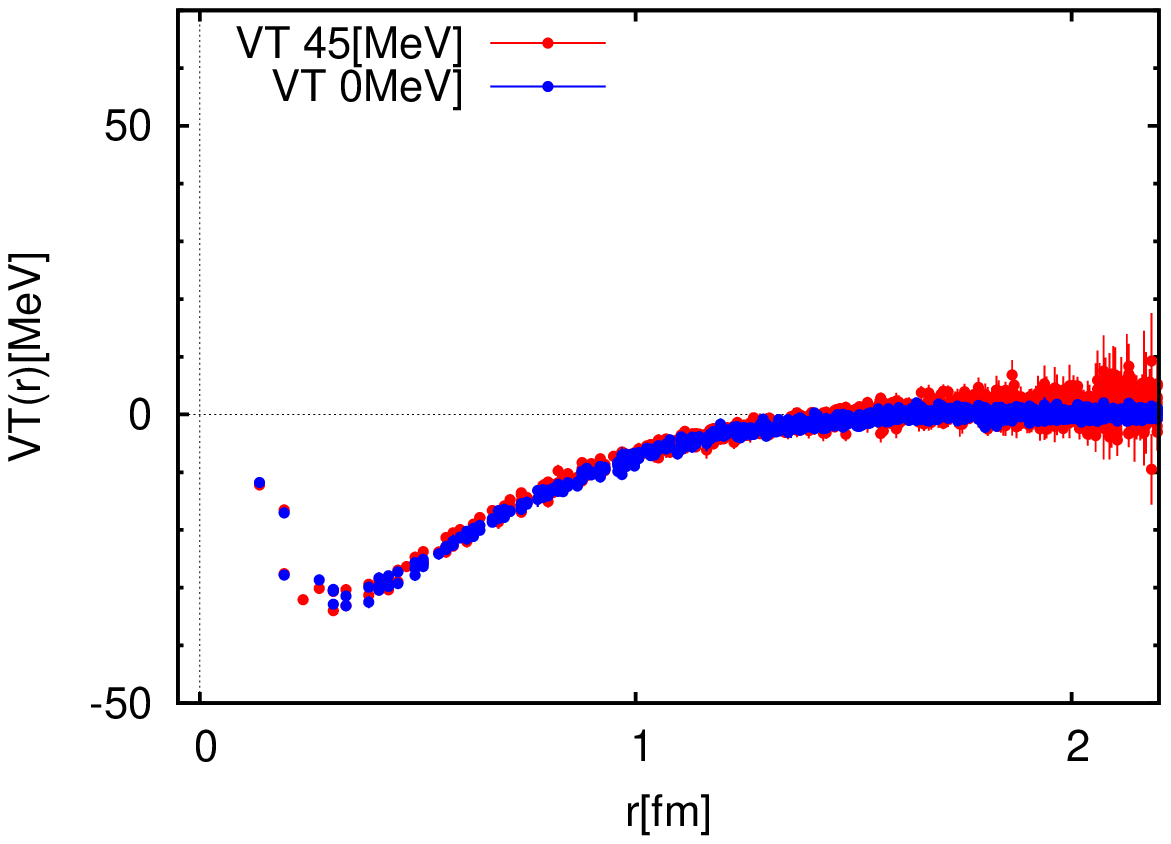}
\caption{(Left)  The spin-triplet  central  potential $V_C(r)^{(1,0)}$
  obtained from the orbital  $A_1^+-T_2^+$ coupled channel in quenched
  QCD at $m_\pi\simeq 529$  MeV. (Right) The tensor potential $V_T(r)$
  from  the  orbital  $A_1^+-T_2^+$  coupled channel.  For  these  two
  figures, symbols  are same as  in Fig.~\ref{fig:E-depA}(Left). Taken
  from Ref.~\protect\citen{Murano:2011nz}.  }
\label{fig:E-depB}
\end{center}
\end{figure}
The potentials so far are derived in the leading order of the velocity
expansion. It is therefore important to investigate the convergence of
the velocity expansion. 
If the non-locality of the
$NN$ potentials  were absent, the leading order  approximation for the
potentials would give exact  results at all energies below the inelastic threshold. The non-locality
of the potentials therefore  becomes manifest in the energy dependence
of the potentials.

To study the energy dependence,  the leading order local potentials at
$E\simeq 45$ MeV, realized by anti-periodic boundary conditions in the
spatial  directions, are  calculated in  quenched QCD  at $m_\pi\simeq
529$            MeV            and            $L\simeq            4.4$
fm\cite{Aoki:2008wy,Murano:2010tc,Murano:2010hh,Murano:2011nz}.
 In this calculation,  four types of momentum-wall sources, defined by 
 \be
 q(t_0; f)
 \equiv
 \sum_{\bf x} q({\bf x},t_0) f({\bf x}) \label{eq:momentum-wall}
 \ee 
 are employed,
 where $f({\bf x}) =\cos( (\pm x  \pm y + z)\pi/L)$. Note that $f({\bf
   x})=1$ corresponds to the wall source used in the periodic boundary
 condition.  These momentum-wall sources induce $L = T_2^+$ as well as
 $L=A_1^+$ states.
 
In Fig.~\ref{fig:E-depA}(Left),  the spin-singlet potential $V_C(r)^{(S,I)=(0,1)}$ obtained from the $L=A_1^+$ state  at $E\simeq 45$ MeV (red circles) is compared with that at $E\simeq 0$ MeV (blue circles),  while a comparison is made in Fig.~\ref{fig:E-depB}
 for the spin-triplet potentials, $V_C(r)^{(S,I)=(1,0)}$(left) and $V_T(r)$ (right).  Good agreements between results at two energies  indicate that higher order contributions are rather small in this energy interval.   In other words, these local potentials obtained at $E\simeq 0$ MeV can be safely used to describe the $NN$ scattering phase shift  in the range between $E=0$ MeV and $E=45$ MeV at this pion mass in quenched QCD. 
  
Non-locality of the potential may become manifest also in its angular momentum dependence, since the orbital angular momentum $L=\br\times\bp$ contains a derivative.
In Fig.~\ref{fig:E-depA} (Right), the spin-singlet potential $V_C(r)^{(S,I)=(0,1)}$ obtained  from the $L=T_2^+$ state, whose main component has $L=2$, is compared to the one  from the $L=A_1^+$ state, whose main component has $L=0$.
In this comparison, local potentials are determined at the same energy, $E\simeq 45$ MeV, but different orbital angular momentum.
Although the statistical errors are rather large in the case of  $L=T_2^+$, a good agreement between the two is again observed, suggesting that the $L$ dependence of the potential is small at least for the spin-singlet case. 

By these comparisons, it is observed that both energy and orbital angular momentum dependencies for local potentials are very weak within statistical errors. We therefore conclude that contributions from higher order terms in the velocity expansion are small and that the LO local potentials in the expansion obtained at $E\simeq 0$ MeV and $L= 0$ are good approximations for the non-local potentials at least up to the   energy $E\simeq 45$ MeV and orbital angular momentum $L=2$. 

\subsection{Central potential in full QCD}

Needless  to say,  it  is  important to  carry out calculations of  $NN$
potentials in full  QCD on larger volumes at  lighter pion masses. The
PACS-CS  collaboration  is performing  $2+1$  flavor QCD  simulations,
which  cover  the  physical  pion  mass\cite{Aoki:2008sm,Aoki:2009ix}.
Gauge configurations  are generated with the Iwasaki  gauge action and  
$O(a)$-improved   Wilson   quark  action   on   a
$32^3\times 64$  lattice.  The lattice spacing $a$  is determined from
$m_\pi$,  $m_K$  and $m_\Omega$  as  $a\simeq  0.089$  fm, leading  to
$L\simeq 2.9$ fm.  Three ensembles of gauge configurations are used to
calculate $NN$ potentials at $(m_\pi, m_N)\simeq $(701 MeV, 1583 MeV),
(570 MeV, 1412 MeV) and (411 MeV,1215 MeV )\cite{Ishii:2009zr} .
To overcome a difficulty to achieve ground state saturations in full QCD simulations,
the time-dependent HAL QCD method in Sec.~\ref{sec:t-dep} 
is employed\cite{HALQCD:2012aa}.

\begin{figure}[bt]
\begin{center}
\includegraphics[width=0.33\textwidth,angle=270]{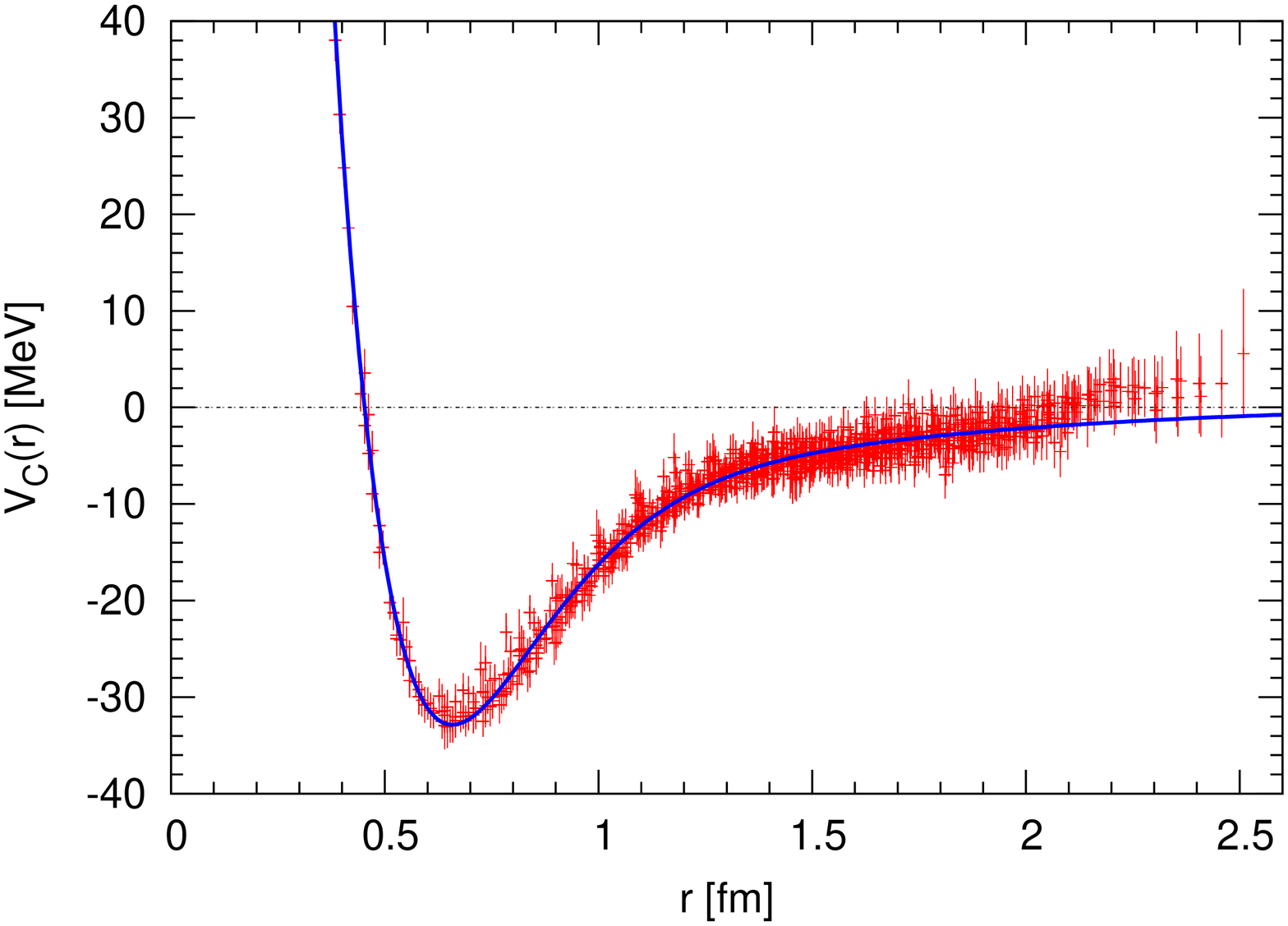}
\includegraphics[width=0.33\textwidth,angle=270]{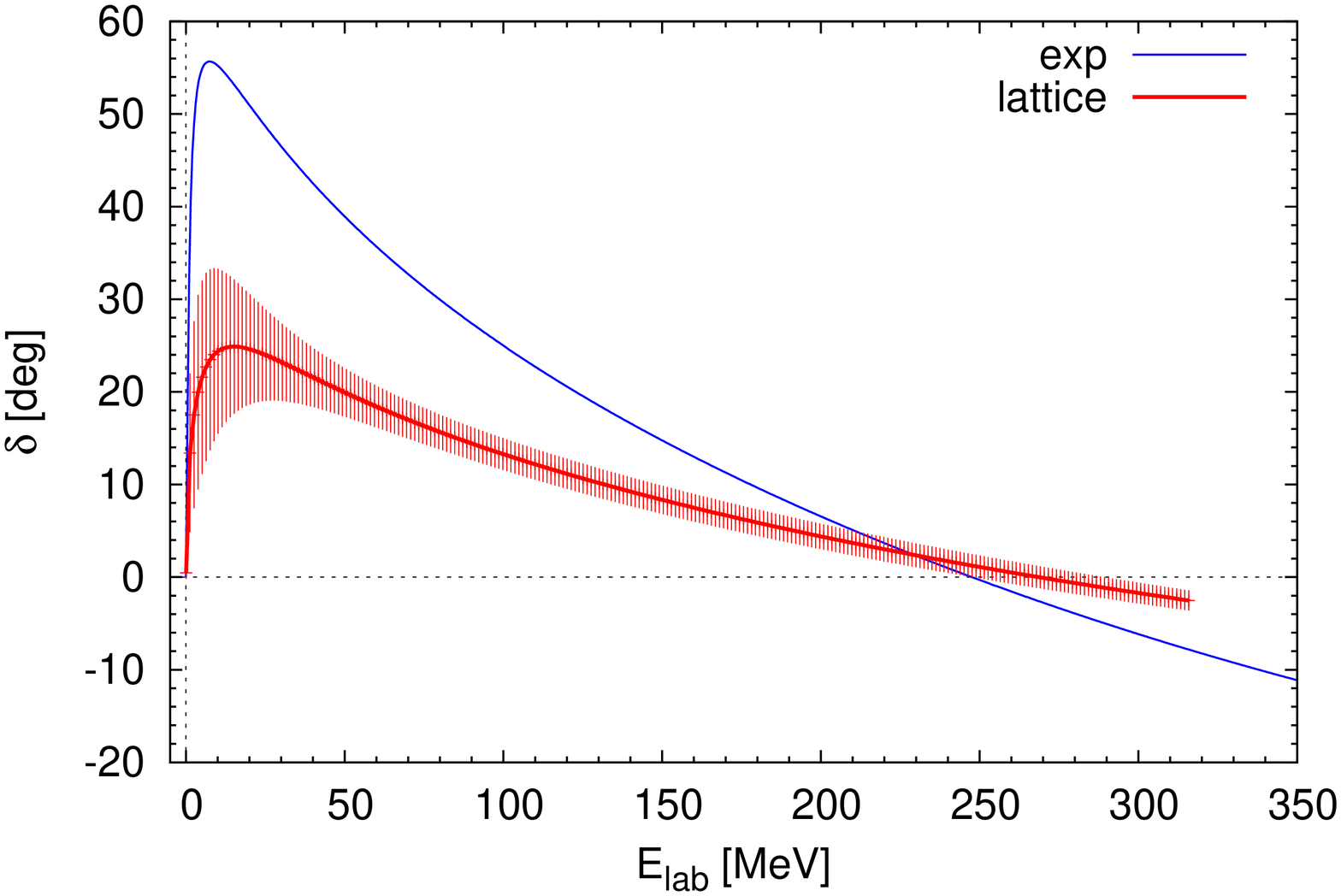}
\caption{(left)  The  multi-Gaussian fit  of  the  central
  potential $V_{\rm C}(r)$ with $N_{\rm  Gauss} = 5$. 
    (right) The  scattering  phase in  $^1S_0$  channel in  the
  laboratory  frame obtained  from the  lattice $NN$  potential, together
  with experimental data~\citen{nn-online}.}
\label{fig:full}
\end{center}
\end{figure}

Fig.~\ref{fig:full}(Left)  shows the  spin-singlet $NN$  central potential $V_C(r)$  obtained at  $E\simeq 0$ from the  PACS-CS configurations with  $m_\pi\simeq 701$ MeV and $m_N\simeq 1583$ MeV. This central potential $V_C(r)$ is fitted with multi-Gaussian function that $g(r) =\sum_{n=1}^{N_{\rm Gauss}} V_n \exp(-\nu_n r^2)$ with fit parameters $V_n$ and $\nu_n ( > 0)$.
A solid line in the figure represents a fit result with $N_{\rm Gauss}=5$. 

We solve the Schr\"odinger equation in $^1S_0$ channel with this fitted potential $V_C(r)$, in order to calculate the scattering phase shift.
Fig.~\ref{fig:full}(Right) shows the scattering phase $\delta(k)$ in the laboratory frame, together with the experimental data\cite{nn-online} for a comparison. A qualitative feature of the experimental data is well reproduced by the lattice potential, though the strength is weaker, most likely due to the heavier pion mass, $m_\pi\simeq 701$ MeV. The scattering length obtained from the derivative of the phase shift at $k=0$ becomes $a(^1S_0)=\lim_{k\rightarrow 0}\tan \delta(k)/k = 1.6(1.1)$ fm, which is compared to the experimental value $a^{\rm exp}(^1S_0)\simeq 20$ fm.    

\subsection{Nuclear force in odd parity sector and the spin-orbit force in full QCD}
\label{eq:parity_odd}

In this subsection, we consider  the potentials in odd parity sectors. 
Together  with  the  nuclear   forces  in  even  parity  sectors,  the
information   on  odd   parity  sectors   is  necessary   in  studying
many-nucleon systems with Schr\"odinger equations.
In  particular,  we  are  interested  in the  spin-orbit (LS)  force, which
  gives  rise  to  a part of the 
spin-orbit  coupling  in  the  average  single-particle  potential  of
nuclei.
It is also expected to induce superfluidity in neutron stars by providing
an attraction between two neutrons in $^3P_2$ channel\cite{Tamagaki1993}.

The LS force appear at NLO of the derivative expansion as
\be
\left[H_0+V_C(r)^{(S,I)} + V_T(r) S_{12} + V_{LS}(r)  {\bL}\cdot{\bS}
 \right]\varphi^W(\br;J^{-},I) = E_k \varphi^W(\br;J^{-},I)
\ee
To  obtain  three unknown potentials, $V_C$, $V_T$ and $V_{LS}$,
we need three independent  NBS  wave functions.
We therefore generalize the two-nucleon source for  odd parity sectors, 
by imposing a momentum on the composite nucleon fields as
\begin{eqnarray}
  \mathcal{J}_{\alpha\beta}(t_0;f^{(i)})
  \equiv
  N_{\alpha}(t_0;f^{(i)})
  N_{\beta} (t_0;f^{(i) *}) \quad \mbox{for} \quad i=\pm1,\pm2, \pm3,
  \label{eq:source-ls2}
\end{eqnarray}
where $N$ denotes a composite nucleon source field carrying a momentum,
\begin{eqnarray}
  N_{\alpha} (t_0;f^{(i)})
  \equiv
  \sum_{{\bf x}_1, {\bf x}_2, {\bf x}_3}
  \epsilon_{abc}
  \left(u_a^T({\bf x}_1)C\gamma_5 d_b({\bf x}_2)\right)
  q_{c,\alpha}({\bf x}_3)
  f^{(i)}({\bf x}_3) ,
\end{eqnarray}
with   $f^{(\pm j)}({\bf x})=\exp[\pm 2\pi  i x_j/L]$.
The star ``*'' in  the r.h.s.  of Eq.~(\ref{eq:source-ls2}) represents
the complex conjugation, which is  used to invert the direction of the
plane wave.
A   cubic   group  analysis   shows   that   the  two-nucleon   source
Eq.~(\ref{eq:source-ls2})    contains    the   orbital    contribution
$A_1^+\oplus E^+  \oplus T_1^-$,  whose main components are  S-wave,
D-wave and P-wave, respectively.
Thus the  two-nucleon source Eq.~(\ref{eq:source-ls2})  covers all the
two-nucleon channels with $J \le  2$.

For   the spin-triplet    odd-parity   sector,   Eq.~(\ref{eq:source-ls2})
generates the  lowest-lying NBS wave functions for  $(J^P,I) = (A_1^-,
1), (T_1^-, 1),  (E^-, 1)$ and $(T_2^-, 1)$,  which roughly correspond
to $J^P = 0^-, 1^-, 2^-$ and $2^-$, respectively.
Among these, we consider Schr\"odinger equations 
for three  NBS wave functions in $J^P=A_1^-, T_1^-,
T_2^-$ as
\begin{equation}
  \left[
    H_0
    + V_C(r)
    + V_T(r) S_{12}
    + V_{LS}(r) {\bL}\cdot{\bS}
  \right]
  \varphi^W(\br;J^P)
  =
  E_0(J^P)
  \varphi^W(\br;J^P) , 
   \label{eq:ls-schrodinger}
\end{equation}
where $E_0(J^P)= k^2/m_N$  from the lowest-lying energy $W = 2\sqrt{m_N^2  + k^2}$ for the $J^P$  sector. 
In order to obtain $V_C(r)$, $V_T(r)$ and $V_{LS}(r)$ in odd parity sectors,
Eqs.~(\ref{eq:ls-schrodinger}) for $J^P = A_1^-,T_1^-,T_2^-$, which correspond to $^3P_0$, $^3P_1$ and $^3P_2+^3F_2$, are solved.

Numerical calculations  are performed  by using  2 flavor  QCD gauge
configurations  on   $16^3\times  32$  lattice   generated  by  CP-PACS
Collaboration \cite{AliKhan:2001tx,CPPACS},
with Iwasaki gauge action  at $\beta = 1.95$ and $O(a)$ improved
Wilson (clover) quark action at  $\kappa = 0.1375$. This setup leads to
the lattice  spacing $a^{-1}  = 1.27$ GeV  ($a \simeq 1.555$  fm), the
pion  mass $m_{\pi} \simeq  1136$ MeV,  the nucleon  mass $m_{N}\simeq
2165$ MeV. The spatial extension amounts to $L=16a \simeq 2.5$ fm.

\begin{figure}[tb]
\begin{center}
\includegraphics[width=0.32\textwidth,angle=0]{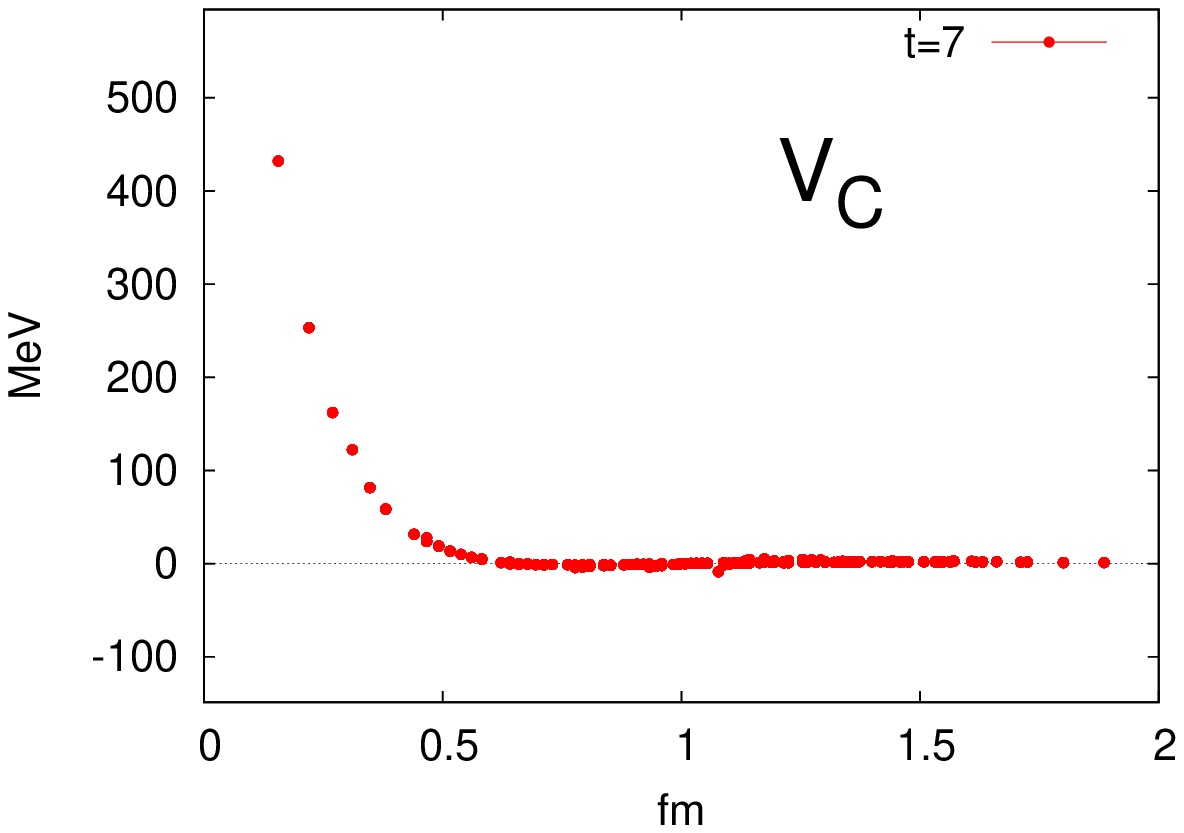}
\includegraphics[width=0.32\textwidth,angle=0]{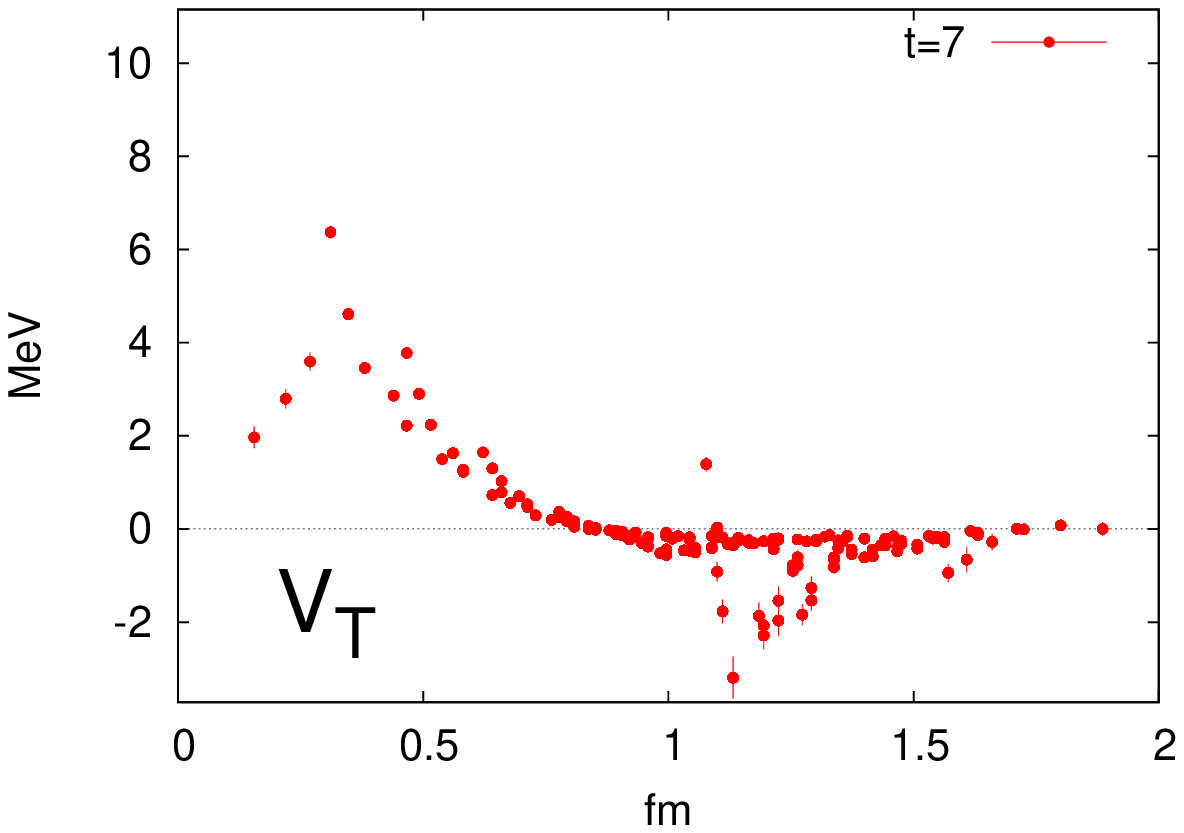}
\includegraphics[width=0.32\textwidth,angle=0]{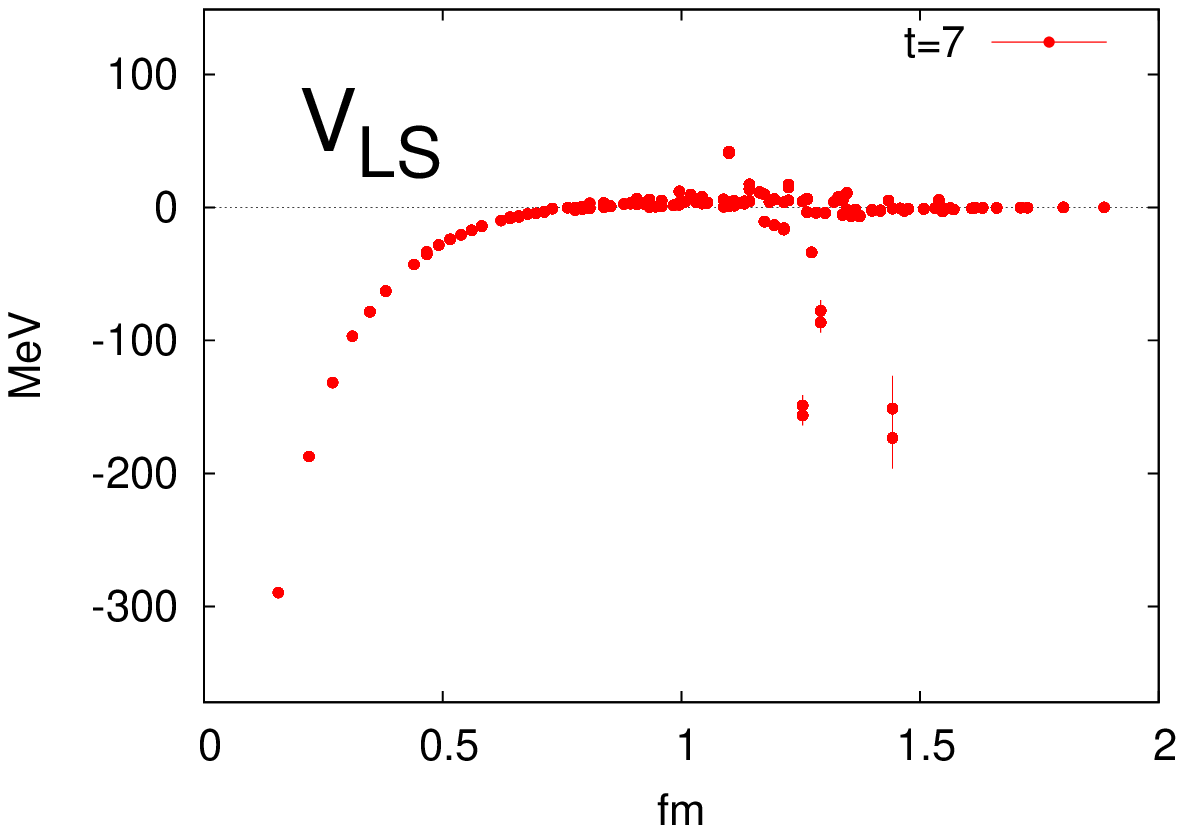}
\caption{Potentials in odd parity sector obtained from $^3P_0$, $^3P_1$ and $^3P_2+^3F_2$ NBS wave functions calculated at $m_\pi=1136$ MeV.
Left, middle, and right figures show central, tensor and spin-orbit force in parity odd sector, respectively. 
}
\label{fig:LS}
\end{center}
\end{figure}

Fig.~\ref{fig:LS} shows preliminary results of the central potential $V_C(r)$,
tensor potential  $V_{T}(r)$ and  the spin-orbit force  $V_{LS}(r)$ in
the spin-triplet odd parity sector.
They  have  the following  qualitative  features.   (1)  $V_C(r)$  has
repulsive core at  short distance.  (2) $V_T(r)$ is  positive and very small.
(3)  $V_{LS}(r)$  is large  and  negative  at  short distance.   These
features   qualitatively  agree   with  those   of  phenomenological
potentials \cite{Wiringa:1994wb}.

\section{Hyperon Interactions}
\label{sec:hyperon}
Study of hyperon-nucleon ($YN$) and hyperon-hyperon ($YY$) 
interactions is 
one of the challenges in contemporary nuclear physics. 
These potentials give a key to   
understand nuclear many-body systems with strangeness. 
Also they are essential to explore 
the structure of the neutron star core, where  strangeness degree of 
freedom is expected to appear. 
At present, 
experimental data on $YN$ and $YY$ scatterings 
are not sufficient to make precise constraints on  the hyperon potentials,
while 
spectroscopic studies of $\Lambda$ hypernuclei,  
performed by employing various reactions 
such as $(\pi^+,K^+)$, $(K^-,\pi^-)$ and $(e,e^\prime K^+)$
\cite{Gibson:1995an,Hashimoto:2006aw},
give some information on the $\Lambda N$ interactions.
Under these circumstances,
studies on the basis of  lattice QCD is quite important
as an alternative method to access $YN$ and $YY$ interactions.
In this section we mainly consider potentials in the strangeness $S=-1$ sector,
obtained from  $2+1$ flavor lattice QCD simulations with  PACS-CS gauge 
configurations.
A study on   
potentials between octet barons
in the flavor SU(3) limit and 
coupled channel analysis on 
potentials
in the strangeness $S=-2$ sector beyond the SU(3) limit
will be discussed in the next two sections.


\subsection{$\Lambda N$ and $\Sigma N$ potentials in full QCD}

The $\Lambda N$ 
and 
the $\Sigma N$ ($I=3/2$) are the lowest states in the strangeness
$S=-1$ systems with  $I=1/2$ and $I=3/2$, respectively. 
Therefore potentials for these states can be calculated as in the case of $NN$ potential. 
In Ref.~\citen{Nemura:2012fm}, 
the $\Lambda N$ potential and the $\Sigma N$ potential with $I=3/2$ 
are calculated by using 2+1 flavor full QCD gauge configurations with the
 original time-independent HAL QCD method.
 In the following, we show improved results 
 on a $32^3\times 64$ lattice at  $a= 0.091(1)$ fm with 
 the time-dependent HAL QCD method discussed in Sec.~\ref{sec:t-dep}.

\begin{figure}[t]
 \centering \leavevmode
 \begin{minipage}[t]{0.49\textwidth}
  \includegraphics[width=.99\textwidth]
  {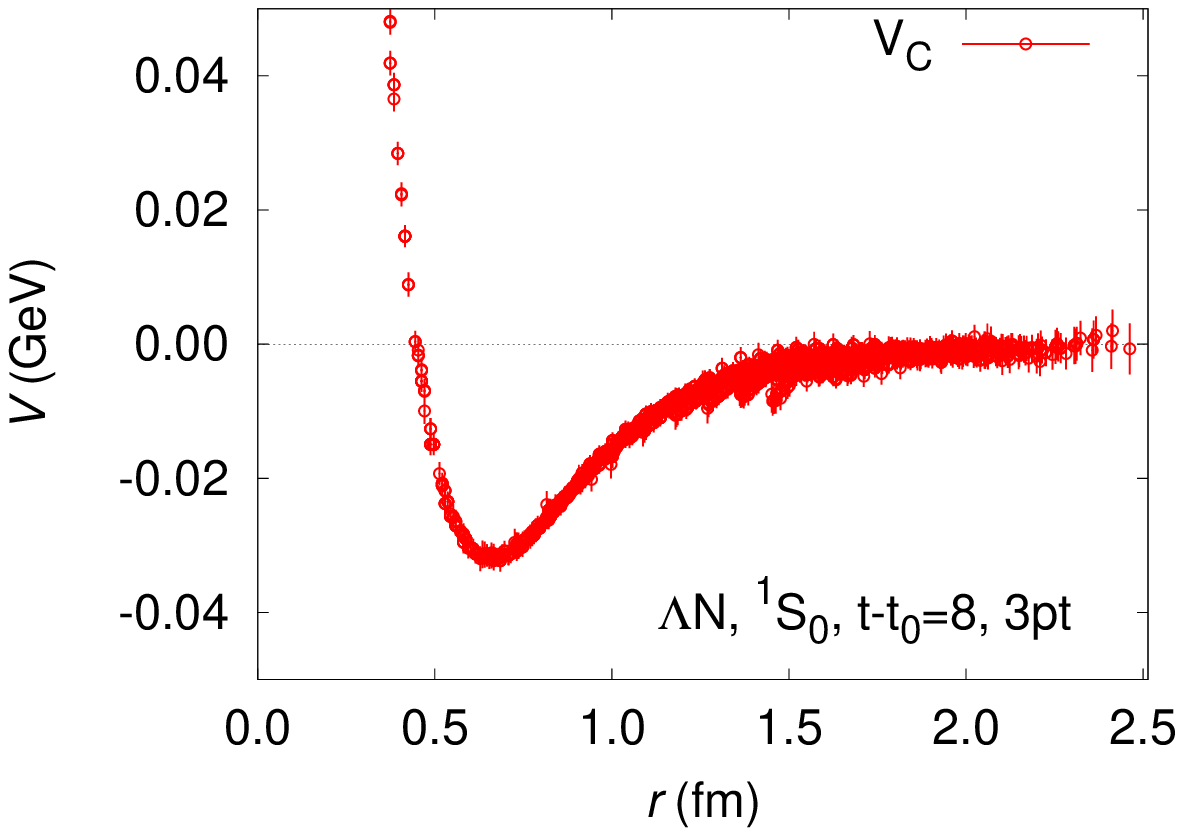}
  \footnotesize
 \end{minipage}~
 \begin{minipage}[t]{0.49\textwidth}
  \includegraphics[width=.99\textwidth]
  {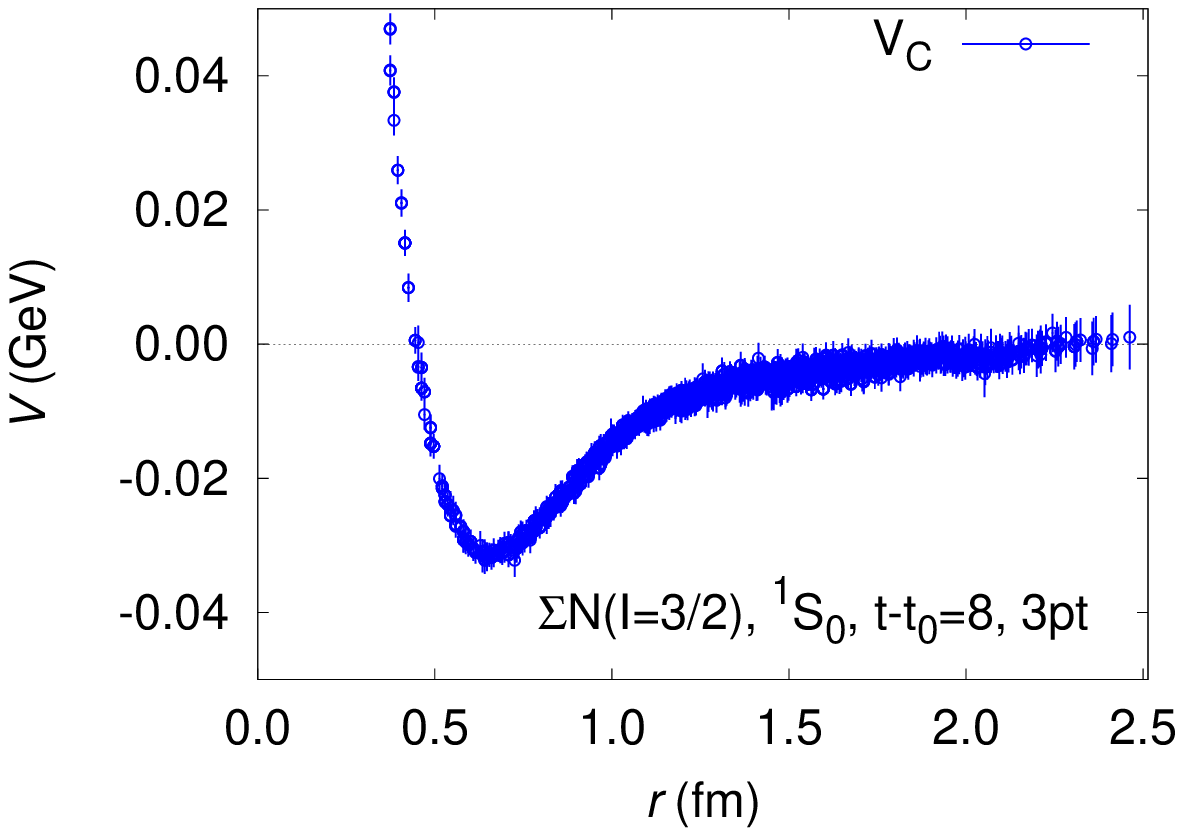}
 \end{minipage}
 \caption{
 Left:
 The central potential 
 in the $^1S_0$ channel of the $\Lambda N$ system
 in $2+1$ flavor QCD as a function of $r$.
 Right:
  The central potential 
 in the $^1S_0$ channel of the $\Sigma N(I=3/2)$ system
 as a function of $r$.
 }
 \label{LN_and_SN2I3_VC_1S0}
\end{figure}
The $\Lambda N$ (left panel) and the 
$\Sigma N(I=3/2)$ (right panel) potentials 
in the $^1S_0$ channel are shown in
Fig.~\ref{LN_and_SN2I3_VC_1S0}. 
In the 2+1 flavor QCD, 
while the $\Sigma N$ ($I=3/2$) potential still belongs directly 
to the $\mathbf{27}(I=3/2)$ representation thanks to the isospin symmetry,
an energy eigenstate of a $\Lambda N$ system in the $^1S_0$ channel 
 is a mixture of $\mathbf{27} (I=1/2)$ and $\mathbf{8}_s$ in the flavor representation,
so that these two potentials are not necessarily equal. 
In the present $2+1$ flavor QCD calculation shown in Fig.~\ref{LN_and_SN2I3_VC_1S0}, 
 these potentials look similar due to small flavor-SU(3) breaking:
 For example, our hadron masses are
 $(m_\pi, m_K, m_N, m_\Lambda, m_\Sigma)
=(0.7006(4), 0.7879(4), 1.574(3), 1.635(3), 1.650(3))$ GeV.
\begin{figure}[t]
 \centering \leavevmode
 \begin{minipage}[t]{0.49\textwidth}
  \includegraphics[width=.99\textwidth]
  {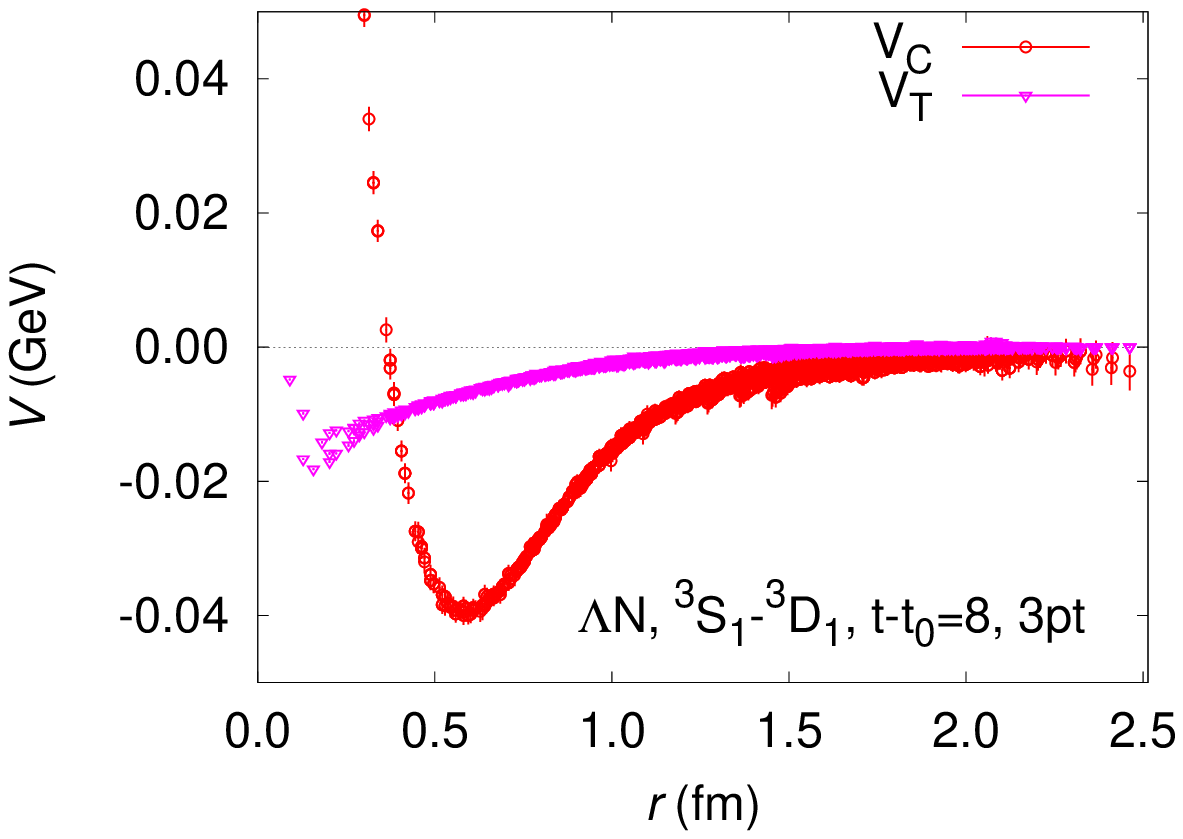}
  \footnotesize
 \end{minipage}
 \hfill
 \begin{minipage}[t]{0.49\textwidth}
  \includegraphics[width=.99\textwidth]
  {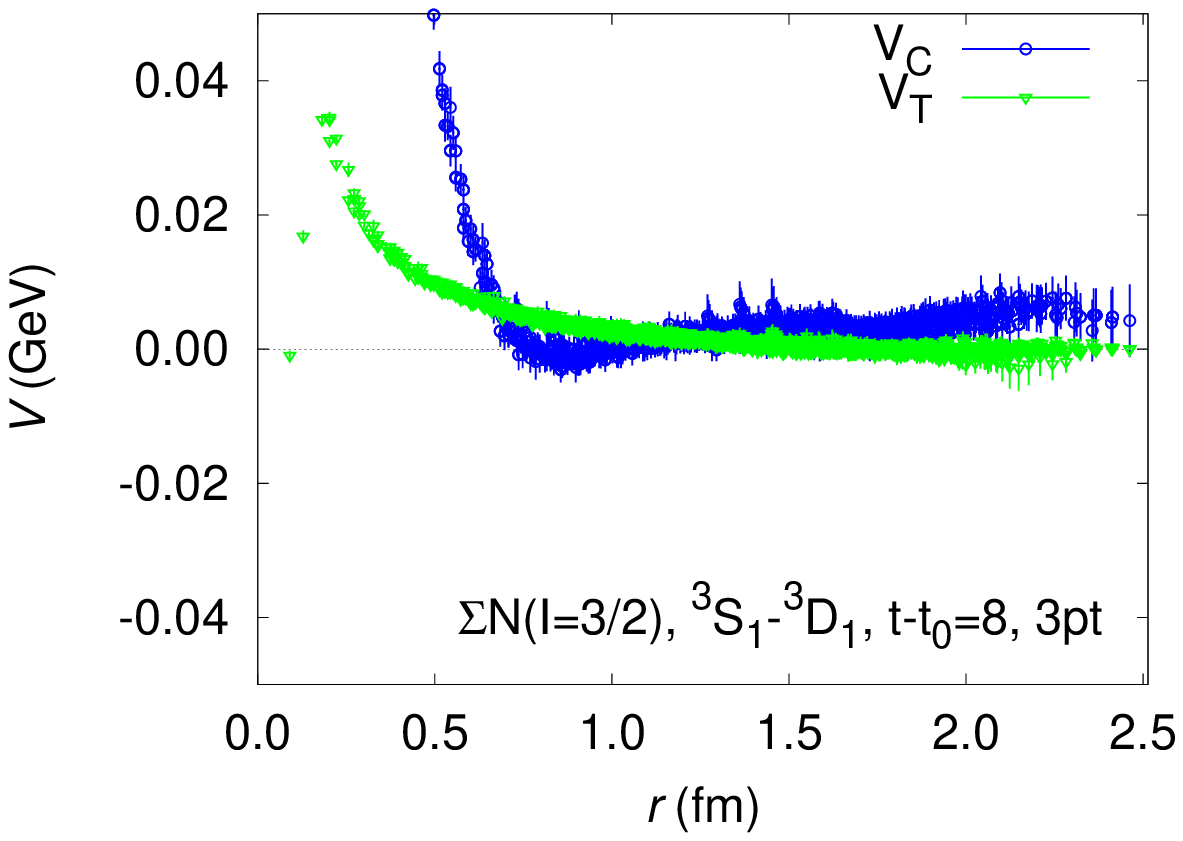}
 \end{minipage}
 \caption{
 Left:
 The central potential (circle) and the tensor potential (triangle) 
 in the $^3S_1-^3D_1$ channel of the 
 $\Lambda N$ system as a function of $r$. 
 Right:
  The central potential (circle) and the tensor potential (triangle) 
 in the $^3S_1-^3D_1$ channel of the
 $\Sigma N(I=3/2)$ system as a function of $r$. 
}
 \label{LN_and_SN2I3_VCT_3E1}
\end{figure}
The left panel of 
Fig.~\ref{LN_and_SN2I3_VCT_3E1} shows the central  potential (circle) 
and the tensor potential (triangle) of the $\Lambda N$ system 
in the $^3S_1-^3D_1$ channel, 
whose eigenstate is a mixture of $\overline{\mathbf{10}}$ and $\mathbf{8}_a$.
The attractive well at  distance $r\approx 0.6$~fm 
is deeper than that of the $\Lambda N$ central potential in the $^1S_0$ channel, 
while the tensor potential itself (triangle) is weaker than the tensor potential 
in the $NN$ system\cite{Ishii:2010th}. 

The right panel of 
Fig.~\ref{LN_and_SN2I3_VCT_3E1} shows the central potential (circle) 
and the tensor potential (triangle) of the $\Sigma N (I=3/2)$ system
in the $^3S_1-^3D_1$ channel. 
Due to the isospin symmetry,
this channel belongs solely to the flavor $\mathbf{10}$ representation without 
mixture of $\overline{\mathbf{10}}$ or $\mathbf{8}_a$
As seen from the figure, 
there is no clear attractive well in the central potential (circle).
This repulsive nature of the $\Sigma N (I=3/2, ^3S_1-^3D_1)$ 
central potential is consistent with the prediction from the 
naive quark model\cite{Oka:2000wj}. 
The tensor force is a little stronger that that of the $\Lambda N$ system but is still 
 weaker in magnitude than that of the $NN$ system.  


\subsection{$\Xi N$ potential in quenched QCD }
\begin{figure}[bt]
\begin{center}
\includegraphics[width=0.45\textwidth,angle=0]{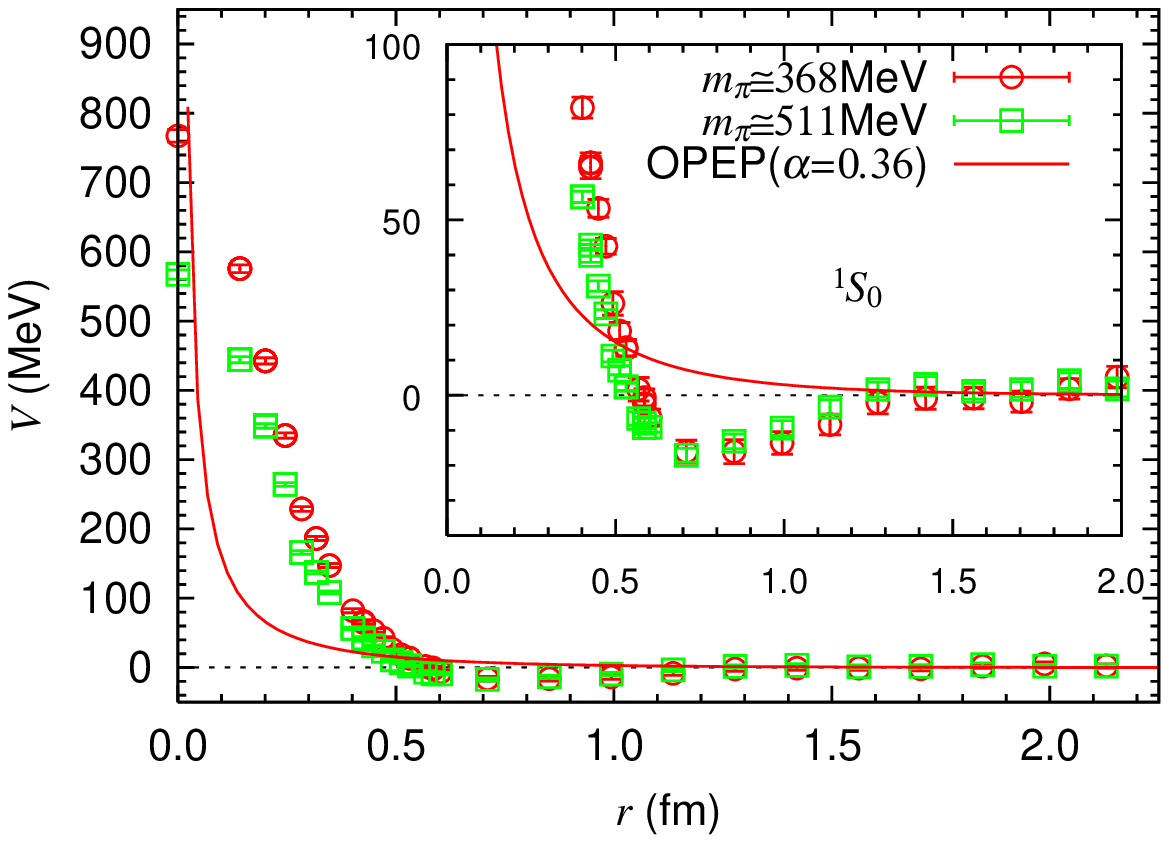}
\includegraphics[width=0.45\textwidth,angle=0]{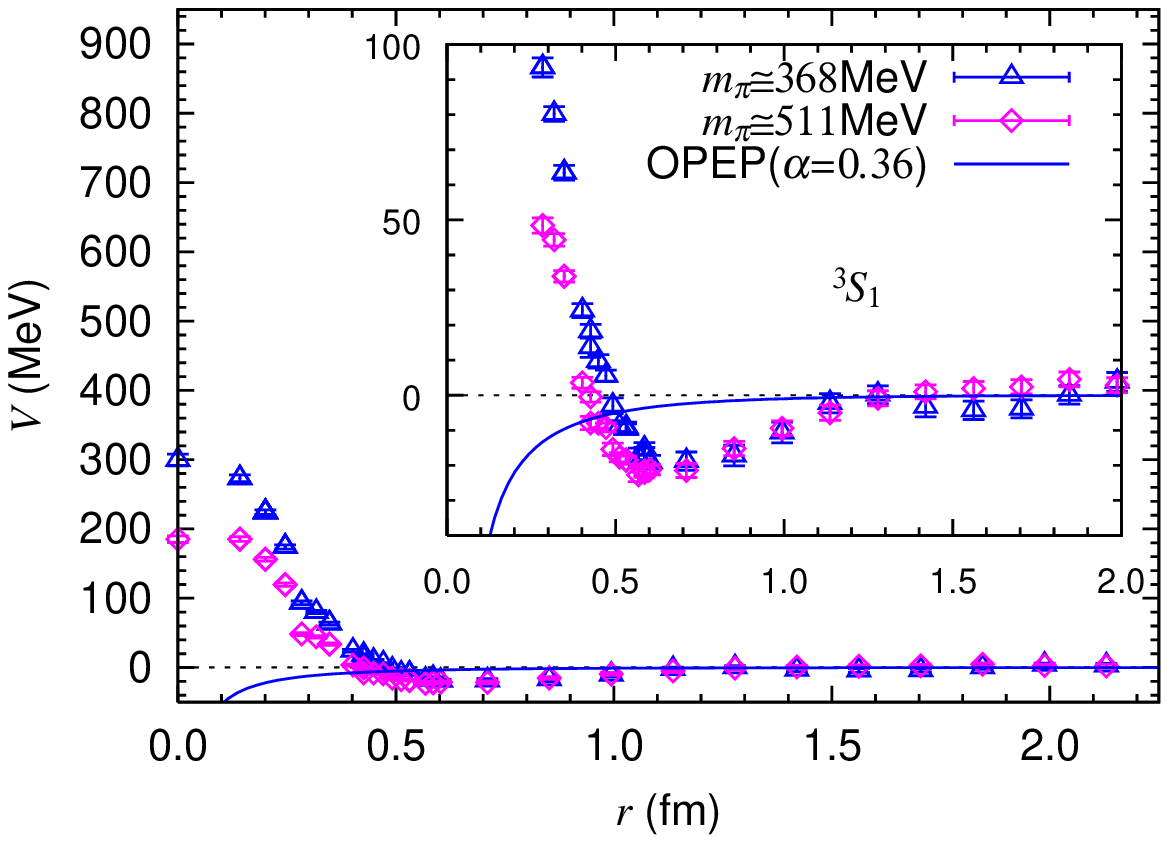}
\caption{ (Left)
 The  spin-singlet central potential for $p\Xi^0$ obtained from the orbital $A_1^+$ channel 
 at $m_\pi\simeq 368$ MeV
 (circle) and $m_\pi\simeq 511$ MeV (box). 
 The central part of the OPEP ($F/(F+D)=0.36)$ in
 Eq.~(\ref{eq:opep}) is also given by solid line. 
 (Right) The spin-triplet effective central potential from the orbital $A_1^+$ channel
 at $m_\pi\simeq 368$ MeV (triangle) and $m_\pi\simeq 511$ MeV (diamond), together with the OPEP (solid line). Taken from Ref.~\citen{Nemura:2008sp}.}
\label{fig:NXi}
\end{center}
\end{figure}
Experimentally, not much information is
 available on the $N\Xi$ interaction except for a few studies: a recent report gives the upper limit of elastic and inelastic cross sections\cite{XN_Ahn} while earlier publications suggest weak attractions of $\Xi-$ nuclear interactions\cite{Nakazawa,Fukuda,Khaustov}. The $\Xi-$nucleus interactions will be soon studied as one of the day-one experiments at J-PARC\cite{JPARC} via $(K^-,K^+)$ reaction with a nuclear target. 
Ref.~\citen{Nemura:2008sp} gives the first result of the potential in $I=1$ $N\Xi$ system, 
 which does not show strong decay into other channels.
Lattice parameters are the same as for the quenched $NN$ potential in Sec.\ref{ssec:CPQ},
but the method to determine the lattice spacing in Ref.~\citen{Nemura:2008sp} is a little different from the one in Sec.\ref{ssec:CPQ}. 
The potential is calculated at $(m_\pi, m_N, m_\Xi) = $ (511(1) MeV, 1300(4) MeV, 1419(4) MeV)  and   (368(1) MeV, 1167(7) MeV, 1383(6) MeV) with the interpolation operators
\be
p_\alpha (x) =\varepsilon_{abc} (u_a^T(x)C\gamma_5 d_b(x)) u_{c,\alpha}(x), \quad
\Xi_\alpha^0(x) = \varepsilon_{abc} (u_a^T(x)C\gamma_5 s_b(x)) s_{c,\alpha}(x) .
\ee 
Since both $p$ and $\Xi^0$ have $(I, I_z)=(1/2,1/2)$, the $p\Xi^0$ system has $I=1$ with the strangeness $S=-2$.

The left (right) of Fig.~\ref{fig:NXi} gives the (effective) central potential of the $p\Xi^0$ system obtained from the $L=A_1^+$ representation for the spin-singlet (triplet) at $m_\pi = 511$ MeV and 368 MeV. Potentials in the $I=1$ $N\Xi$ system for both channels show a repulsive core at $r\le 0.5$ fm surrounded by an attractive well, similar to the $NN$ systems. In contrast to the $NN$ case, however, the repulsive core of the $p\Xi^0$ potential in the spin-singlet channel is substantially stronger than in the triplet channel. The attraction in the medium to long distance region ( 0.6 fm $ \le r \le 1.2$ fm ) is similar  in both channels. 
The height of the repulsive core increases as the light quark mass decreases, while a significant difference is not seen for the attraction in the medium to long distance within statistical errors.
Potentials in Fig.~\ref{fig:NXi} are weakly attractive 
on the whole in both spin channels at both pion masses, in spite of the repulsive core at short distance, and the attraction in the triplet is a little stronger than that in the singlet.

The solid lines in Fig.~\ref{fig:NXi} are the one-pion exchange potential (OPEP), given by
\be
V_C^\pi = -(1-2\alpha)\frac{g_{\pi NN}^2}{4\pi} \frac{(\btau_N\cdot\btau_\Xi)
(\bsigma_N\cdot\bsigma_\Xi)}{3}\left(\frac{m_\pi}{2m_N}\right)^2 \frac{e^{-m_\pi r}}{r}
\label{eq:opep}
\ee
with $(m_\pi, m_N) =(368 {\rm MeV}, 1167 {\rm MeV})$,
where the pseudo-vector $\pi\Xi\Xi$ coupling $f_{\pi\Xi\Xi}$ is related to the  $\pi NN$ coupling as $f_{\pi\Xi\Xi}= -f_{\pi NN} (1-2\alpha)$ with the parameter $\alpha = F/(F+D)$, and $g_{\pi NN} = f_{\pi NN}\frac{m_\pi}{2m_N}$. The empirical vales, $\alpha \simeq 0.36$ and $g_{\pi NN}/(4\pi) \simeq 14.0$, are used for the plot.
Unlike the $NN$ potential, the OPEP in the present case has  opposite sign between the spin-singlet channel and spin-triplet channel. In addition, the absolute magnitude is smaller due to the factor $1-2\alpha$. No clear signature of the OPEP at long distance ($r\ge 1.2$ fm) is yet
observed in Fig.~\ref{fig:NXi} within statistical errors.

\section{Baryon interaction in the flavor SU(3) limit}
\label{sec:su3limit}

\subsection{Potentials in the flavor SU(3) limit}

In order to reveal the nature of the hyperon interactions in  various channels,
it is more convenient to consider an idealized flavor SU(3) symmetric world, 
where $u,d$ and $s$ quarks are all degenerate with a common finite mass.
In this limit, one can capture essential features of the interaction,
in particular, the short range force without contamination from the quark mass difference.

In the flavor SU(3) limit, the ground state baryon belongs to the flavor-octet with spin $1/2$,
and two-baryon states with a given angular momentum can be labeled by the irreducible representation of SU(3) as
\be
{\bf 8}\otimes {\bf 8} = \underbrace{{\bf 27}\oplus {\bf 8}\oplus {\bf 1}}_{\rm symmetric}\oplus 
 \underbrace{\overline{\bf 10}\oplus {\bf 10}\oplus {\bf 8}}_{\rm anti-symmetric},
\ee
where "symmetric" and "anti-symmetric" stand for the symmetry under the exchange of the flavor for two baryons.
For the system with orbital S-wave, the Pauli principle for baryons imposes {\bf 27}, {\bf 8} and {\bf 1}
to be spin-singlet ($^1S_0$), while  $\overline{\bf 10}$, {\bf 10}  and {\bf 8} to be spin-triplet ($^3S_1-^3D_1$).
Calculations in the SU(3) limit allow us to extract potentials for these six flavor irreducible multiplets
as follows.

\begin{figure}[tb]
\begin{center}
 \includegraphics[width=0.45\textwidth]{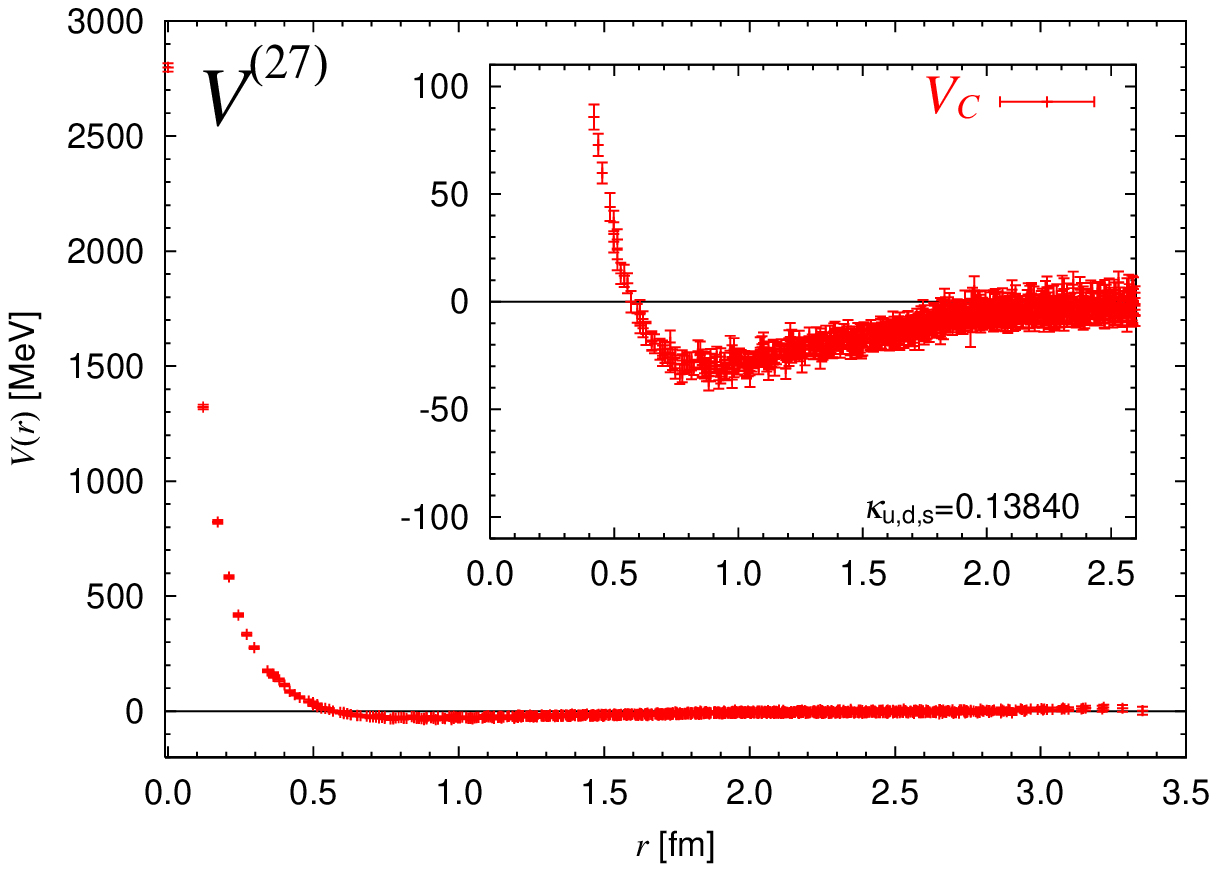}
 \includegraphics[width=0.45\textwidth]{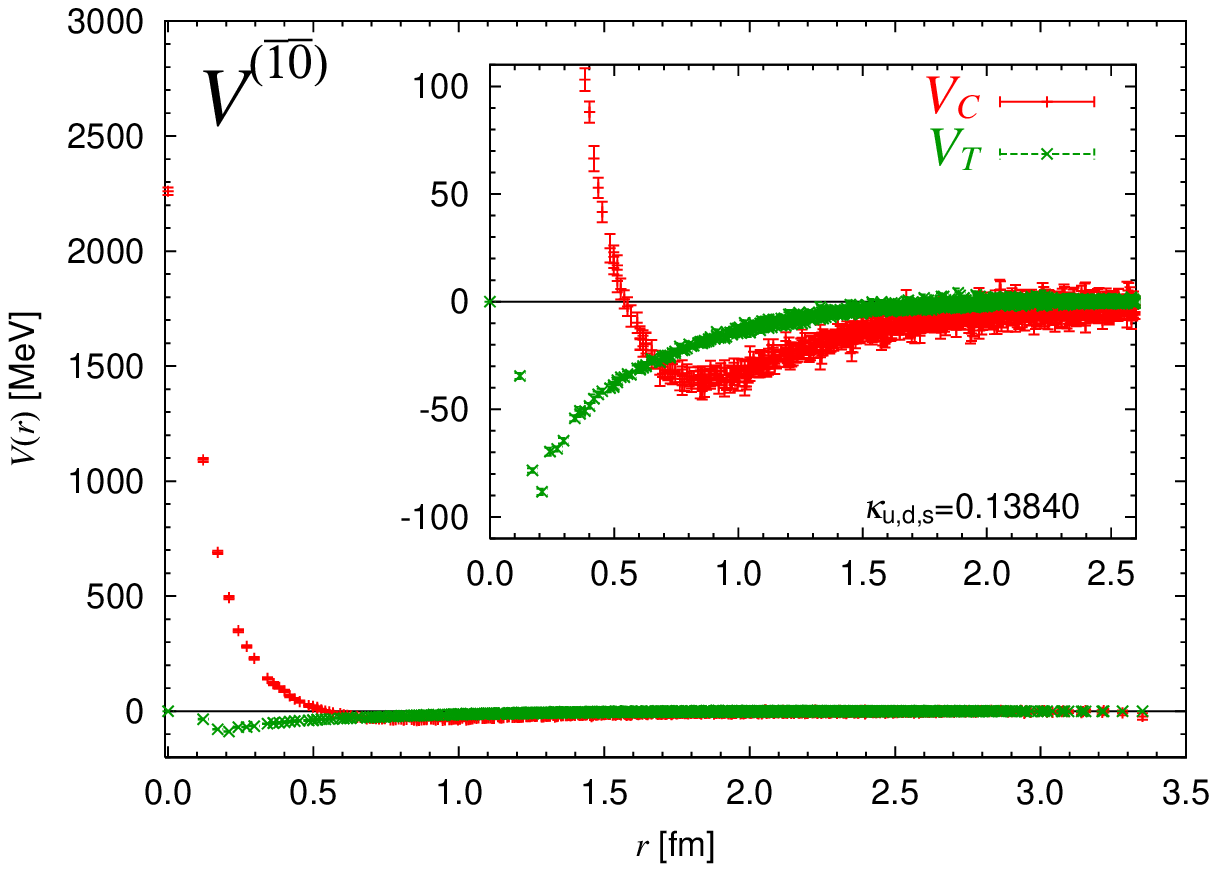}
 \end{center}
 \caption{
  The $BB$ potentials in {\bf 27} (Left) and $\overline{\bf 10}$ (Right) representations
  extracted from the lattice QCD simulation at $M_{\rm ps}=469$ MeV.
  Taken from Ref.~\protect\citen{Inoue:2011ai}.
 } 
 \label{fig:su3limitA}
\end{figure} 

A two-baryon operator ${BB}^{(X)}$ which belongs to one definite flavor representation $X$,
can be given in terms of the baryon base operator with the corresponding Clebsch-Gordan (CG) coefficients $C^X_{ij}$
as ${BB^{(X)}} =\sum_{ij} C_{ij}^X {B_i B_j}$.
By using this operator at source and/or sink, the NBS wave function for two-baryon system
in the flavor representation can be obtained.
Potentials in the flavor base, $V^{(X)}(r)$, are extracted form such wave functions
in the same manner for nuclear forces explained in Sec. \ref{sec:lattice}.

\begin{figure}[tb]
\begin{center} 
 \includegraphics[width=0.45\textwidth]{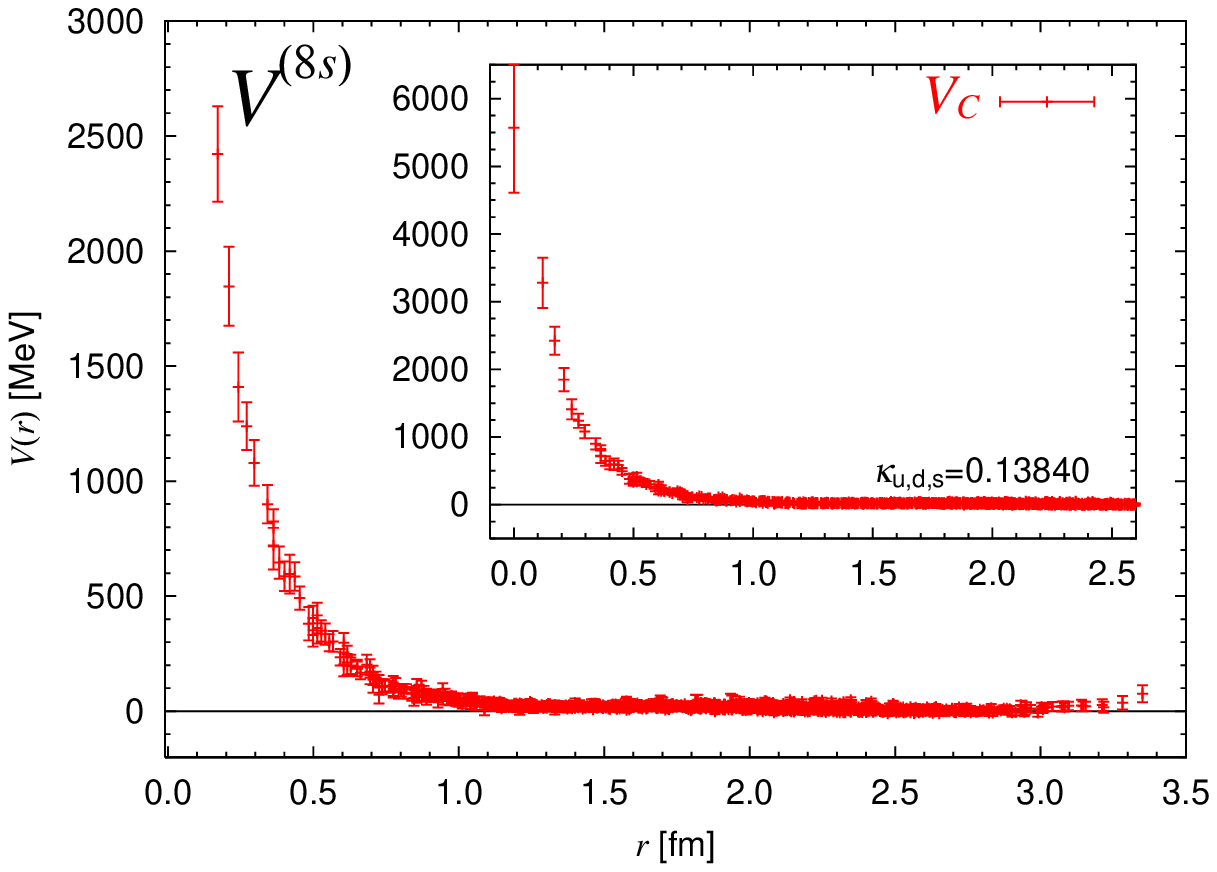}
 \includegraphics[width=0.45\textwidth]{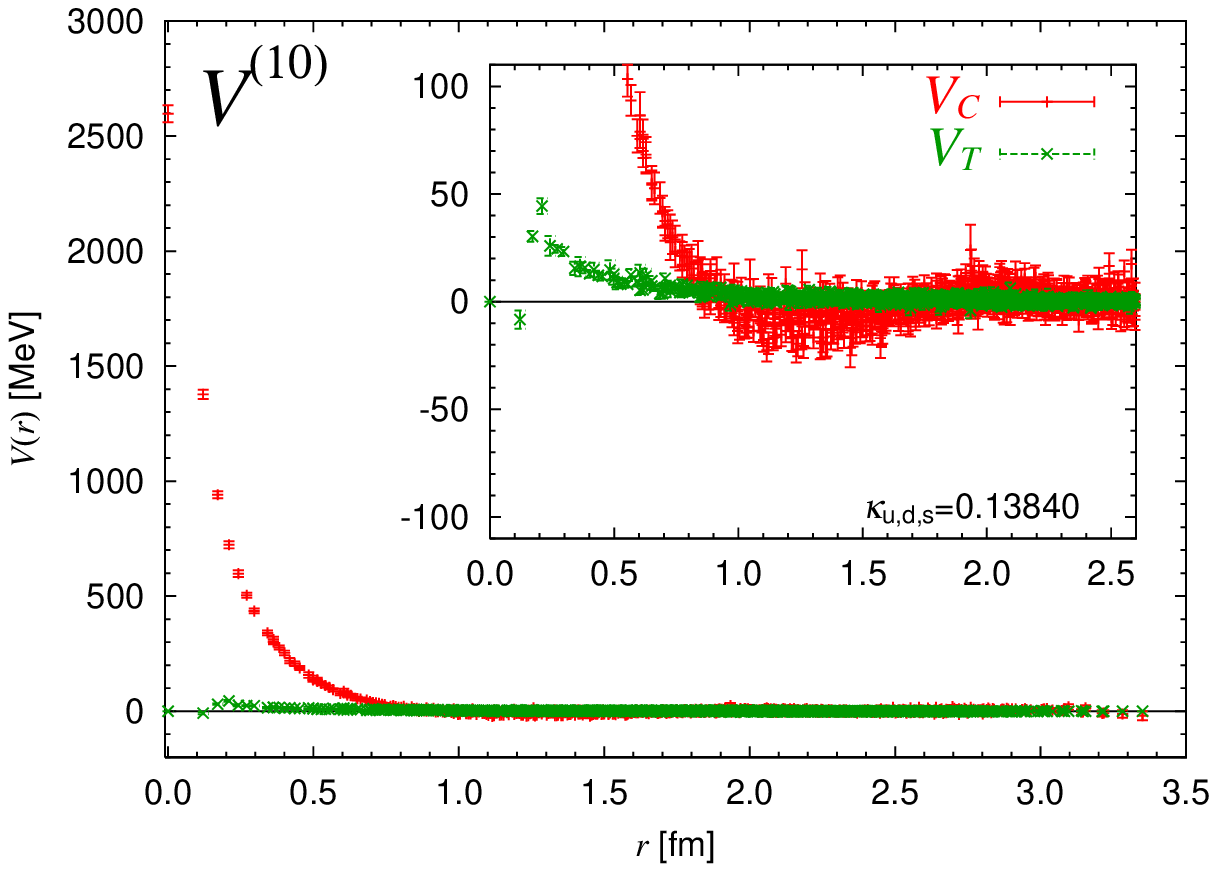}
 \includegraphics[width=0.45\textwidth]{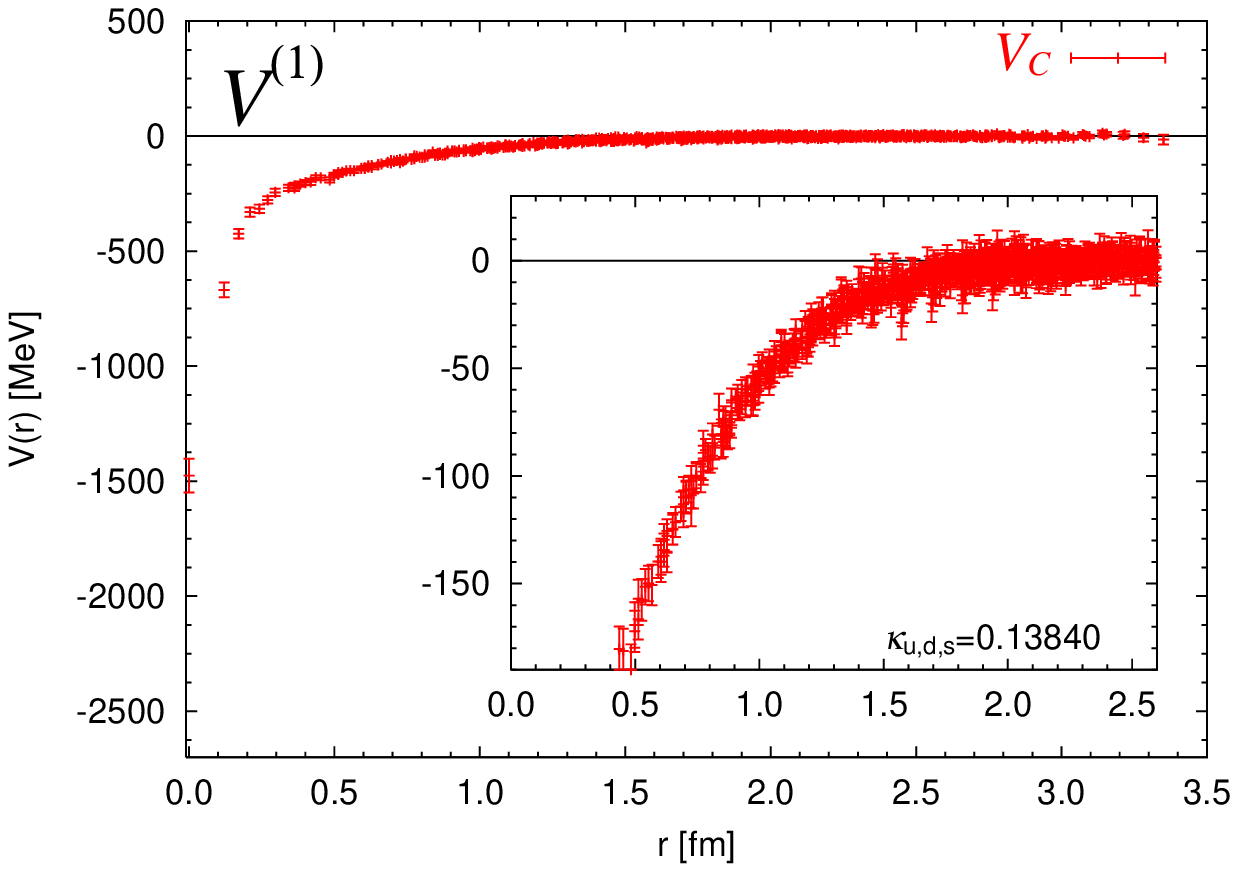}
 \includegraphics[width=0.45\textwidth]{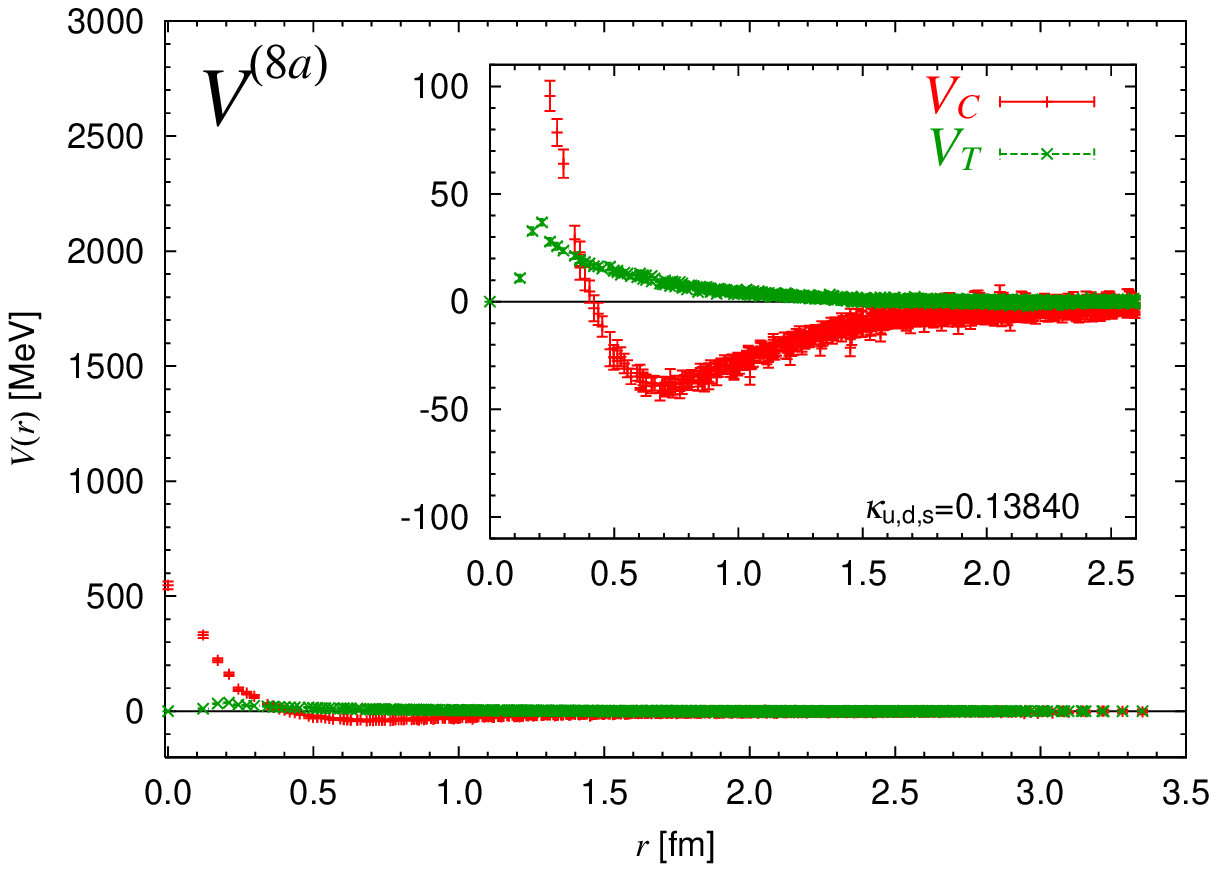}
\caption{
 The $BB$ potentials in ${\bf 8}_s$ (Upper-Left), {\bf 10} (Upper-Right),
 {\bf 1} (Lower-Left) and ${\bf 8}_a$ (Lower-Right)
 extracted from the lattice QCD simulation at $M_{\rm ps}=469$ MeV.
 Taken from Ref.~\protect\citen{Inoue:2011ai}.  
 } 
\label{fig:su3limitB}
\end{center}
\end{figure} 

In Ref.\citen{Inoue:2010hs}, the (effective) central potentials are calculated
 in the original time-independent HAL QCD method
by using the 3 flavor full QCD gauge configuration\cite{CPPACS-JLQCD} 
on a $16^3\times 32$ lattice at $a\simeq 0.12$ fm, 
and at two values of quark hopping parameter corresponding to
$(M_{\rm ps}, M_B) = (1014(1) {\rm MeV}, 2026(3) {\rm MeV})$ and (835(1) MeV, 1752(3) MeV),
where $M_{\rm ps}$ and $M_B$ denote the octet pseudo-scalar (PS) meson mass and the octet baryon mass, respectively.
In Refs.~\citen{Inoue:2010es,Inoue:2011tk,Inoue:2011ai}, on the other hand, the central and tensor potentials are calculated in the 
time-dependent HAL QCD method discussed in Sec.\ref{sec:t-dep}
by using the 3 flavor full QCD gauge configuration generated by HAL QCD Collaboration
on a $32^3\times 32$ lattice at $a\simeq 0.12$ fm,
and at five values of quark hopping parameter which correspond to
$(M_{\rm ps}, M_B) = (1170.9(7){\rm MeV}, 2274(2){\rm MeV} )$, (1015(1) MeV, 2030(2) MeV ),
(837(1) MeV, 1748(1) MeV ), (673(1) MeV, 1485(2) MeV ), and 
(468.6(7) MeV, 1161(2) MeV ).
Figs.~\ref{fig:su3limitA} and \ref{fig:su3limitB} show the flavor basis potentials 
for  $M_{\rm ps} = 469$ MeV~\cite{Inoue:2011ai}. 
The left panels show central potentials for the spin-singlet channel from the $J^P=A_1^+$ state,
while the right panels give central ($V_C$) and tensor ($V_T$) potentials for the spin-triplet channel from the $J^P=T_1^+$ state.

\begin{table}[tb]
   \begin{center}
   \begin{tabular}{| c|l |}
   \hline \hline
   flavor multiplet      & baryon pair (isospin)                      \\
   \hline 
   {\bf 27}       & \{$NN$\}(I=1), \{$N\Sigma$\}(I=3/2), \{$\Sigma\Sigma$\}(I=2),  \\
                  & \{$\Sigma\Xi$\}(I=3/2), \{$\Xi\Xi$\}(I=1) \\
   {\bf 8}$_s$      & none                             \\
   ~{\bf 1}       & none                             \\  
  \hline 
  {\bf 10}$^*$    & [$NN$](I=0), [$\Sigma\Xi$](I=3/2)    \\   
  {\bf 10}        & [$N\Sigma$](I=3/2), [$\Xi\Xi$](I=0)\\
  {\bf 8}$_a$     & [$N\Xi$](I=0)       \\
   \hline \hline
\end{tabular}
\caption{\label{tab:pairs}Baryon pairs which belongs to an irreducible flavor SU(3) representation,
 where $\{BB'\}$ and $[BB']$ denotes $BB' + B'B$ and $BB' - B'B$, respectively.}
 \end{center}
\end{table}

As listed in Table~\ref{tab:pairs}, some of octet-baryon pairs solely belong to an irreducible representation of flavor SU(3).
For example, symmetric $NN$ belongs to ${\bf 27}$ representation.
Therefore, $V^{\bf (27)}(r)$ can be considered as flavor SU(3) symmetric limit of the $NN$ spin-singlet $(^1S_0)$ potential.
Similarly $V^{(\overline{\bf 10})}$, $V^{({\bf 10})}$ and $V^{({\bf 8}_a)}$ can be considered
as flavor SU(3) symmetric limit of some $BB$ potentials of the particle basis,
while $V^{({\bf 1})}$ and $V^{({\bf 8}_s)}$ are always mixtures of different $BB$ potentials in the particle basis.

Fig.~\ref{fig:su3limitA} shows $V^{\bf (27)}_C(r)$ and $V^{\bf (\overline{10})}_{C,T}(r)$,
which correspond to spin-singlet and spin-triplet $NN$ potentials, respectively.
Both central potentials have a repulsive core at short distance with an attractive pocket around 0.8 fm.
These qualitative features are consistent with the results found for the $NN$ potential in previous section.
The upper-right panel of Fig.~\ref{fig:su3limitB} shows that $V^{\bf (10)}_C(r)$ has a stronger repulsive core
and a weaker attractive pocket compared to $V^{\bf (27)}_C(r)$ and $V^{\bf (\overline{10})}_C(r)$.
Furthermore $V^{({\bf 8}_s)}_C(r)$ in the upper-left panel of Fig.~\ref{fig:su3limitB} has a very strong repulsive core
among all 6 channels, while $V^{({\bf 8}_a)}_C(r)$ in the lower-right panel has a very weak repulsive core. 
In contrast to all other cases,  $V^{\bf (1)}_C(r)$ has attraction at short distances instead of repulsion,
as shown in the lower-left panel. 

Above features are consistent with what has been observed in a SU(6) quark model~\cite{Oka:2000wj}. 
In particular, the potential in the ${\bf 8}_s$ channel in this quark model becomes strongly repulsive
at short distance since the six quarks cannot occupy the same orbital state due to the Pauli exclusion for quarks.
On the other hand, the potential in the {\bf 1} channel does not suffer from the quark Pauli exclusion at all,
and can become attractive due to the short-range gluon exchange.
Such agreements between the lattice data and the quark model suggest
that the quark Pauli exclusion plays an essential role for the repulsive core in $BB$ systems.  

The potential for the flavor singlet is entirely attractive even at very short distance.
This might produce a bound state, the $H$-dibaryon, in this channel.
We will discuss this possibility in the next subsection.

\begin{figure}[tb]
\begin{center}
 \includegraphics[width=0.45\textwidth]{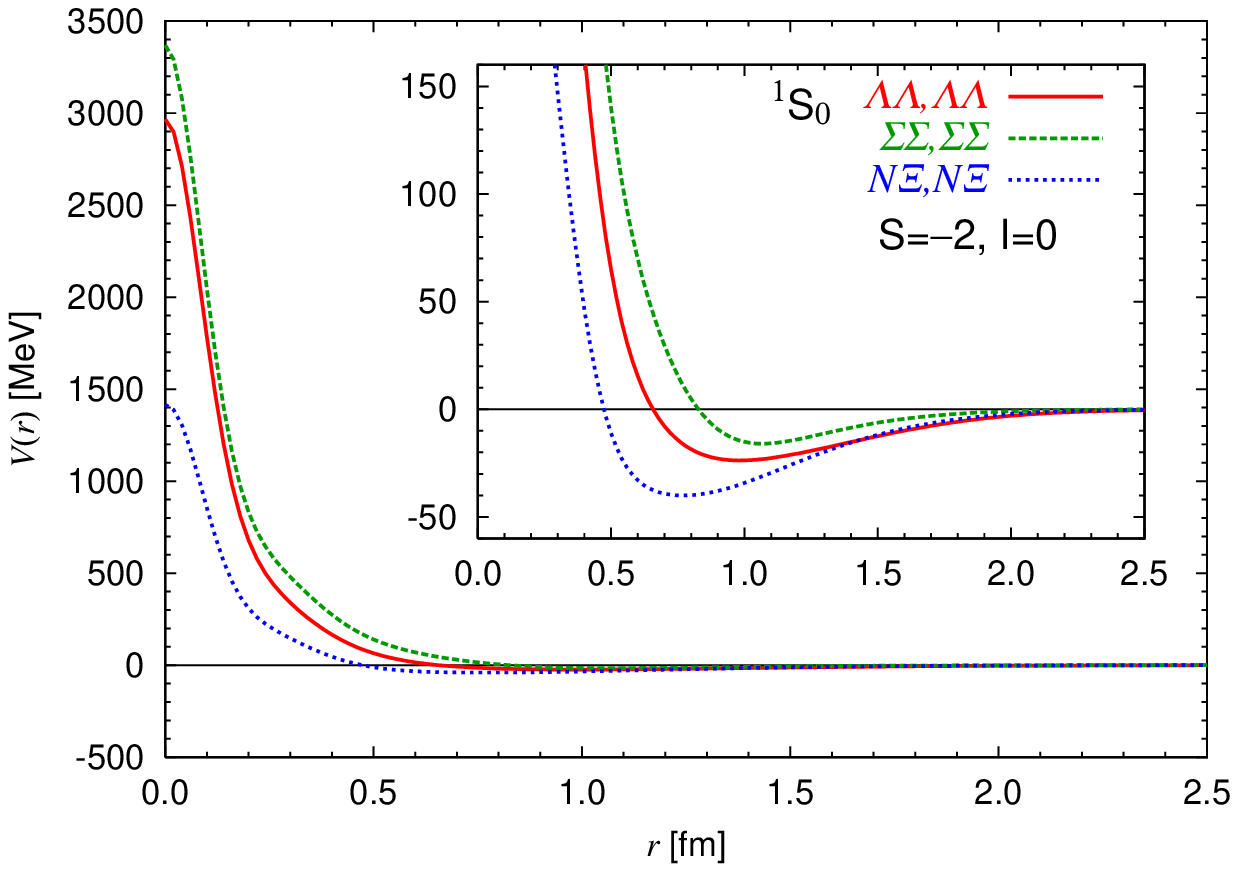}
 \includegraphics[width=0.45\textwidth]{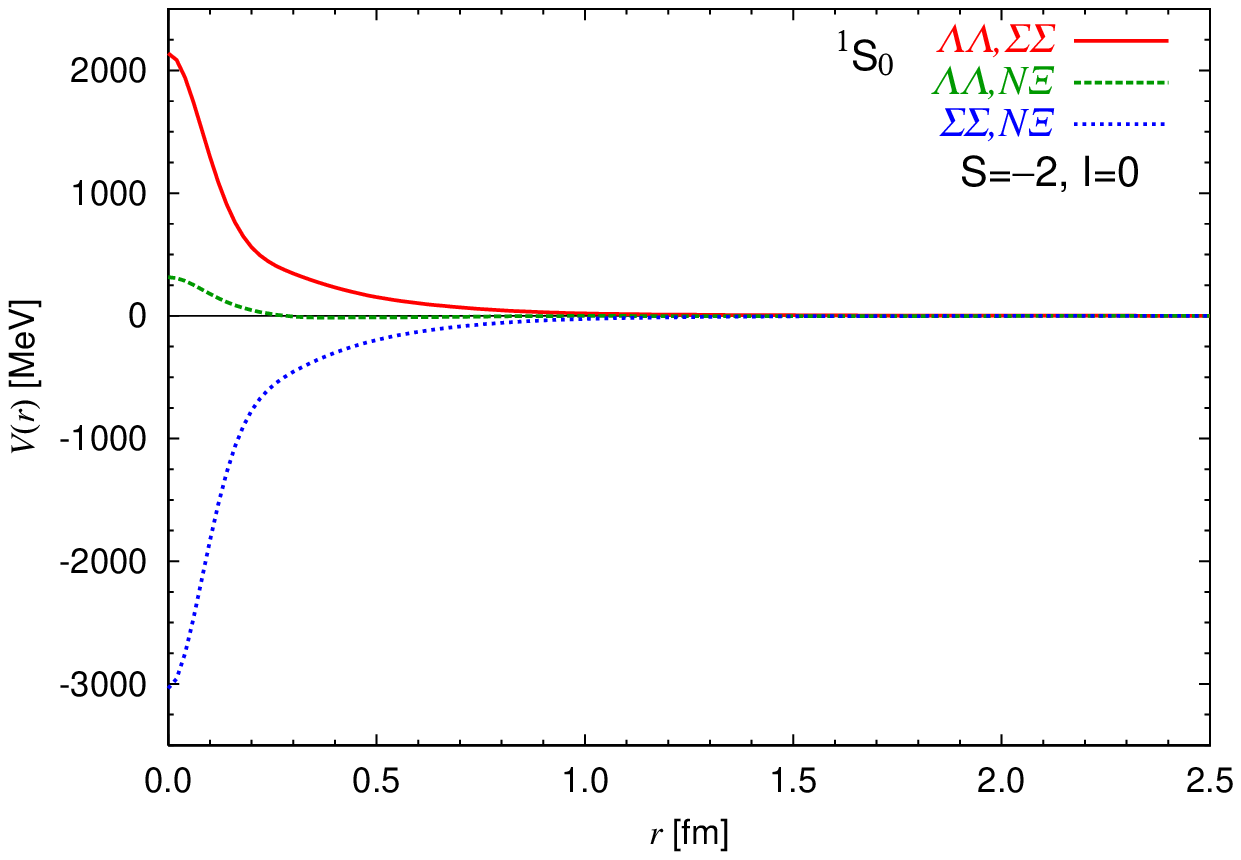}
 \caption{\label{fig:pot_lamlam}
  $BB$ potentials in particle basis for the S=$-$2, I=0, $^1S_0$ sector. 
  Three diagonal(off-diagonal) potentials are shown in left(right) panel.}
 \end{center}
\end{figure}

In the flavor SU(3) limit, the $BB$ potentials in the particle basis can be obtained from those in flavor basis by a unitary rotation as
\be
V_{ij}(r) = \sum_X U_{i X} V^{(X)}(r)U_{X j}^*
\label{eq:rotation}
\ee
where $U_{iX}$ is a unitary matrix which rotates the flavor basis $\ketv X>$ 
to the particle basis $\ketv i>$ as $\ketv i> = U_{i X} \ketv X>$, and given in terms of the CG coefficients.
The explicit forms of the unitary matrix $U$ are found in Ref.~\citen{Inoue:2010hs}.  

In Fig.~\ref{fig:pot_lamlam}, we show $BB$ potentials for S=$-$2, I=0, $^1S_0$ sector at $M_{\rm PS}= 469$ MeV, as a characteristic example.
The flavor base potentials are fitted 
by the analytic function composed of
an attractive Gaussian core plus a long range (Yukawa)$^2$ attraction,
\begin{equation}
V(r) = b_1 e^{-b_2\,r^2} + b_3(1 - e^{-b_4\,r^2})^2\left( \frac{e^{-b_5\,r}}{r} \right)^2,
\label{eq:fit_su3}
\end{equation}
with five parameters $b_{1,2,3,4,5}$.
The left panel of Fig.~\ref{fig:pot_lamlam} shows the diagonal potentials.
One observes that all three diagonal potentials have a repulsive core.
The repulsion is most strong in the $\Sigma\Sigma$(I=0) channel,
reflecting its largest CG coefficient of the ${\bf 8}_s$ state among three channels,
while the attraction in the ${\bf 1}$ state is reflected most in the $N\Xi$(I=0) potential
due to its largest CG coefficient. 
The right panel of Fig.~\ref{fig:pot_lamlam} shows the off-diagonal potentials, which
are comparable in magnitude to the diagonal ones, 
except for the $\Lambda\Lambda$-$N\Xi$ transition potential.
Since the off-diagonal parts are not negligible in the particle basis,
a fully coupled channel analysis is necessary to study observables in this system.
This is important when we study the real world with the flavor $SU(3)$ breaking, 
where only the particle base is meaningful.

\begin{figure}[tb]
\begin{center}
 \includegraphics[width=0.32\textwidth]{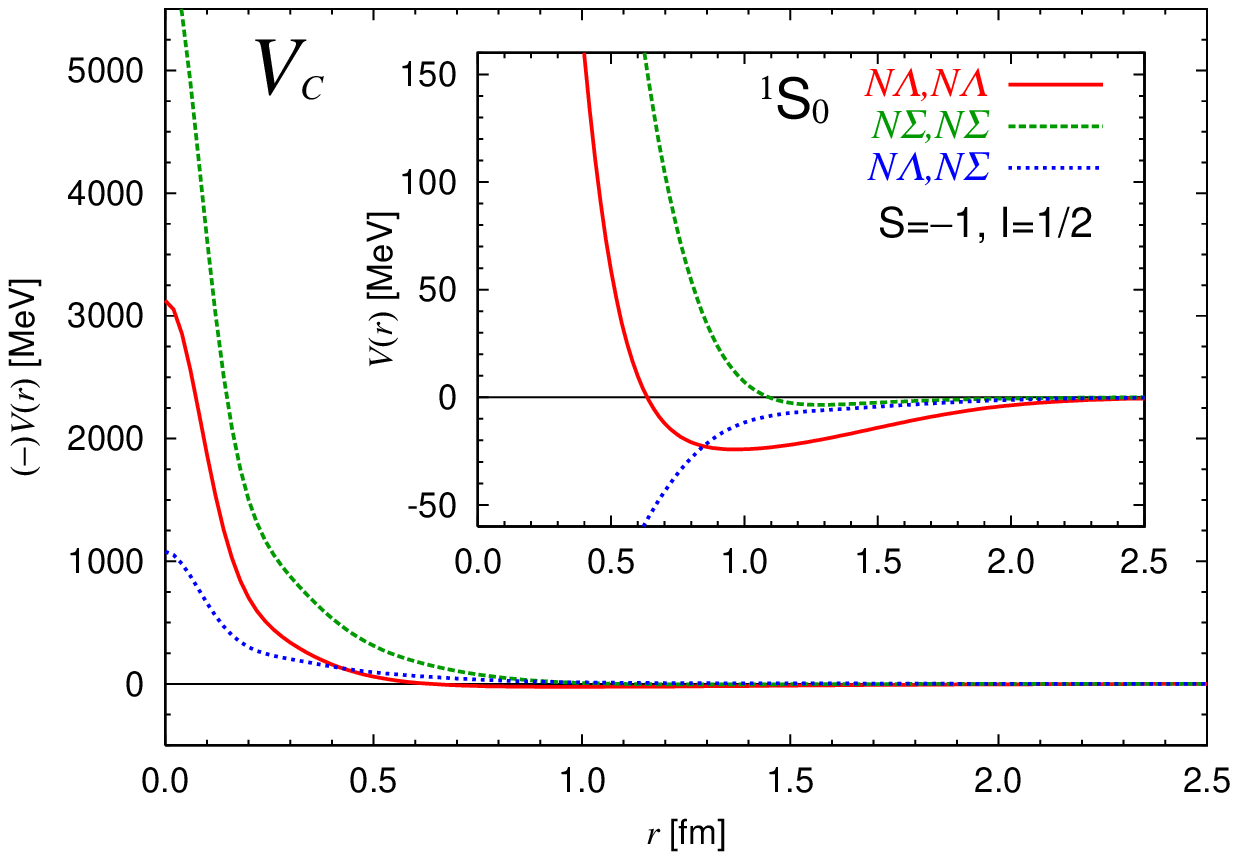}
 \includegraphics[width=0.32\textwidth]{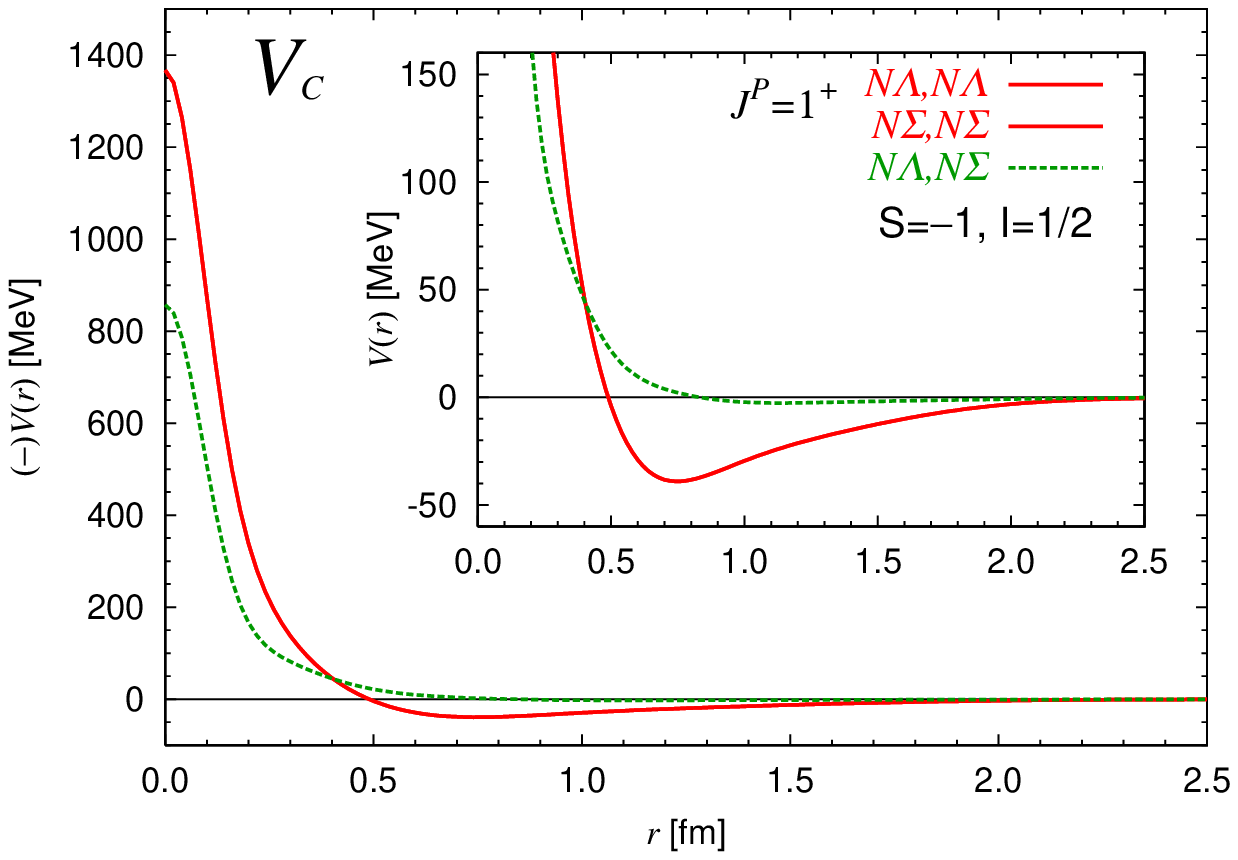}
 \includegraphics[width=0.32\textwidth]{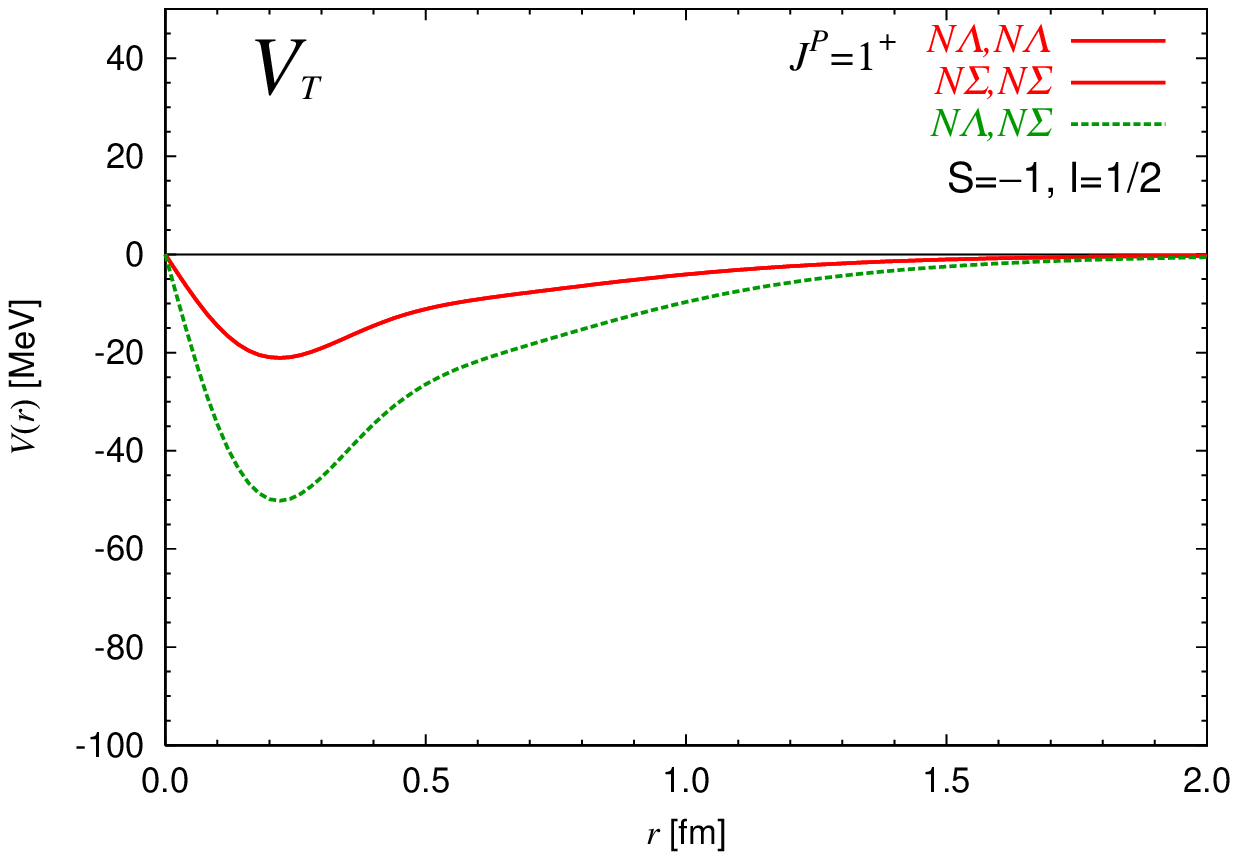}
 \caption{\label{fig:pot_nsignlam}
  $BB$ potentials in particle basis for the S=$-$1, I=1/2 sector,
  $^1S_0$ (Left),  spin-triplet $V_C$ (Center) and spin-triplet $V_T$ (Right), 
  extracted from the lattice QCD simulation at $M_{\rm ps}=469$ MeV.
 }
 \end{center}
\end{figure}

As another example, we show $BB$ potentials for S=$-$1, I=1/2 sector at $M_{\rm PS}= 469$ MeV in Fig.~\ref{fig:pot_nsignlam}.
The left panel of Fig.~\ref{fig:pot_nsignlam} shows the potential in the $^1S_0$ channel,
while the center (right) panel of Fig.~\ref{fig:pot_nsignlam} shows
the central (tensor) potential for $J^P=T_1^+$ spin-triplet channel.
We observe that the off-diagonal $N\Lambda$-$N\Sigma$ potentials are significantly large, especially in the spin-triplet channel, so that
the full $N\Lambda$-$N\Sigma$ coupled channel analysis is also necessary to study observables.
In addition, the repulsive cores in the spin-single channel are much stronger than that in the spin-triplet channel, due to the strong repulsion of the ${\bf 8}_s$ state, which couples only to the spin-singlet channel.

Although all quark masses of 3 flavors are degenerate and rather heavy in the present simulations,
these particle base potentials, shown in Fig.~\ref{fig:pot_lamlam} and Fig.~\ref{fig:pot_nsignlam},
may provide useful information for the behavior of hyperons in hyper-nuclei and
in neutron stars~\cite{Hashimoto:2006aw,SchaffnerBielich:2010am}.
$BB$ potentials in all other sectors can be found in Ref.~\citen{Inoue:2010hs}. 

\subsection{Bound $H$ dibaryon in the flavor SU(3) limit}
In this subsection, we investigate the potential of the flavor singlet channel
in order to see whether the bound $H$ dibaryon exists or not in the flavor SU(3) limit case.

Potentials  for the flavor irreducible channels in the SU(3) limit have been calculated in Ref.~\citen{Inoue:2010es,Inoue:2011tk,Inoue:2011ai}
on $16^3\times 32$, $24^3\times 32$ and $32^3\times 32$ lattices at $a=0.121(2)$ fm
and five values of the quark mass, as mentioned before.

\begin{figure}[tb]
\begin{center}
\includegraphics[width=0.45\textwidth]{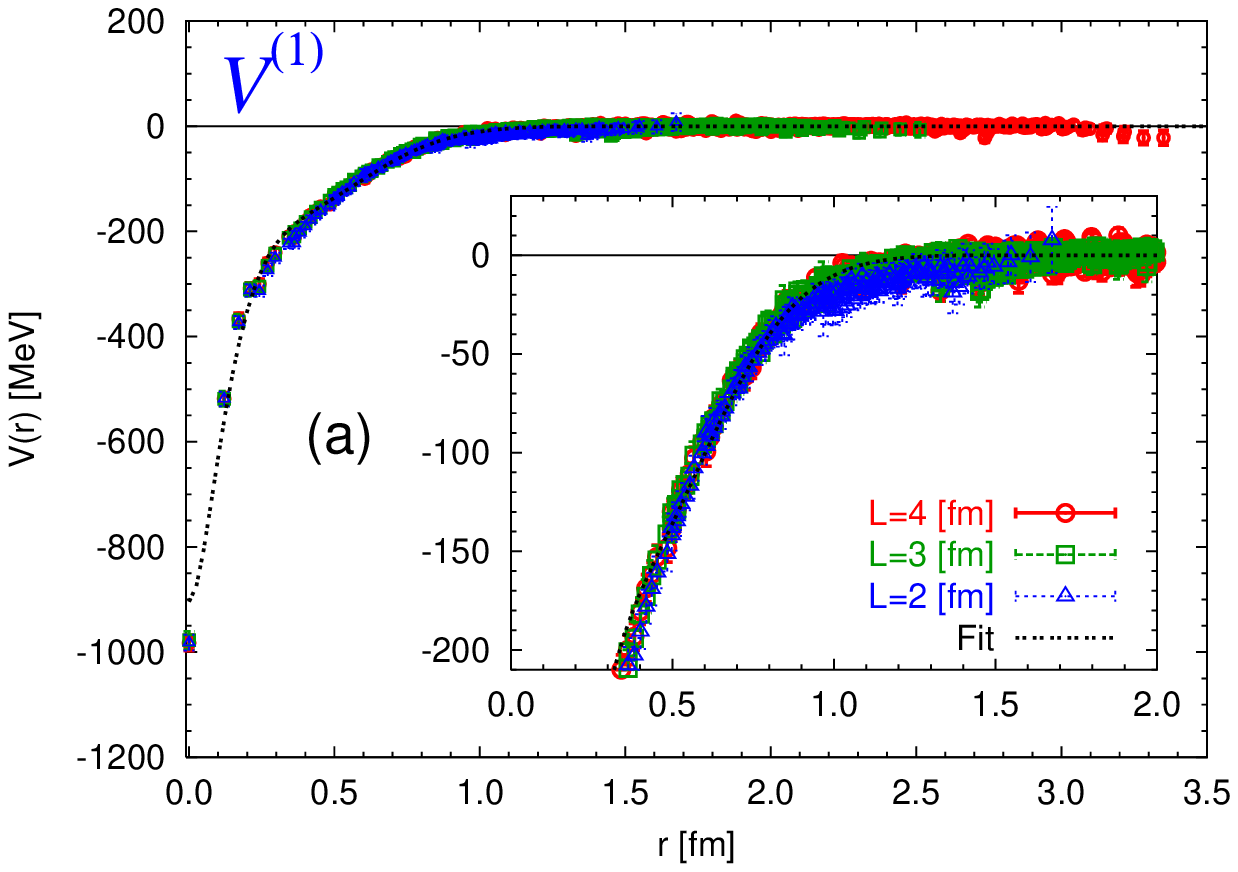}
\includegraphics[width=0.45\textwidth]{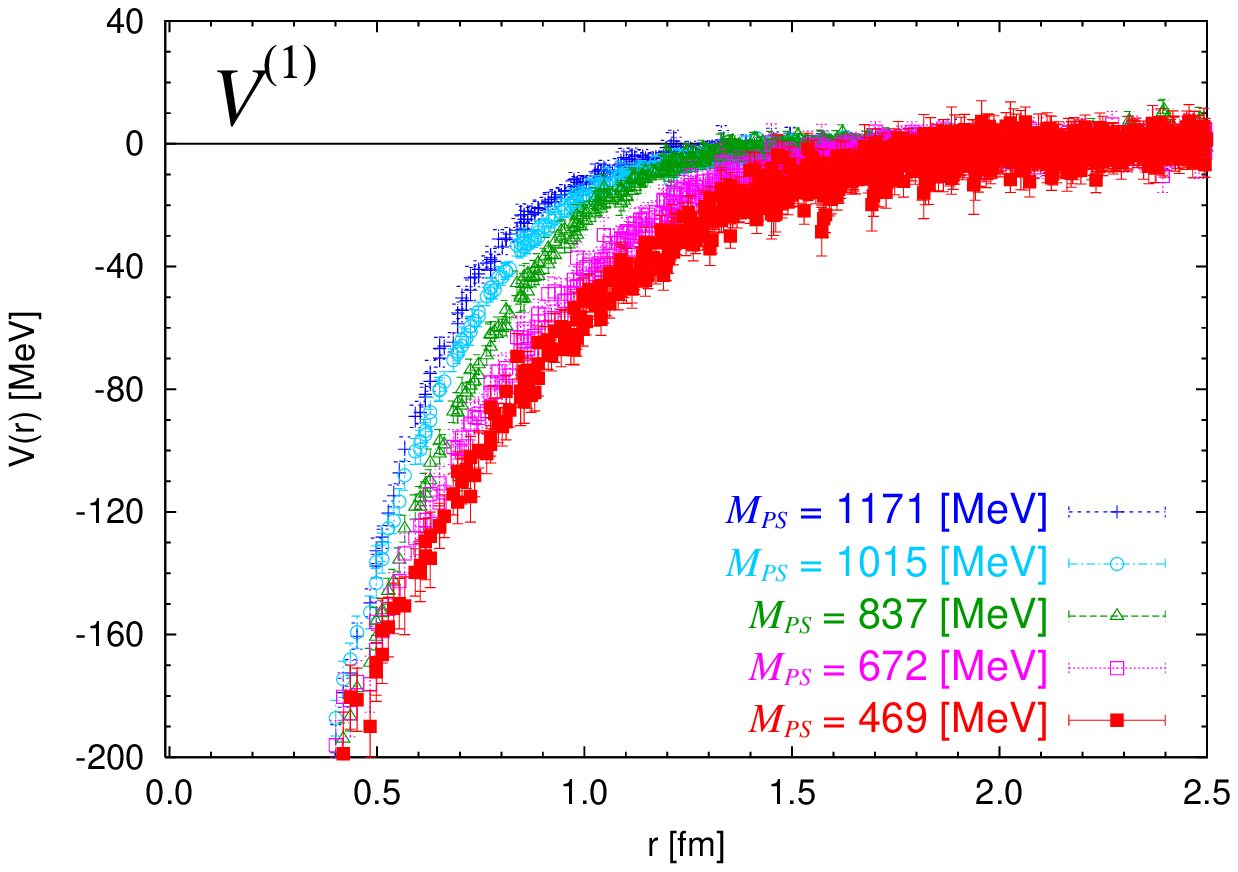}
\end{center}
\caption{ The flavor-singlet potential $V_C^{(1)}(r)$ at $(t-t_0)/a = 10$. 
 (Left) Results for $L=1.94,\, 2.90,\, 3.87$ fm at $M_{\rm ps}=1015$ MeV.
 (Right) Results for $L=3.87$ fm at  $M_{\rm ps}=1171,\, 1015,\, 837,\, 673,\, 469$ MeV.}
\label{fig:singlet}
\end{figure}

Shown in Fig.~\ref{fig:singlet}(Left) and Fig.~\ref{fig:singlet}(Right)
are the volume and the quark mass dependencies 
of the central potential in the flavor-singlet channel $V_C^{(1)}(r)$ at $(t-t_0)/a=10$,
where the potentials do not have appreciable change with respect to the choice of $t$.
The flavor-singlet potential is shown to have an ``attractive core" and to be well localized in space.
Because of the latter property, no significant volume dependence of the potential is observed
within the statistical errors, as seen in Fig.~\ref{fig:singlet}(Left).
As the quark mass decreases in Fig.~\ref{fig:singlet}(Right), the long range part of the attraction tends to increase. 

The resultant potential is fitted by the form in Eq. (\ref{eq:fit_su3}). 
With the five parameters, $b_{1,2,3,4,5}$, the lattice results can be fitted reasonably well
with $\chi^2/{\rm dof} \simeq 1$.
The fitted result for $L=3.87$ fm is shown by the dashed line in Fig.~\ref{fig:singlet}(Left). 

\begin{figure}[t]
\begin{center}
\includegraphics[width=0.475\textwidth]{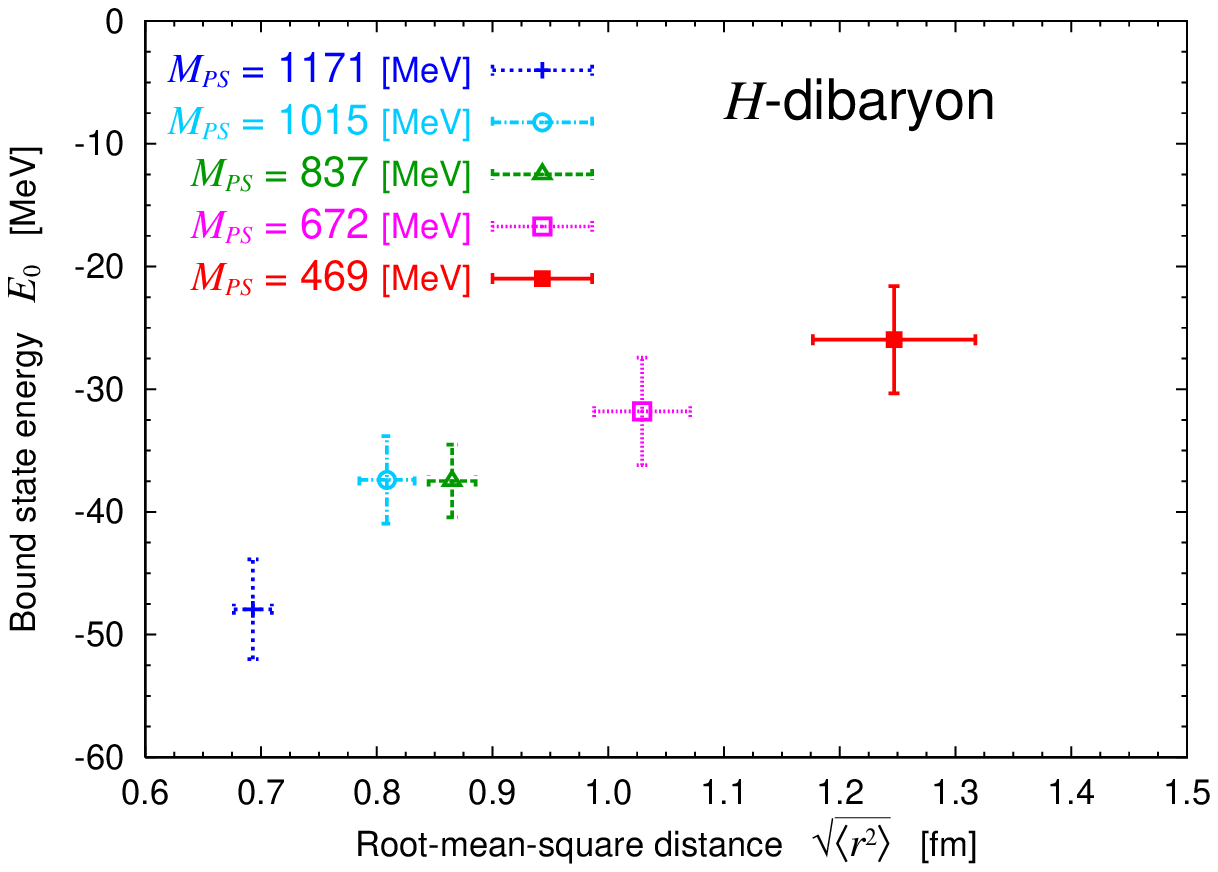}
\includegraphics[width=0.475\textwidth]{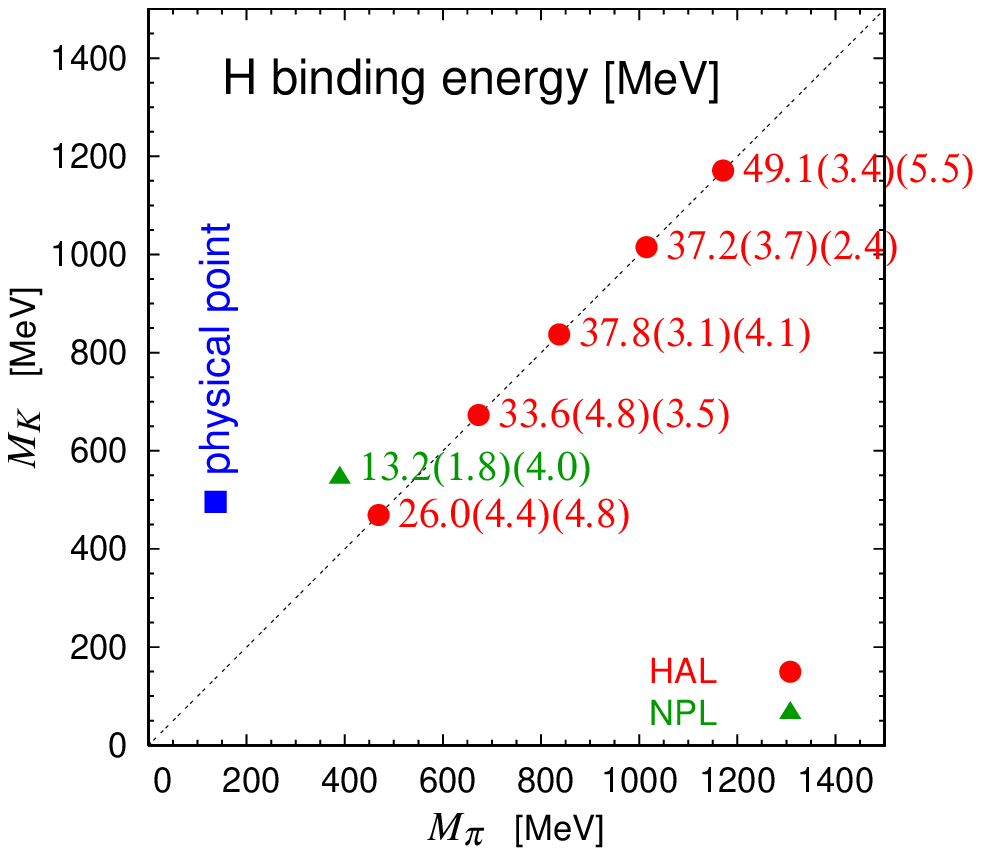}
\caption{(Left) The energy $E_0$ and the root-mean-square distance $\sqrt{\langle r^2\rangle}$ of the bound state in the flavor singlet channel at each quark mass. Bars represent statistical errors only.
(Right)  Summary of the $H$-dibaryon binding energy in recent full QCD simulations.
 HAL stands for the present results and NPL stands for the result in Ref.~\protect\cite{Beane:2011iw}.}
\label{fig:H-dibaryon1}
\end{center}
\end{figure}

Solving the Schr\"{o}dinger equation with the fitted potential in infinite volume,
the energies and the wave functions are obtained at the present quark masses in the flavor SU(3) limit.
It turns out that, at each quark mass, there is only one bound state with binding energy of 20--50 MeV.  
Fig.~\ref{fig:H-dibaryon1}(Left) shows the energy and the root-mean-squared (rms) distance of the bound state
at each quark mass obtained from the potential at $L=3.87$ fm and $(t-t_0)/a=10$,
where errors are estimated by the jackknife method. 
Despite the fact that the potential becomes more attractive as quark mass decrease,
the resultant binding energies of the $H$-dibaryon decrease in the present range of the quark masses,
since 
the increase of the attraction toward the lighter quark
mass is compensated by the increase of the kinetic energy for the lighter baryon mass. 
It is noted that there appears no bound state for the potential of the 27-plet channel  or  the $\overline{10}$-plet channel ("deuteron" ) in the present range of the quark masses.

By including systematic errors caused by the choice of sink-time $t$ in $R(\br,t-t_0)$,
the final results of the binding energy $B_H$ and the rms distance $\sqrt{\langle r^2\rangle}$ are summarized below,
where the 1st and 2nd parentheses correspond to statistical and systematic errors, respectively.
\begin{eqnarray}
 M_{\rm ps}&=& 1171  ~\mbox{MeV} :~ \  {B}_H = 49.1 (3.4)(5.5) ~\mbox{MeV}~~  \sqrt{\langle r^2 \rangle} = 0.685(13)(25)  ~\mbox{fm} \nonumber \\
 M_{\rm ps}&=& 1015  ~\mbox{MeV} :~ \  {B}_H = 37.2 (3.7)(2.4) ~\mbox{MeV}~~  \sqrt{\langle r^2 \rangle} = 0.809(23)(10)  ~\mbox{fm} \nonumber \\
 M_{\rm ps}&=& ~~837 ~\mbox{MeV} :~ \  {B}_H = 37.8 (3.1)(4.2) ~\mbox{MeV}~~  \sqrt{\langle r^2 \rangle} = 0.865(20)(25)  ~\mbox{fm} \nonumber \\
 M_{\rm ps}&=& ~~672 ~\mbox{MeV} :~ \  {B}_H = 33.6 (4.8)(3.5) ~\mbox{MeV}~~  \sqrt{\langle r^2 \rangle} = 1.029(41)(23)  ~\mbox{fm} \nonumber \\
 M_{\rm ps}&=& ~~469 ~\mbox{MeV} :~ \  {B}_H = 26.0 (4.4)(4.8) ~\mbox{MeV}~~  \sqrt{\langle r^2 \rangle} = 1.247(70)(59)  ~\mbox{fm} . \nonumber 
\end{eqnarray} 

Recently, the existence of $H$-dibaryon is also investigated by a direct calculation
of its binding energy in 2+1 full QCD simulations~\cite{Beane:2010hg,Beane:2011iw}, 
where $B_H=13.2(1.8)(4.0)$ MeV is reported in the $L\rightarrow \infty$ extrapolation
at $m_\pi \simeq 389$ MeV, $m_{K} \simeq 544$ MeV.
Fig.~\ref{fig:H-dibaryon1}(Right) gives a summary of the $H$-dibaryon binding energy from full QCD simulations recently reported.

Since the binding energy is comparable to the splitting between physical hyperon masses and not so sensitive to quark mass, 
there may be a possibility of weakly bound or resonant $H$-dibaryon even in the real world with lighter quark masses
and the flavor SU(3) breaking. Our phenomenological trial analysis using 3-flavor lattice QCD results,
suggests a resonant $H$-dibaryon above $\Lambda\Lambda$ but bellow $N\Xi$ thresholds~\cite{Inoue:2011ai}. 
To make a definite conclusion on this point, 
however, the $\Lambda\Lambda-N\Xi-\Sigma\Sigma$ coupled channel analysis is necessary for $H$
in the (2+1)-flavor lattice QCD simulations, as will be discussed in Sec.~\ref{sec:inelastic}.

\section{Hadronic interactions above inelastic threshold}
\label{sec:inelastic}
In this section, we discuss a method to investigate hadron interactions above inelastic threshold
by generalizing the 
 HAL QCD method. Then, we apply the method to  coupled channel potentials in the $S=-2$ and $I=0$ sector. 

\subsection{Coupled-channel approach to  inelastic scattering}

For simplicity, let us discuss a case of $A+B\rightarrow C+D$ scattering where $A,B,C,D$ represent some 1-particle states. This is a simplified version of the octet baryon scattering in the strangeness $S=-2$ and isospin $I=0$ channel, where $\Lambda\Lambda$, $N\Xi$ and $\Sigma\Sigma$ appear as asymptotic states of the strong interaction if the total energy is larger than $2m_\Sigma$.
We here assume $m_A + m_B < m_C + m_D < W $, where $W=E_k^A + E_k^B$ is the total energy of the system, and $E_k^X =\sqrt{m_X^2 +\bk^2}$. In this case, the QCD eigenstate with the quantum numbers of the $AB$ state  and center of mass energy $W$  is expressed  as
\begin{eqnarray}
\vert W \rangle &=& c_{AB} \vert AB,W\rangle + c_{CD} \vert CD, W\rangle +\cdots\\
\vert AB, W\rangle &=&  \vert A, \bk \rangle_{\rm in}\otimes \vert B, -\bk \rangle_{\rm in}, \quad
\vert CD, W\rangle =  \vert C, \bq \rangle_{\rm in}\otimes \vert D, -\bq \rangle_{\rm in},
\end{eqnarray}
where $W=E_k^A+E_k^B = E_q^C+E_q^D$. Then
we define the following NBS wave functions,
\begin{eqnarray}
\varphi_{AB}(\br,\bk )e^{-Wt} &=& \langle 0 \vert T\{ \varphi_A(\bx+\br,t) \varphi_B(\bx,t) \}\vert W \rangle, \\
\varphi_{CD}(\br,\bq )e^{-Wt} &=&  \langle 0 \vert T\{ \varphi_C(\bx+\br,t) \varphi_D(\bx,t) \}\vert W \rangle .
\end{eqnarray}
Using the partial wave decomposition such that\footnote{Here we ignore spins for simplicity.}
\bea
\varphi_{X}(\br,\bk ) &=& 4\pi\sum_{l,m} i^\ell \varphi^{\ell}_{X}(r,k)Y_{\ell m}(\Omega_{\brs}) \overline{Y_{\ell m}(\Omega_{\bks})} 
\eea
for $X=AB$ or $CD$,
it can be shown\cite{Aoki:2011gt} that  these wave functions satisfy
\be
(\nabla^2 + \bk^2 )\varphi_{AB}(\br,\bk) =0, \quad
(\nabla^2 + \bq^2 )\varphi_{CD}(\br,\bq) =0
\ee
for $r\rightarrow \infty$.

Let us now consider QCD in the finite volume $V$ where
 $ \vert AB, W \rangle $ and $ \vert CD, W \rangle $ are no longer eigenstates of the Hamiltonian. 
True eigenvalues are shifted from $W$ to $W_i = W + O(V^{-1})$ ($i=1,2$).
By the diagonalization method in lattice QCD, 
it is relatively easy to determine $W_1$ and $W_2$. With these values L\"uscher's finite volume formula gives two conditions, which,
however, are insufficient to determine three observables, two scattering phase shifts $\delta_\ell^1$, $\delta_\ell^2$ and one mixing angle $\theta$. 
We here explain a new  approach proposed in Refs.\citen{ Ishii:2011tq, Aoki:2011gt}  to overcome this difficulty.
Let us consider the NBS wave functions at two different values of energy, $W_1$ and $W_2$,
in the finite volume:
\bea
\varphi_{X}(\bx,\bp^X_i )e^{-W_i t} &=& \langle 0 \vert T\{ \varphi_{X_1}(\bx+\br,t) \varphi_{X_2}(\br,t) \}\vert W_i \rangle , \quad i=1,2 .
\eea
where $X(=X_1X_2)=AB$ or $CD$ with $\bp^{AB}=\bk$ or $\bp^{CD}=\bq$.
We then define the coupled channel non-local potentials from the coupled channel Schr\"odinger equation as
\begin{eqnarray}
\left[\frac{(p^X_i)^2}{2\mu_{X}} - H_0\right] \varphi_{X}(\bx,\bp^X_i) &=&\sum_Y
\int d^3 y\ U_{X,Y}(\bx, \by)\ \varphi_{Y}(\by,\bp^Y_i)
\end{eqnarray}
for $i=1,2$ where the reduced mass is defined by $1/\mu_X =1/m_{X_1}+1/m_{X_2}$.
In the leading order of the velocity expansion, we have
\begin{eqnarray}
K_{X}(\bx,\bp^X_i)\equiv \left[\frac{(p_i^X)^2}{2\mu_{X}} - H_0\right] \varphi_{X}(\bx,\bp^X_i) &=&\sum_Y
 V_{X,Y}(\bx)\ \varphi_{Y}(\bx,\bp^Y_i)
\end{eqnarray}
These equations for $i=1,2$  can be solved as
\begin{eqnarray}
\left(
\begin{array}{ll}
V_{AB,AB}(\bx) &  V_{AB,CD}(\bx) \\
V_{CD,AB}(\bx) & V_{CD,CD}(\bx)\\
\end{array}
\right) &=&
\left(
\begin{array}{ll}
K_{AB}(\bx,\bk_1) & K_{AB}(\bx,\bk_2)\\
K_{CD}(\bx,\bq_1) & K_{CD}(\bx,\bq_2)\\
\end{array}
\right) \nonumber \\
&\times &
\left(
\begin{array}{ll}
\varphi_{AB}(\bx,\bk_1) & \varphi_{AB}(\bx,\bk_2) \\
\varphi_{CD}(\bx,\bq_1) & \varphi_{CD}(\bx,\bq_2) \\
\end{array}
\right)^{-1}. 
\end{eqnarray}

Once we obtain the coupled channel local potentials $V_{X,Y}(\bx)$, we solve the coupled channel Schr\"odinger equation in {\it infinite} volume with some appropriate boundary condition such that the incoming wave has a definite $\ell$ and consists of the $AB$ state only, in order to extract
three observables for each $\ell$ ($\delta_\ell^1(W)$, $\delta_\ell^2(W)$ and $\theta_\ell(W)$)  
at all values of $W$.
Of course, since $V_{X,Y}$ is the leading order approximation in the velocity expansion of $U_{X,Y}(\bx,\by)$,  results for three observables $\delta_\ell^1(W)$, $\delta_\ell^2(W)$ 
and $\theta(W)$  at $W\not=W_1, W_2$ are also approximate ones and might be different from the exact values.  By performing an  additional extraction of $V_{X,Y}(\bx)$ at $(W_3,W_4)\not=( W_1,W_2)$, we can test  how good the leading order approximation is.

The method considered above can be generalized to inelastic scattering where a number of particles is not conserved such that $A+B\rightarrow A+B$ and $A+B\rightarrow A+B+C$.
See  Ref. \citen{Aoki:2011gt} for more details.

\subsection{Coupled-channel potentials in $(S,I)=(-2,0)$ channel}
\label{sec:coupled_channel}
\begin{table}[b]
\begin{center}
  \begin{tabular}{|c|cccccc|}
  \hline \hline
    & $m_\pi$ & $m_K$ & $m_N$ & $m_\Lambda$ & $m_\Sigma$ & $m_\Xi$ \\
  \hline
  Set 1 & $875(1)$ & $916(1)$ & $1806(3)$ & $1835(3)$ & $1841(3)$ & $1867(2)$ \\
  Set 2 & $749(1)$ & $828(1)$ & $1616(3)$ & $1671(2)$ & $1685(2)$ & $1734(2)$ \\
  Set 3 & $661(1)$ & $768(1)$ & $1482(3)$ & $1557(3)$ & $1576(3)$ & $1640(3)$ \\
  \hline \hline
   \end{tabular}
    \caption{Hadron masses in units of MeV  and number of configurations for each set
    adopted in Sec.\ref{sec:coupled_channel}.}
\label{tab:hadron_2+1}
 \end{center}
\end{table}
As an application of the method in the previous subsection, let us consider  $BB$ potentials for the $S=-2$ and $I=0$ system, i.e., the coupled $\Lambda \Lambda$-$N \Xi$-$\Sigma \Sigma$ system.
It is interesting to investigate this system since it involves the flavor single state, which is free from the Pauli blocking of quark degrees of freedom at  short distance.
Since mass differences of these $BB$ systems are quite small, 
all 3 states appear in NBS wave functions. 
The coupled channel method in the previous subsection can be applied to
treat such complicated systems.

At the leading order of the velocity expansion, 
the coupled channel $3\times 3$ potential matrix in this case is given by
\be
V_{X,Y}(\bx) = \sum_{i=1}^3 K_{X}^i (\bx,\bp_i^X) \left[\varphi_{Y}^{W_i}(\bx,\bp_{Y}^i)\right]^{-1} ,
\ee
where $i$ is a label for energy $W_i$ and $X, Y = \Lambda\Lambda$, $N\Xi$ or $\Sigma\Sigma$.  
Here the last factor is the inverse of  the $3\times 3$ matrix $\varphi_{A}^{W_i}(\br,\bk_{A}^i)$
with indices $i$ and $A$.

\begin{figure}[tb]
 \begin{center}
\begin{tabular}{cc}
  \includegraphics[width=0.48\textwidth]{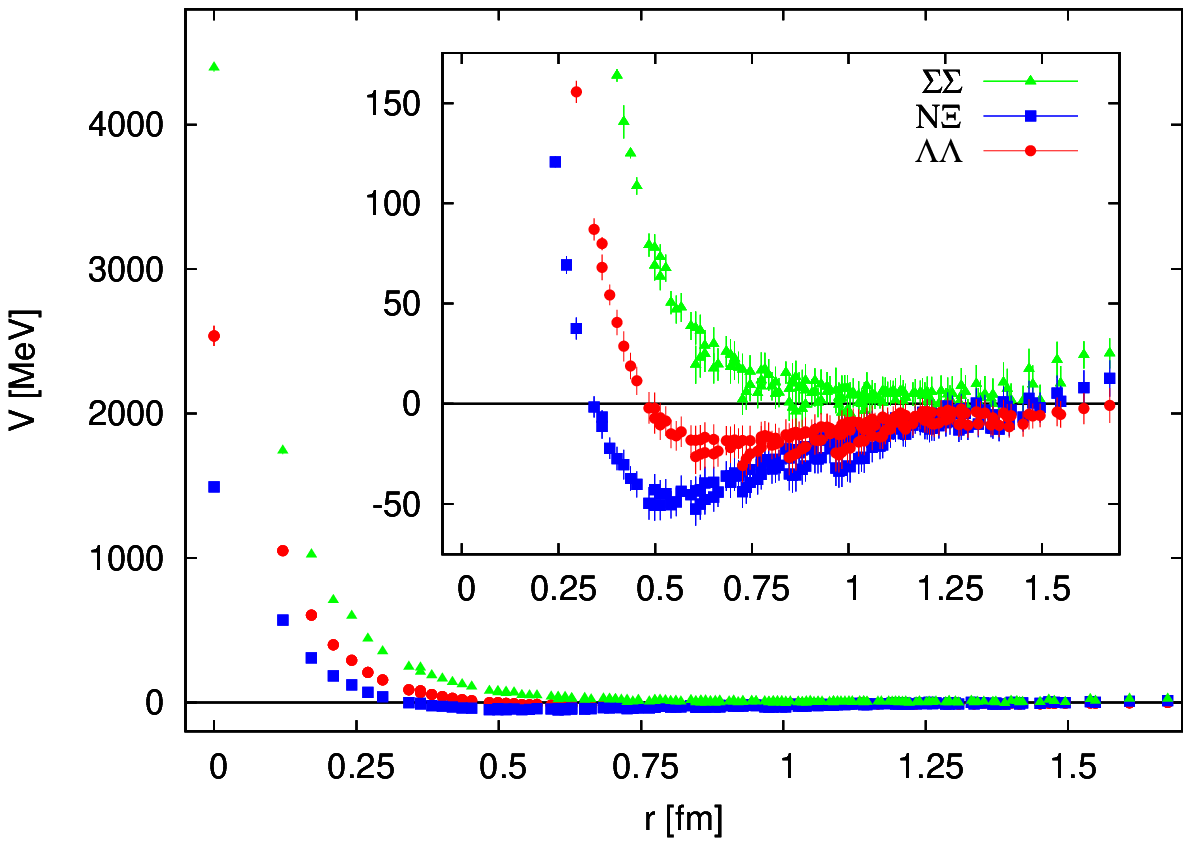} \hspace*{-1em}
&  \includegraphics[width=0.48\textwidth]{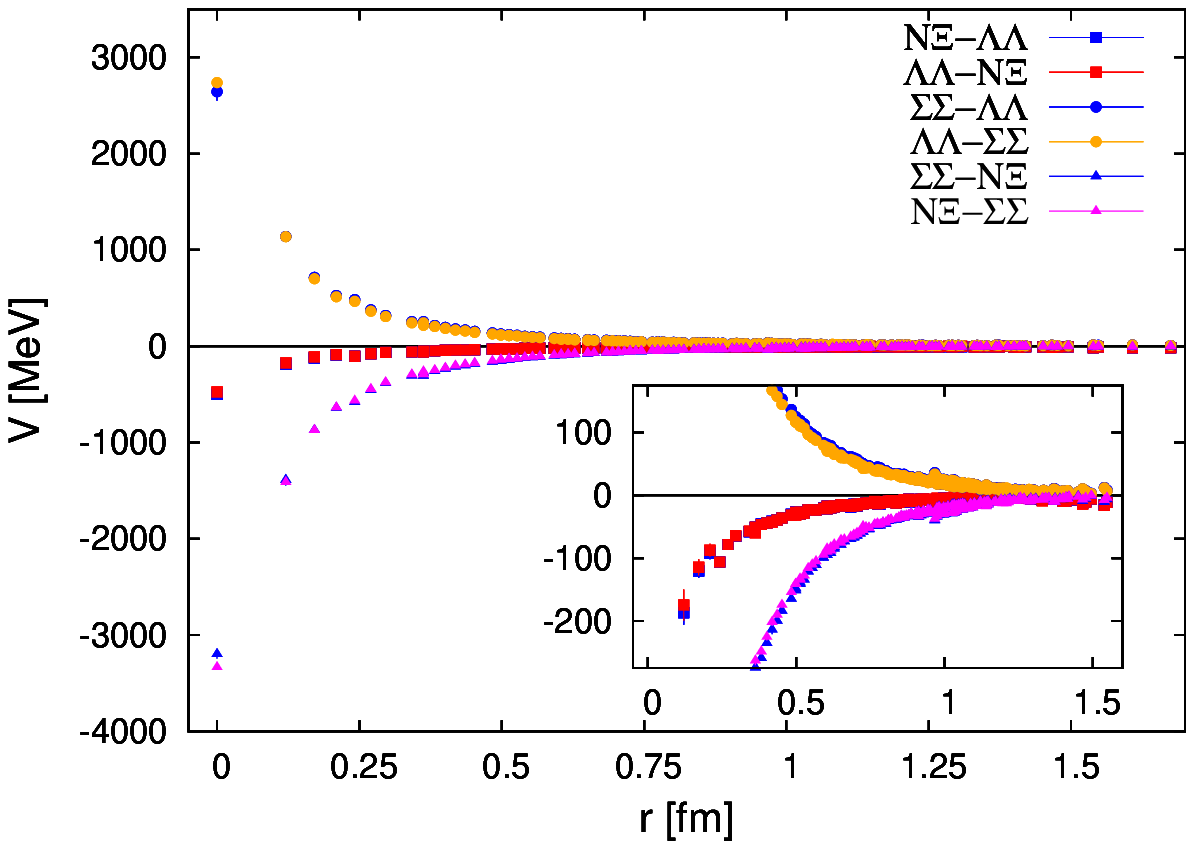} \hspace*{-1em}
\end{tabular}
 \end{center}
\caption{The coupled channel potential matrix  from  the NBS wave function for Set 1. The vertical axis is the potential strength in units of MeV, while the horizontal axis is the relative distance between two baryons in units of fm.}
\label{FIG:PTpbSet1}
\end{figure}
Gauge configurations generated by CP-PACS/JLQCD Collaborations on a $16^3\times 32$ lattice at $a\simeq 0.12$ fm ( therefore $L\simeq 1.9$ fm) in 2+1-flavor full QCD simulations are employed   to calculate the coupled channel potentials at three different values of the light quark mass with the fixed bare strange quark mass\cite{Sasaki:2010bh}.  Quark propagators are calculated with the spatial wall source at $t_0$ with the Dirichlet boundary condition in time at $t=t_0+16$.  
Corresponding hadron masses are given in Table~\ref{tab:hadron_2+1}.

\begin{figure}[tb]
 \begin{center}
\begin{tabular}{ccc}
  \includegraphics[width=0.32\textwidth]{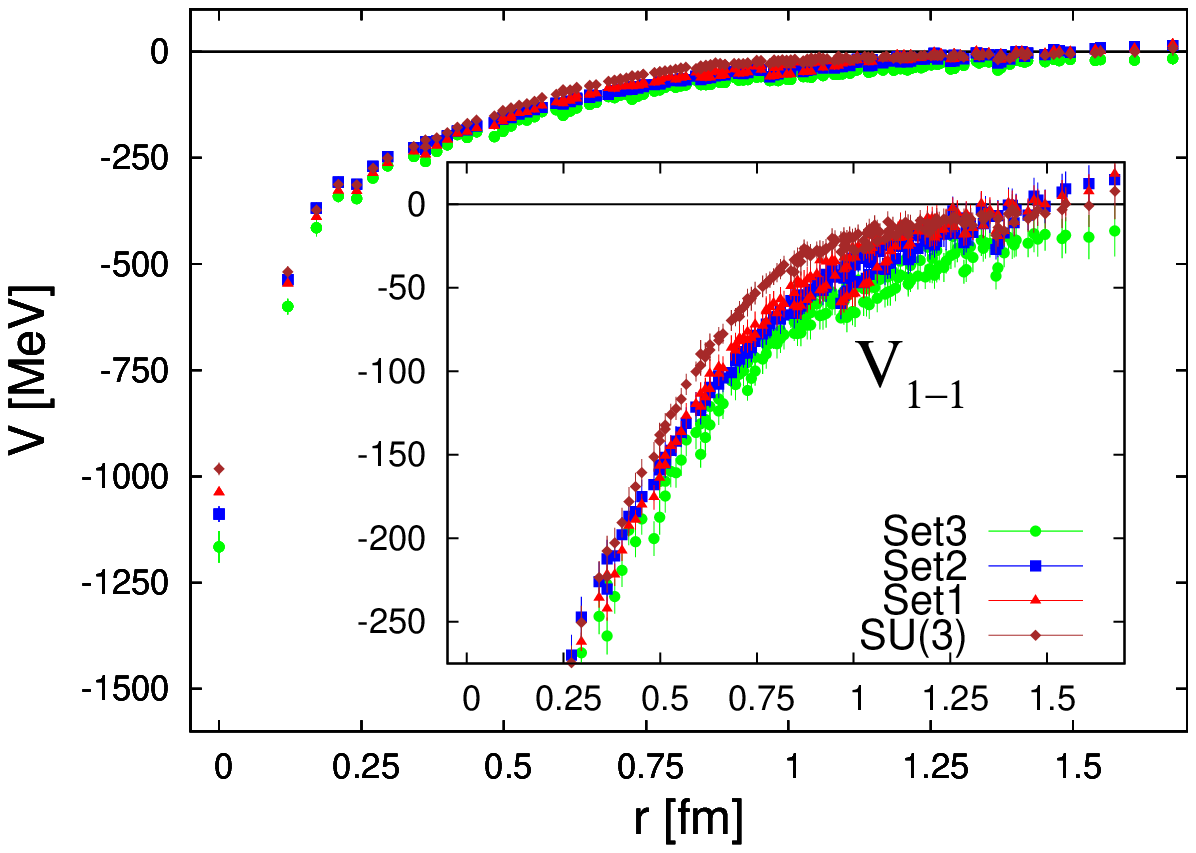} \hspace*{-1em}
& \includegraphics[width=0.32\textwidth]{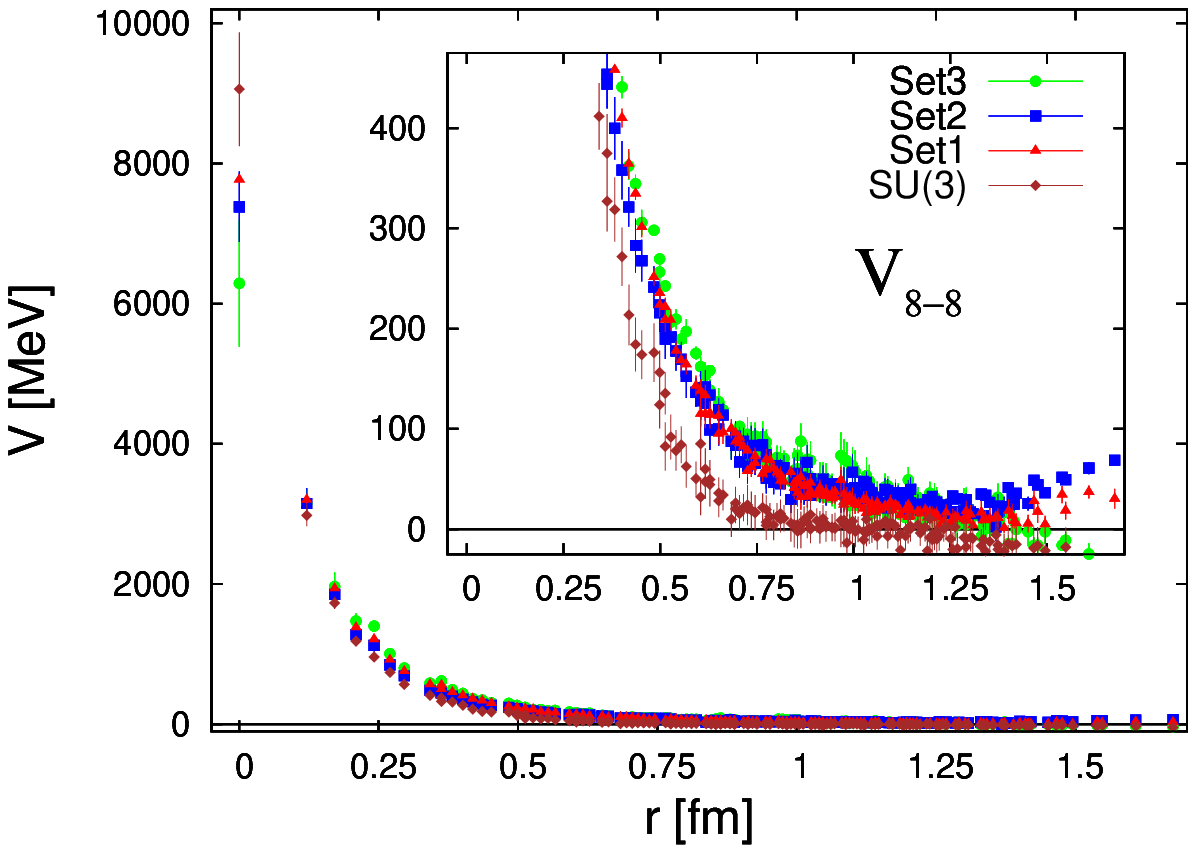} \hspace*{-1em}
& \includegraphics[width=0.32\textwidth]{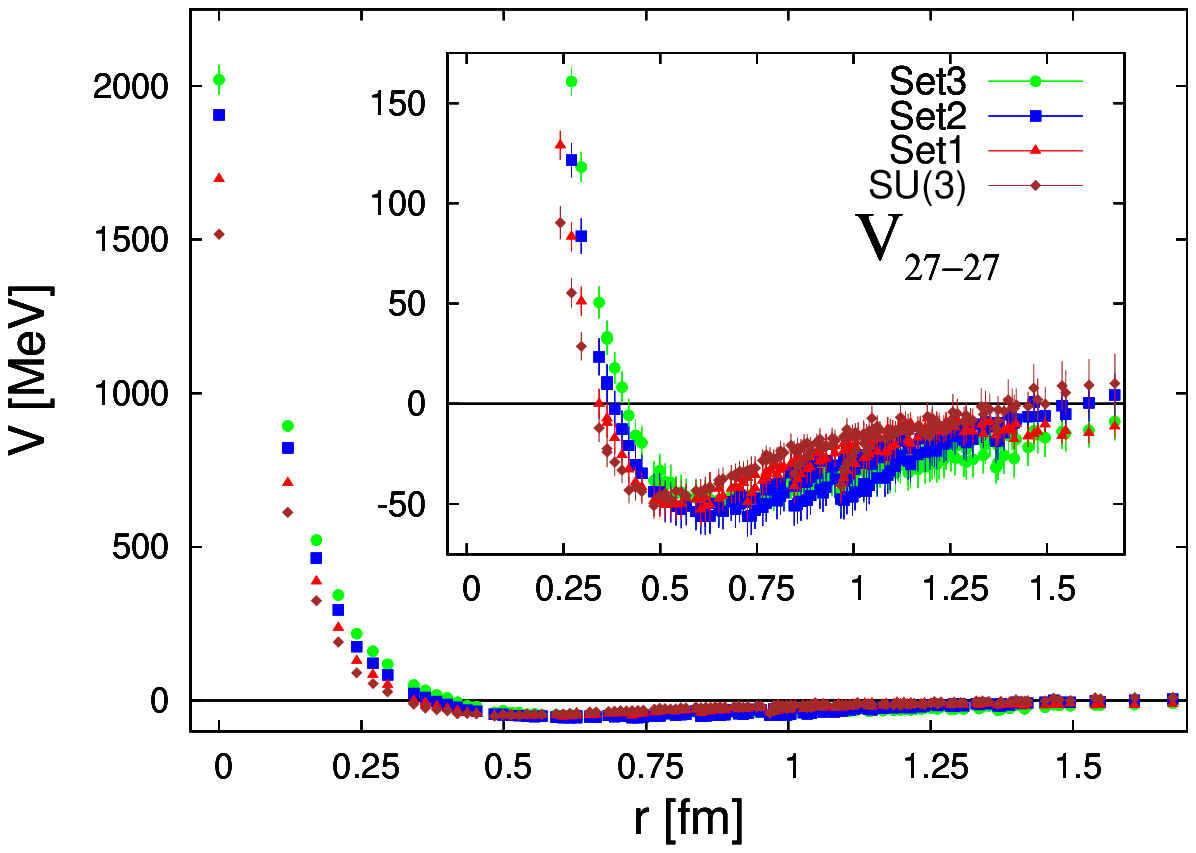} \hspace*{-1em}
\\
  \includegraphics[width=0.32\textwidth]{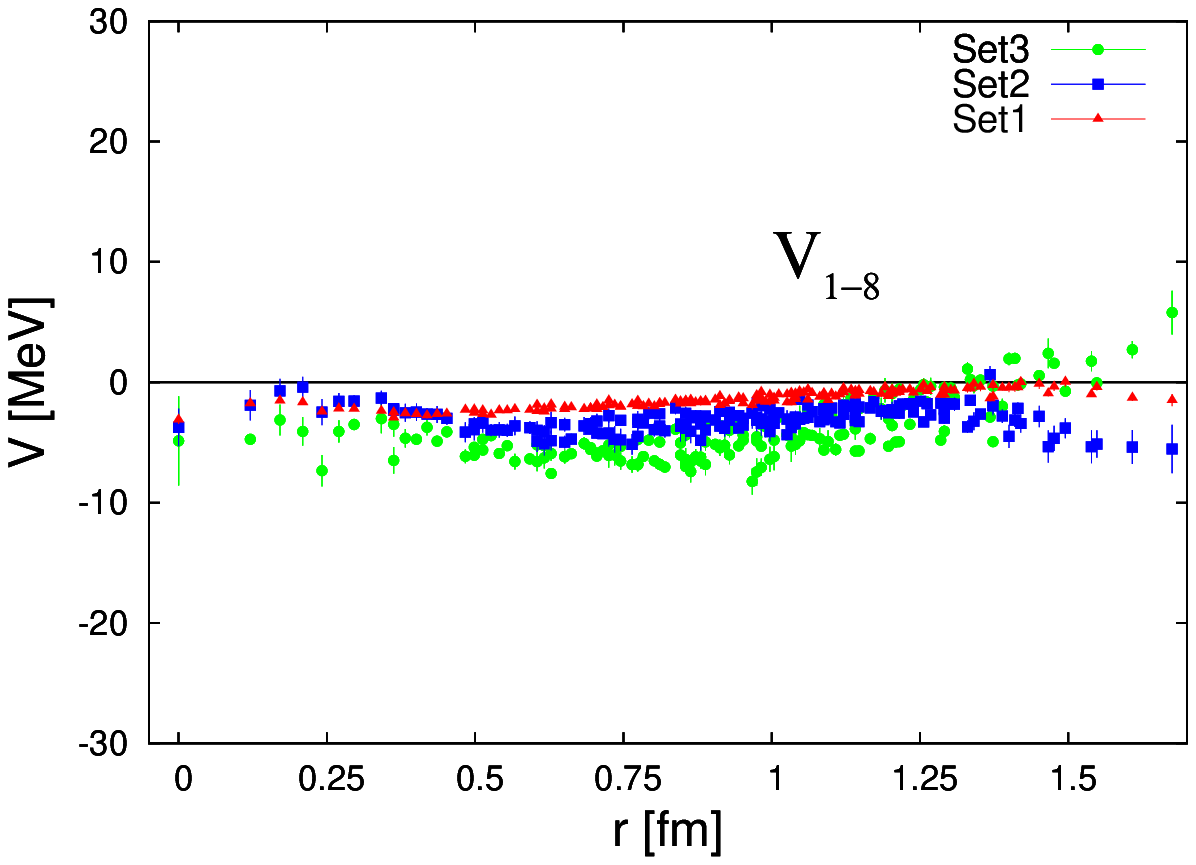} \hspace*{-1em}
& \includegraphics[width=0.32\textwidth]{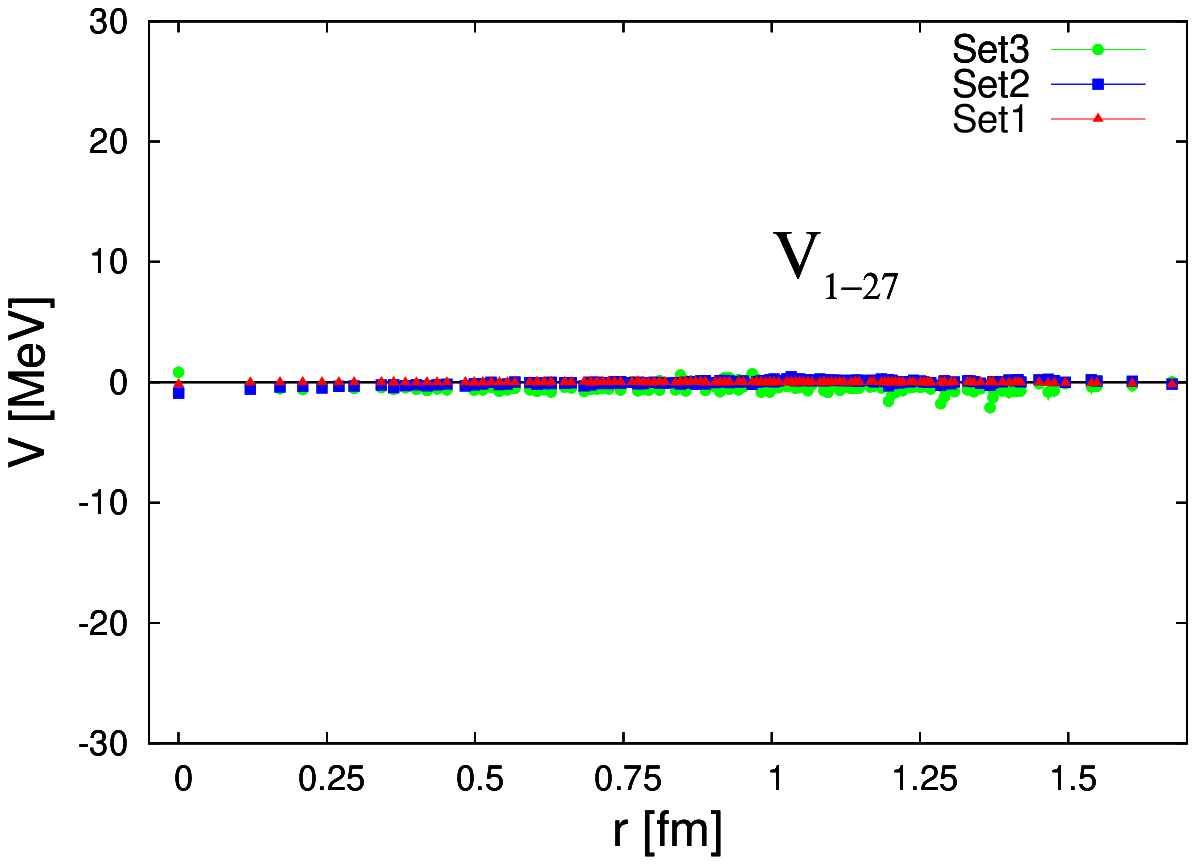} \hspace*{-1em}
& \includegraphics[width=0.32\textwidth]{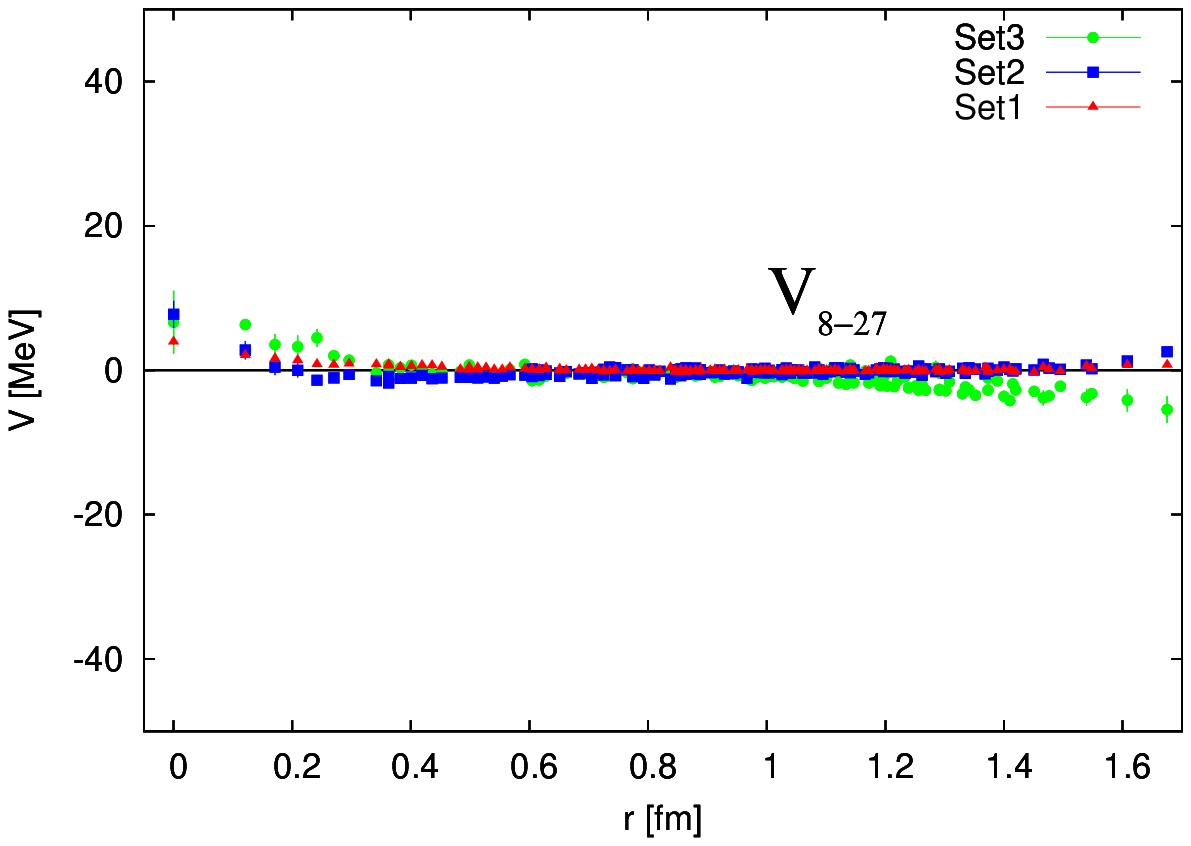} \hspace*{-1em}
\end{tabular}
 \end{center}
\caption{Transition potentials in the flavor SU(3) IR basis. Red, blue and green symbols correspond to results of Set1, Set2 and Set3, respectively. The result of the flavor SU(3) symmetric limit at the same strange quark mass is  also plotted with brown symbols~\protect\citen{Inoue:2010hs}.}
\label{FIG:potall}
\end{figure}
The coupled channel potential matrix $V_{AB,CD}$ from the NBS wave function for Set 1 is shown in Fig.~\ref{FIG:PTpbSet1}. 
All diagonal components of the potential matrix have a repulsion at short distance.  
The strength of the repulsion in each channel, however, varies, reflecting properties of its main component in the irreducible representation of the flavor SU(3).
In particular, the $\Sigma \Sigma$ potential has the strongest repulsive core of these three components.
It is important to note that  off-diagonal parts of the potential matrix 
 satisfy the hermiticity relation $V_{AB,CD} = V_{CD,AB}$ within statistical errors.
 In addition  the off-diagonal parts of $\Lambda \Lambda$ to $N \Xi$ transition $V_{\Lambda \Lambda,N \Xi}$ is much smaller than the other two off-diagonal potentials, $V_{\Lambda \Lambda,\Sigma \Sigma}$ and $V_{N \Xi,\Sigma \Sigma}$. 

In order to compare the results of the potential matrix calculated in three configuration sets, the potentials from the particle basis are transformed by the unitary rotation $U$ to those in the flavor SU(3) irreducible representation (IR) basis  as
\begin{eqnarray}
 V^{IR} = U^\dagger V U = 
 \left( \begin{array}{ccc}
 V_{1,1} & V_{1,8} & V_{1,27} \\
 V_{8,1} & V_{8,8} &V_{8,27} \\
 V_{27,1} & V_{27,8} & V_{27,27} \\
 \end{array} \right) .
\end{eqnarray}
The potential matrix in the IR basis give a good measure of  flavor SU(3) breaking effects since it is diagonal in the SU(3) symmetric limit.

In Fig.~\ref{FIG:potall}, the results of the potential matrix in the IR basis are compared among different configuration sets, together with the one  in the flavor SU(3) symmetric limit.
As the pion mass decreases, the attraction in $V_{1,1}$ potential increases in whole range.
While the $V_{8,8}$ potential in the flavor SU(3) limit deviates from others,   we do not observe a clear pion mass dependence of the potentials among these flavor breaking cases.
In the $V_{27,27}$ potential, we observe the growth of attraction range of potential and the enhancement of repulsive core.
The $V_{1,27}$ and $V_{8,27}$ transition potentials are consistent with zero within statistical errors.
On the other hand, it is noteworthy that the flavor SU(3) symmetry breaking effect becomes manifest in the $V_{1,8}$ transition potential.

\section{Three-Nucleon Forces}
\label{sec:3NF}

In this section, 
we expand the scope of our studies
from two-nucleon (2N) systems 
to $A$-body nucleon systems.
Generally speaking, 
there could exist not only 2-body forces 
but also $3, 4, \cdots A$-body forces
in such systems.
In particular, the determination of three-nucleon forces (3NF)
attracts a great deal of interest,
since it has been revealed that 
3NF play an important and nontrivial role 
in various phenomena.
Some examples include 
(a) binding energies of light nuclei~\cite{Pieper:2007ax},
(b) deuteron-proton elastic scattering experiments~\cite{Sekiguchi:2011ku},
(c) the anomaly in the oxygen isotopes near the neutron drip-line~\cite{Otsuka:2009cs}, and 
(d) the nuclear equation of state (EoS)
  at high density  relevant to the physics of neutron stars~\cite{Akmal:1998cf}.
  Universal short-range repulsion for three baryons (nucleons and hyperons) 
  has also been suggested in relation to the maximum mass of 
   neutron stars with hyperon core~\cite{Nishizaki:2002ih, Takatsuka:2008zz}.

Despite of its phenomenological importance,
microscopic understanding  of 3NF is  still limited.
Pioneered by Fujita and Miyazawa~\cite{Fujita:1957zz},
the long range part of 3NF has been modeled
 by the two-pion exchange (2$\pi$E)~\cite{Coon:2001pv},
particularly with the $\Delta$-resonance excitation.
This 2$\pi$E-3NF component is known to have 
an attractive nature at long distance.
An additional repulsive component of 3NF at short distance is 
often introduced in a purely phenomenological way~\cite{Pieper:2001ap}.
 An approach based on  the chiral effective field theory (EFT)
 is quite useful to   classify 
 the  two-, three- and more-nucleon forces and
has been studied intensively~\cite{Weinberg:1992yk}.
.
A completely different approach 
based on holographic QCD is 
recently proposed, which obtains repulsive 3NF at short distance~\cite{Hashimoto:2010ue}.

To go beyond phenomenology,
it is most desirable to determine 3NF directly
from the fundamental degrees of freedom (DoF), the quarks and the gluons,
on the basis of quantum chromodynamics (QCD).
In this section, 
we present an exploratory study of 
first-principle lattice QCD calculation of 3NF
in the quantum numbers of $(I, J^P)=(1/2,1/2^+)$ (the triton channel).
For details of this study, see Ref.~\citen{Doi:2011gq}. 

\subsection{Formalism}
\label{subsec:3NF_formalism}

We consider 
the NBS wave function $\psi_{3N}(\br,\brho)$ 
extracted from the six-point correlator as
\begin{eqnarray}
\label{eq:6pt_3N}
G_{3N} (\br,\brho,t-t_0) 
&\equiv& 
\frac{1}{L^3}
\sum_{\bR}
\langle 0 |
          (N(\bx_1) N(\bx_2) N (\bx_3))(t) \
\overline{(N'       N'        N')}(t_0)
| 0 \rangle ,\ \ \ \\
&\xrightarrow[t \gg t_0]{}& A_{3N} \psi_{3N} (\br,\brho) e^{-E_{3N}(t-t_0)} ,
A_{3N} = \langle E_{3N} | \overline{(N' N' N')}(0) | 0 \rangle ,  
\nonumber
\\
\label{eq:NBS_3N}
\psi_{3N}(\br,\brho) &\equiv& 
\langle 0 | N(\bx_1) N(\bx_2) N(\bx_3)(0) | E_{3N}\rangle ,
\end{eqnarray}
where
$E_{3N}$ and $|E_{3N}\rangle$ denote
the energy and the state vector of the 3N ground state, respectively, 
$N$ ($N'$) the nucleon operator in the sink (source),
and
$\bR \equiv ( \bx_1 + \bx_2 + \bx_3 )/3$,
$\br \equiv \bx_1 - \bx_2$, 
$\brho \equiv \bx_3 - (\bx_1 + \bx_2)/2$
the Jacobi coordinates.
We consider the following 
Schr\"odinger equation of the 3N system 
with the derivative expansion of the potentials,
\begin{eqnarray}
\biggl[ 
- \frac{1}{2\mu_r} \nabla^2_{r} - \frac{1}{2\mu_\rho} \nabla^2_{\rho} 
+ \sum_{i<j} V_{2N} (\br_{ij})
+ V_{3NF} (\br, \brho)
\biggr] \psi_{3N}(\br, \brho)
= E_{3N} \psi_{3N}(\br, \brho) , \nonumber \\
\label{eq:Sch_3N}
\end{eqnarray}
where
$V_{2N}(\br_{ij})$ with $\br_{ij} \equiv \bx_i - \bx_j$
denotes the 2NF between $(i,j)$-pair,
$V_{3NF}(\br,\brho)$ the 3NF,
$\mu_r = m_N/2$, $\mu_\rho = 2m_N/3$ the reduced masses.
3NF can be determined as follows.
We first calculate 
$\psi_{3N}(\br, \brho)$ 
and obtain the total potential of the 3N system 
through Eq.~(\ref{eq:Sch_3N}).
Once we obtain all necessary $V_{2N}(\br_{ij})$
by performing (separate) lattice simulations
for genuine 2N systems,
we can extract $V_{3NF}(\br,\brho)$
by subtracting $\sum_{i<j} V_{2N}(\br_{ij})$ from the total potential. 
The extension to four- and more-nucleon forces can be immediately understood.
Note that potentials determined in this way reproduce the 
energy of the system by construction.

An important remark is that 3NF are always determined in combination with
2NF, and 3NF alone do not make too much sense.
Therefore a comparison between lattice 3NF and phenomenological 3NF can be done 
only at a qualitative level.
Rather, our purpose is to determine two-, three-, (more-) nucleon forces 
systematically, and to provide them as a consistent set.

One of the difficulties in the 3NF study from lattice QCD
is that computational costs 
become exceptionally enormous.
Since there are 9 valence quarks, 
the DoF of color and spinor are significantly enlarged.
In addition,
the number of diagrams in the Wick contraction
tends to diverge with a factor of $N_u ! \times N_d !$,
where $N_u$ ($N_d$) are numbers of up (down) quarks in the system.
We here develop several techniques to reduce these computational costs.
We first take an advantage of symmetries
(such as isospin symmetry) to reduce the number of Wick contractions.
Second,
we utilize a freedom for the choice of a nucleon interpolating operator.
In particular, a potential is independent of the choice of a nucleon operator at the source, $N'$,
and 
we employ the non-relativistic operator as $N'=N_{nr} \equiv \epsilon_{abc}(q_a^T C \gamma_5 P_{nr} q_b) P_{nr} q_c$
with $P_{nr} = (1+\gamma_4)/2$,
which reduces the spinor DoF.
Similar techniques are
(independently) developed in Ref.~\citen{Yamazaki:2009ua}.
On the other hand, 
a potential  depends on the choice of a nucleon operator at the sink, $N$.
As discussed in Sec.~\ref{sec:remark},
choosing $N$ corresponds to choosing the ``scheme''
to calculate nuclear forces.
Note that physical observables calculated from these different potentials 
such as phase shifts and binding energies are unique.
In order to determine 3NF and 2NF in the same ``scheme'',
we employ the same sink operator 
$N=N_{std} \equiv \epsilon_{abc}(q_a^T C \gamma_5 q_b) q_c$ 
in the 3NF study,
as employed in 2NF.
Recall that the choice of $N=N_{std}$ is 
shown to have good convergence in the derivative expansion in Sec.~\ref{sec:convergence},
and can be considered to be a good ``scheme'' for lattice nuclear forces.

We next consider the geometry of the 3N.
Since the spacial DoF for general ($\br,\brho$) is too large,
it is necessary to find an efficient way to restrict the geometry.
In this exploratory study, 
we propose to use the ``linear setup''with $\brho={\bf 0}$,
with which 3N are aligned linearly with equal spacings of 
$r_2 \equiv |\br|/2$.
In this setup,
the third nucleon is attached
to $(1,2)$-nucleon pair with only S-wave.
Considering the total 3N quantum numbers of 
$(I, J^P)=(1/2,1/2^+)$,
the triton channel, 
the wave function can be completely spanned by
only three bases, which can be labeled
by the quantum numbers of $(1,2)$-pair as
$^1S_0$, $^3S_1$, $^3D_1$.
Therefore, the Schr\"odinger equation
leads to 
the $3\times 3$ coupled channel equations
with the bases of 
$\psi_{^1S_0}$, $\psi_{^3S_1}$, $\psi_{^3D_1}$.
The reduction of the dimension of bases 
is expected to improve the S/N as well.
It is worth mentioning that
considering the linear setup is not an approximation:
Among various geometric components of 
the wave function of the ground state, we calculate the (exact) linear setup component
as
a convenient choice to study 3NF.
While we can access only a part of 3NF from it,
we plan to extend the calculation to more general geometries
step by step,
toward the complete determination of the full 3NF.

Finally,
we emphasize that 
3NF study requires the precise determination of 2NF.
This is not surprising because the interactions in 3N systems are
mostly dominated by 2NF, and thus small uncertainties in 2NF could 
easily obscure the signal of 3NF.
Note that we generally need precise 2NF in both of parity-even and parity-odd channels,
since a 2N-pair inside the 3N system 
could be either of positive or negative parity.
On this point, we have shown that 2NF in parity-even channel can be determined 
with good precision in lattice QCD.
On the other hand, the determination of 2NF in parity-odd channel is 
much more difficult.
While the formulation for parity-odd 2NF is developed in Sec.~\ref{eq:parity_odd},
the results are found to suffer from larger statistical errors than parity-even 2NF.
This is considered to be a general tendency,
since one has to inject a non-zero momentum 
in the parity-odd 2NF study.
Therefore, in the 3NF study, 
it is essential to suppress the uncertainties 
originated from parity-odd 2NF.

In order to address this issue,
we propose to consider the following channel~\cite{Doi:2011gq},
\begin{eqnarray}
\psi_S \equiv
\frac{1}{\sqrt{6}}
\Big[
-   \Pu \Nu \Nd + \Pu \Nd \Nu               
                - \Nu \Nd \Pu + \Nd \Nu \Pu 
+   \Nu \Pu \Nd               - \Nd \Pu \Nu
\Big]  ,
\label{eq:psi_S}
\end{eqnarray}
which is anti-symmetric
in spin/isospin spaces 
for any 2N-pair.
Combined with the Pauli-principle,
it is automatically guaranteed that
any 2N-pair couples with even parity only.
Therefore, 
parity-odd 2NF vanish
in $\langle \psi_S | H | \psi_{3N} \rangle$,
where $H$ is the Hamiltonian of the 3N system,
and we can extract 3NF unambiguously 
without referring to parity-odd 2NF.
Note that no assumption on the choice of 3D-geometry of $\br$, $\brho$
is imposed in this argument,
and we can take an advantage of this feature
for future 3NF calculations with various 3N geometries.

\subsection{Numerical results}
\label{subsec:3NF_latcalc}

\begin{figure}[t]
\begin{minipage}{0.48\textwidth}
\begin{center}
\includegraphics[width=0.95\textwidth]{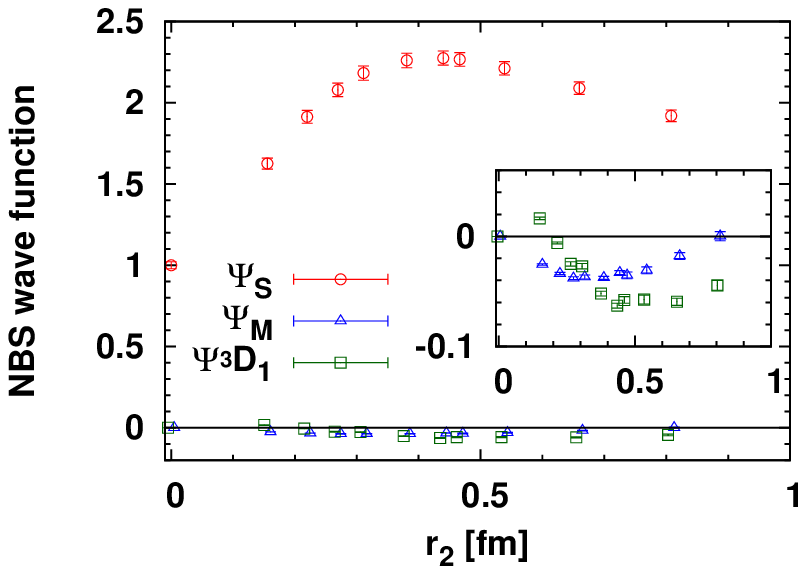}
\caption{
\label{fig:wf}
3N wave functions 
at $(t-t_0)/a=8$.
Circle (red), triangle (blue), square (green) points denote
$\psi_S$, $\psi_M$, $\psi_{\,^3\!D_1}$, respectively.
}
\end{center}
\end{minipage}
\hfill
\begin{minipage}{0.48\textwidth}
\begin{center}
\includegraphics[width=0.95\textwidth]{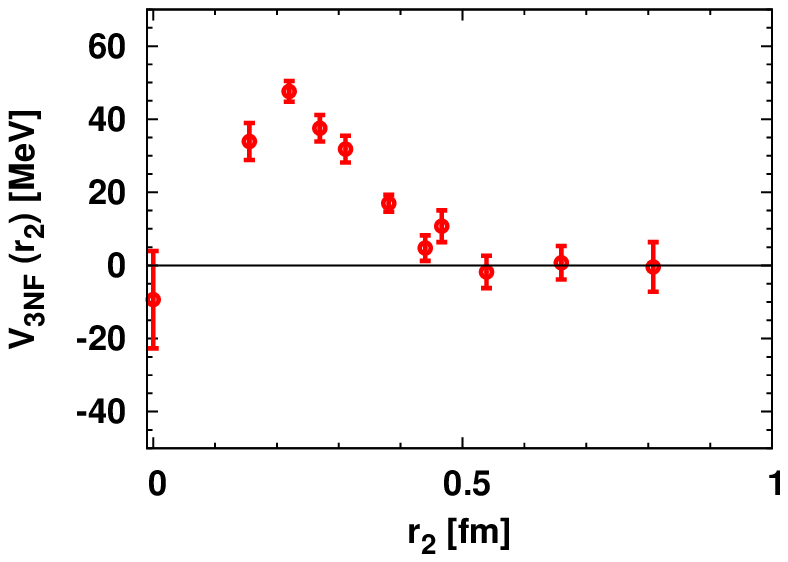}
\caption{
\label{fig:TNR}
The effective scalar-isoscalar 3NF 
in the triton channel with the linear setup
obtained 
at $(t-t_0)/a = 8$.
$r_2$ 
is
the distance between the center and edge
in the linear setup.
}
\end{center}
\end{minipage}
\end{figure}

We employ
$N_f=2$ dynamical 
configurations
with mean field improved clover fermion 
and 
RG-improved
gauge action
generated by CP-PACS Collaboration~\cite{AliKhan:2001tx}.
We use
598 configurations at
$\beta=1.95$ and
the lattice spacing of
$a^{-1} = 1.269(14)$ GeV,
and 
the lattice size of $V = L^3 \times T = 16^3\times 32$
corresponds to
(2.5 fm)$^3$ box in physical spacial size.
For $u$, $d$ quark masses, 
we take the hopping parameter at the unitary point
as
$\kappa_{ud} = 0.13750$,
which corresponds to
$m_\pi = 1.13$ GeV, 
$m_N = 2.15$ GeV and
$m_\Delta = 2.31$ GeV.
We use the wall quark source with Coulomb gauge fixing,
and periodic (Dirichlet) boundary condition is imposed
in spacial (temporal) direction.
In order to enhance the statistics,
we perform the measurement 
at 32 source time slices for each configuration,
and the forward and backward propagations are averaged.
The results from both of 
total angular momentum $J_z=\pm 1/2$ are averaged as well.
Due to the enormous computational cost,
we can perform the simulations only at a few sink time slices.
Looking for the range of sink time where the ground state saturation is achieved,
we carry out preparatory simulations for effective 2NF in the 3N system~\cite{Doi:2011gq}
in the triton channel at $2 \leq (t-t_0)/a \leq 11$, 
and find that the results 
are consistent with each other as long as $(t-t_0)/a \geq 7$~\cite{Doi:2011gq}.
Being on the safer side,
we perform linear setup calculations of 3NF at $(t-t_0)/a =$ 8 and 9.
We perform the simulation
at eleven values of the spacial distance $r_2$. 

In Fig.~\ref{fig:wf},
we plot
the radial part of each wave function of
$\psi_S = ( - \psi_{^1S_0} + \psi_{^3S_1} )/\sqrt{2}$,
$\psi_M \equiv ( \psi_{^1S_0} + \psi_{^3S_1} )/\sqrt{2}$
and
$\psi_{^3D_1}$ 
obtained at $(t-t_0)/a = 8$.
Here, we normalize the wave functions
by the central value of $\psi_S(r_2=0)$.
What is noteworthy is that
the wave functions are obtained with good precision,
which is quite nontrivial for the 3N system.
We observe that 
$\psi_S$ overwhelms the wave function,
and $\psi_M$, $\psi_{\,^3\!D_1}$ are much smaller
by one to two orders of magnitude.
This indicates that 
higher partial wave components in $\psi_S$
are also strongly suppressed,
and 
the wave function
is completely dominated by 
the component with which all three nucleons are in S-wave
in this lattice setup.

We determine 3NF
by subtracting 2NF from total potentials in the 3N system.
As discussed before,
we have only one channel, 
$\langle \psi_S | H | \psi_{3N} \rangle
=
\langle \psi_S | H | \psi_S \rangle
+ \langle \psi_S | H | \psi_M \rangle
+ \langle \psi_S | H | \psi_{^3D_1} \rangle
$,
which is free from parity-odd 2NF.
Correspondingly, 
we can determine 
one type of spin/isospin functional form for 3NF. 
In this study,
3NF are
effectively represented in 
a scalar-isoscalar functional form.
This form is an efficient representation,
since $\psi_S$ overwhelms the wave function and thus
$|\langle \psi_S | V_{3NF} | \psi_S \rangle|
\gg |\langle \psi_S | V_{3NF} | \psi_M \rangle|, |\langle \psi_S | V_{3NF} | \psi_{^3D_1} \rangle|$
is expected.
Note also that the scalar-isoscalar functional form is 
often employed for the 
short-range part of 3NF in phenomenological models~\cite{Pieper:2001ap}.

In Fig.~\ref{fig:TNR}, we plot the results
for the effective scalar-isoscalar 3NF at $(t-t_0)/a = 8$.
We here include $r_2$-independent shift by energy,
$\delta_E \simeq 5$~MeV, 
which is determined by 
long-range behavior of potentials
~\cite{Doi:2011gq}.
While $\delta_E$ suffers from $\simleq 10$ MeV systematic error,
it does not affect the following discussions much, since $\delta_E$ merely serves as an overall offset.
In order to check the dependence on the sink time slice,
we compare 3NF from $(t-t_0)/a =$ 8 and 9. 
While the results with $(t-t_0)/a=9$ suffer from quite large errors, 
they are consistent with each other within statistical fluctuations.

Fig.~\ref{fig:TNR} shows that
3NF are small at the long distance region of $r_2$.
This is in accordance with the suppression
of 2$\pi$E-3NF for the heavy pion.
At short distance, 
however,
an indication of repulsive 3NF is observed.
Note that a repulsive short-range 3NF is phenomenologically required 
to explain the properties of high density matter.
Since multi-meson exchanges are strongly suppressed 
for the large quark mass, 
the origin of this short-range 3NF may be attributed to the 
quark and gluon dynamics directly.
In fact, we recall that the short-range repulsive (or attractive) cores
in the generalized two-baryon potentials 
 in the flavor SU(3) limit discussed in Sec.\ref{sec:su3limit}
 are consistent with  the Pauli exclusion principle in the quark level~\cite{Inoue:2010hs, Oka:2000wj}.
In this context, 
it is intuitive to expect that the 3N system is subject to extra Pauli repulsion effect,
which could be an origin of the observed short-range repulsive 3NF.
Further investigation along this line is certainly an interesting subject in future.

We remark here that the 
quark mass dependence of 3NF 
is certainly 
an important issue, 
since the lattice simulations are carried out 
only at single large quark mass.
In the case of 2NF,
short-range cores have the enhanced strength
and broaden range by decreasing the quark mass%
~\cite{Aoki:2009ji}.
We therefore would expect a significant quark mass dependence
exist in short-range 3NF as well.
In addition,
long-range 2$\pi$E-3NF will emerge 
at lighter quark masses, in particular, at the physical point.
Quantitative investigation through
lattice simulations with lighter quark masses
are currently underway.

\section{Meson-baryon interactions}
\label{sec:MBint}
Since our potential method can be naturally extended to meson-baryon systems as well,  we consider meson-baryon interactions in this section.
The first application is a study on kaon-nucleon ($KN$) interactions 
in the $I(J^P) =0(1/2^-)$ and $1(1/2^-)$ channels. 
The elastic $KN$ scattering allows us to study the origin of ``non-resonant'' nuclear forces,
since kaon contains $u \bar{s}$ quarks, and these quarks do not annihilate in non-strange nucleons.
Therefore, the direct productions of conventional baryon resonances are ruled out.
Also, the $KN$ systems in the $I(J^P) =0(1/2^-)$ and $1(1/2^-)$ channels 
may be relevant for a possible exotic state $\Theta^+$~\cite{Nakano:2003qx},  whose existence is still controversial.

It is important to emphasize that the one-pion exchange is absent in the $KN$ systems, so that
the short- and mid-range interactions dominate the elastic $KN$ scattering.
Theoretical studies of the $KN$ interactions so far have been carried out by constituent quark models
and meson-exchange models.
In both models, 
it was found that genuine quark-gluon dynamics become 
important to describe the empirical scattering phase shifts~\cite{Hashimoto:1984th}.

To investigate the $KN$ potentials in 2+1 flavor full QCD,
we have utilized the gauge configurations of 
JLDG(Japan Lattice Data Grid)/ILDG(International Lattice Data Grid)
generated by PACS-CS Collaboration on a $32^3 \times 64$ lattice~\cite{PACSCS}. 
The renormalization group improved Iwasaki gauge action
and non-perturbatively $O(a)$ improved Wilson quark action are used 
at $\beta=1.90$, which corresponds to the lattice spacing $a=0.09$ fm
determined from $\pi$, $K$ and $\Omega$ masses.
The physical size of the lattice is about (2.9 fm)$^3$ and the 
the hopping parameters are taken to be 
$\kappa_u=\kappa_d=0.1370$
and $\kappa_s=0.1364$.
In the present simulation, we adopt the wall source located at $t_0$ with
the Dirichlet boundary condition at time slice $t=t_0+32$ in the temporal direction and
the periodic boundary condition in each spatial direction.
The Coulomb gauge fixing is employed at $t=t_0$.
The number of gauge configurations used in the simulation is 399.
With this setup, we obtain $m_{\pi} = 705(2)$, $m_{K}= 793(2)$ and
$M_N= 1590(8)$ MeV\cite{Ikeda:2011qm}.

\begin{figure}[tb]
\begin{center}
\includegraphics[width=0.48\textwidth]{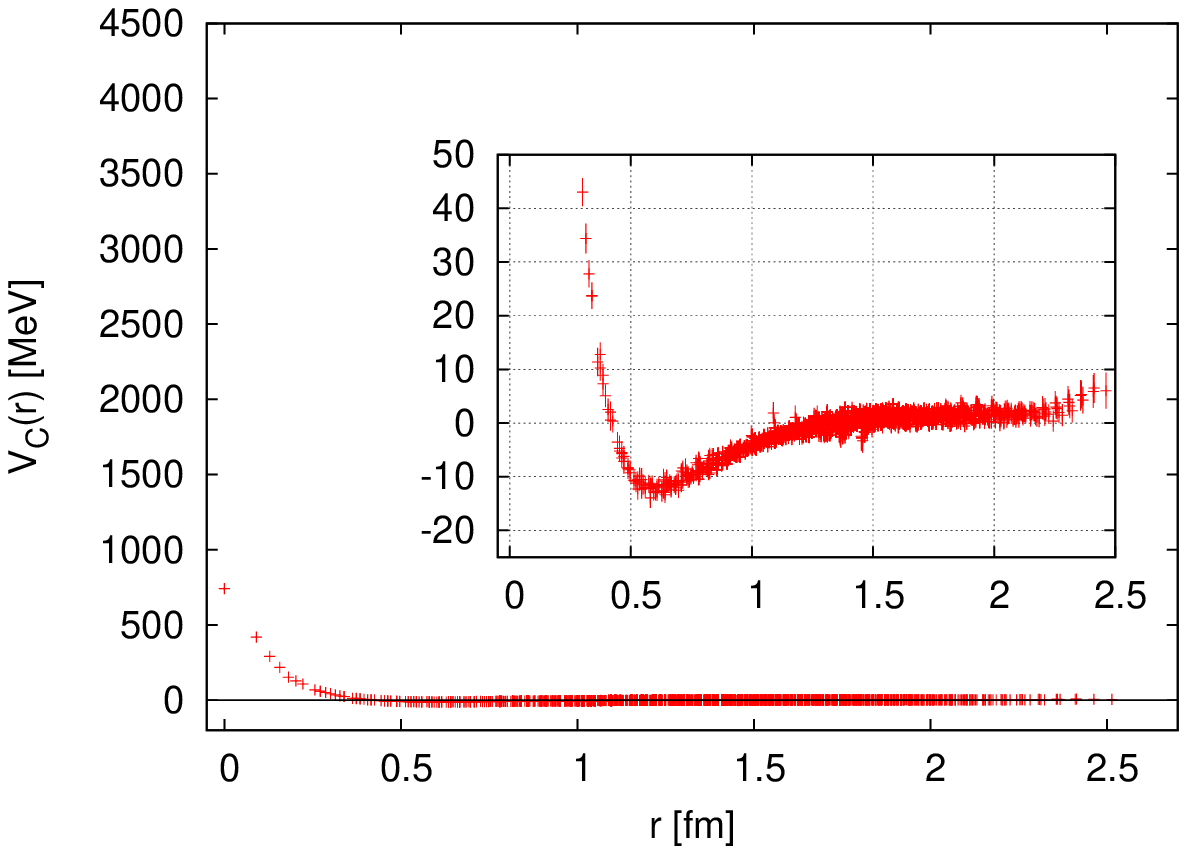}
\includegraphics[width=0.48\textwidth]{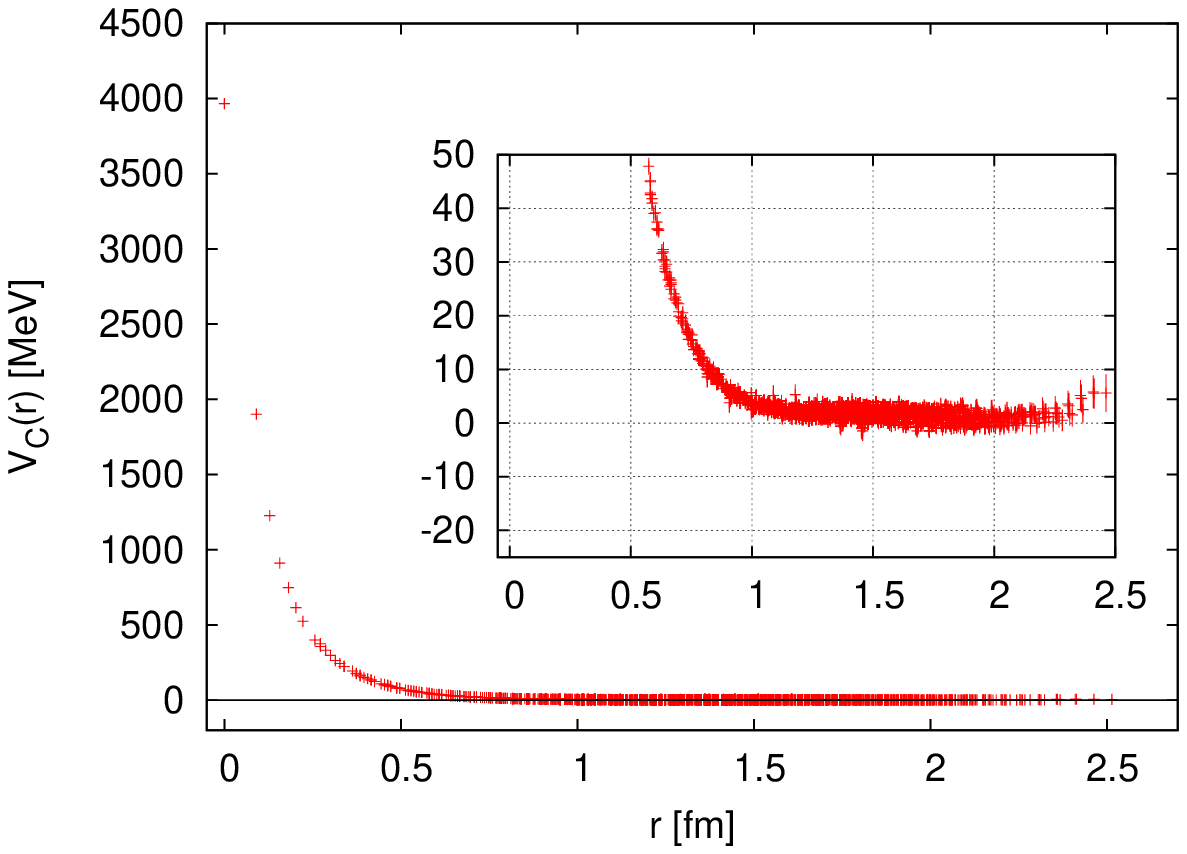}
\end{center}
\caption{The LO S-wave central potentials for the $KN$ states 
in the $I=0$ (left) and the $I=1$ channels (right). }
\label{fig:pot_NK}
\end{figure} 

Fig.~\ref{fig:pot_NK} shows the LO S-wave potential $V(r)$ for the $KN$ state 
in the $I=0$ (left) and $I=1$ (right).
The repulsive interactions are observed at short distance in both channels,
while the attractive well appears at the mid-distance ($0.4<r<0.8$ fm) in the $I=0$ channel,
which is not found in the constituent quark model of hadrons~\cite{Barnes:1992ca}.
These results indicate that there are no bound states
in $I(J^{\pi})=0(1/2^-)$ and $1(1/2^-)$ states at $m_{\pi} \simeq 705$ MeV.

The strong repulsions near origin in the $I=1$ channel can be expected
by the quark Pauli blocking effects.
This was first pointed out by Machida and Namiki~\cite{Machida:65} for the
meson-baryon systems.
In the $I=1$ $KN$ ($K^+ p$) state
whose configuration is 
$K^+ p \sim (u \bar s)\ (uud)$,
one of the $u$-quarks cannot be in the S-state.
The strong repulsion at short distance in this channel found in our simulations 
suggests a manifestation of the quark Pauli blocking.
In addition, the repulsive interactions in the S-wave $I=1$ $KN$ state can be expected
much stronger than that of the $I=0$ $KN$ state in the constituent quark model.
Our simulation shows that the repulsion at short distance for the $KN$ potential 
becomes significantly smaller in the $I=0$ channel than $I=1$ channel. 
This again confirms the expectation from  the quark Pauli blocking effects.

Kawanai and Sasaki investigate charmonium-nucleon ($c \bar{c}$-$N$) interactions 
using the potential method.
Since charmonia do not share the same quark flavor with the nucleon,
the $c \bar{c}$-$N$ interaction is mainly induced by the genuine QCD effect of 
multi-gluon exchanges and does not manifest the repulsive core near origin.
In Ref.~\citen{Kawanai:2010ev},
the charmonium-nucleon potentials are calculated in quenched QCD on $16^3\times 48$ and $32^3\times 48$ lattices 
at $a\simeq 0.94$ fm at three different values of the light quark mass corresponding to 
$(m_\pi, m_N)\simeq (640,1430), (720,1520), (870,1700)$ in unit of MeV 
and one fixed value of the charm quark mass corresponding to $m_{\eta_c} \simeq 2920$ MeV 
and $m_{J/\Psi} \simeq 3000$ MeV.
They have found that the effective central $c\bar{c}$-$N$ potentials clearly exhibit entirely attractive interactions 
without any repulsion at all distances.
Absence of the short range repulsion (the repulsive core) is related to  absence of the Pauli exclusion 
between the heavy quarkonium and the light hadron.

\section{Conclusion}
\label{sec:conclusion}

  In this report, we  reviewed the basic notion of the HAL QCD method for the 
  baryon-baryon ($BB$) potential and its field-theoretical 
  derivation from the  equal-time Nambu-Bethe-Salpeter (NBS) wave function in QCD. 
  The potential $U(\bx, \by)$ (or the integral kernel of the 
  Schr\"{o}dinger type equation) in the HAL QCD method has three characteristic
  features: (i)  non-local in relative coordinate, (ii)  energy independent,
  and (iii) scheme dependent. Each of these features has been discussed in detail
  in this report.    By construction, this potential  correctly reproduces the 
  scattering phase shift defined from the S-matrix in QCD below the inelastic threshold.
    
   One can construct $U(\bx, \by)$,  once all the 
  NBS wave functions for scattering energies below the inelastic
  threshold are obtained. In lattice QCD simulations in a finite box, however,
  it is more practical to adopt  the velocity (derivative) expansion of 
  $U(\bx, \by)$ by its non-locality and determine
  the local potentials $V(\bx)$ order by order. This is also in conformity  with 
  phenomenological potentials widely used in nuclear physics:
  An advantage of the HAL QCD method is that  one can check 
  the accuracy of this velocity expansion by changing the scattering energies on the lattice.
  To avoid the well-known problem of exponential error-growth
  in  the temporal correlation of multi hadrons, we have introduced a time-dependent
  HAL QCD method on the basis of the (imaginary) time Schr\"{o}dinger type equation.
  Due to this improved method, we could achieve better construction of the 
  potential as demonstrated in this report.
     
 The leading order (LO) terms of the velocity expansion correspond to the 
 central and tensor potentials:  Those for the  
  nucleon-nucleon ($NN$), hyperon-nucleon ($YN$) and
    hyperon-hyperon ($YY$) interactions have been investigated
 in full QCD simulations, some of which are recapitulated in this report. 
 The next-to-leading order (NLO) term is the spin-orbit potential: By introducing 
 finite momentum to the nucleons, we could extract the $NN$ spin-orbit force 
  for the first time. The origin of the 
 repulsive core of the $NN$ interaction
   has been also investigated by extending the  SU(2)-flavor to degenerate 
   SU(3)-flavor. The role of the Pauli principle
  in the quark level to describe the short range part of the interaction becomes
 clear. In particular, there arises a short range  ``attractive" core in the 
 flavor singlet channel; we found that 
  it is strong enough to form a bound state, $H$-dibaryon, in the SU(3) limit.

The HAL QCD method can be extended to the case beyond the inelastic threshold.
 This is necessary to treat the $YN$ and $YY$ interactions with SU(3)-flavor
 symmetry breaking. We have presented its application to $(S,I)=(-2,0)$ system
 and derived the 
  coupled channel potentials among $\Lambda\Lambda$, $N\Xi$ and $\Sigma\Sigma$.
 The HAL QCD method
 is also applied to the three-nucleon force relevant for the extra binding of 
 finite nuclei and also for the maximum mass of neutron star, and to 
  the  meson-baryon interactions relevant for the 
   meson-baryon resonances and the pentaquark. 
  
 So far, our full QCD simulations of the $BB$ interactions are performed at non-zero lattice spacing on a finite volume with relatively large quark masses.
  We therefore need careful studies of systematic errors on
  finite volume effect, 
quark mass dependence 
  and the lattice spacing effect.
  Among others, the most 
 important direction  is to carry out 
  (2+1)-flavor simulations on a large volume (e.g. $L=6-9$ fm) at
   physical quark mass ($m_{\pi}=135$ MeV)  to extract the 
    realistic $BB$ and $BBB$  potentials.  Such simulations  are
     planned  at 10 PFlops  ``K~computer''  
   in  Advanced Institute for Computational  Science (AICS), RIKEN.

    If it turns out that the program described in this paper indeed works in lattice QCD
     at the physical quark mass, it would be 
      a major  step toward the understanding 
   of atomic nuclei and neutron stars   from the fundamental law of the strong
   interaction, the quantum chromodynamics.

 \section*{Acknowledgement}
We thank CP-PACS, JLQCD and PACS-CS Collaborations and  ILDG/JLDG for providing us    
the gauge    configurations \cite{AliKhan:2001tx,CPPACS-JLQCD,Aoki:2008sm,Beckett:2009cb,ildg/jldg}.
We are grateful for the  authors and maintainers of CPS++ \cite{cps}, a
modified version of which is used for measurement done in this work.
The numerical simulations have been carried out on
Blue Gene/L at KEK,
T2K at University of Tsukuba and at University of Tokyo,
SR16000 at YITP in Kyoto University,
and SX9 and SX8R at RCNP in Osaka University.
This  research is  supported in  part by  Grant-in-Aid  for Scientific
Research  on  Innovative  Areas(No.2004:20105001,20105003)  and  for
Scientific   Research(C)  23540321, 24740146,  JSPS  21$\cdot$5985   and  SPIRE
(Strategic Program for Innovative REsearch).


\end{document}